\documentclass[12pt]{article}
\usepackage{latexsym}
\usepackage{epsfig,amssymb,euscript,slashed}
\usepackage{amsmath}
\usepackage{cite} 
\usepackage{array,calc,epsfig}
\usepackage[linktocpage,hidelinks]{hyperref}
\usepackage{pstricks}
\usepackage[normalem]{ulem} 

\usepackage{bbm}
\usepackage{xcolor}
\usepackage{fancybox}

\def\be{\begin{equation}}
\def\ee{\end{equation}}
\def\bseq{\begin{subequations}}
\def\eseq{\end{subequations}}

\def\bea{\begin{eqnarray}}
\def\eea{\end{eqnarray}}

\def\bseq{\begin{subequations}}
\def\eseq{\end{subequations}}

\numberwithin{equation}{section} 
\usepackage[]{graphicx}

\def\d {{\rm d}}

\def\cala         {{\cal A}}

\def\calb         {{\cal B}}
\def\calc         {{\cal C}}
\def\cald         {{\cal D}}
\def\cale         {{\cal E}}
\def\calf         {{\cal F}}
\def\calg         {{\cal G}}

\def\calh         {{\cal H}}
\def\cali         {{\cal I}}

\def\calk         {{\cal K}}
\def\call         {{\cal L}}
\def\calm         {{\cal M}}
\def\caln         {{\cal N}}
\def\calo         {{\cal O}}
\def\calp         {{\cal P}}
\def\calq         {{\cal Q}}
\def\calr         {{\cal R}}
\def\cals         {{\cal S}}
\def\calt         {{\cal T}}

\def\calv         {{\cal V}}
\def\calw         {{\cal W}}
\def\calz         {{\cal Z}}

\def\del          {\partial}

\def\ii           {{\rm i}}

\def\Re           {{\rm Re\hskip0.1em}}
\def\Im           {{\rm Im\hskip0.1em}}

\def\sqr#1#2{{\vcenter{\vbox{\hrule height.#2pt
 \hbox{\vrule width.#2pt height#1pt \kern#1pt \vrule width.#2pt}\hrule
 height.#2pt}}}}

\def\d{\text{d}}

\def\slashchar#1{\setbox0=\hbox{$#1$}           
\dimen0=\wd0                                 
\setbox1=\hbox{/} \dimen1=\wd1               
\ifdim\dimen0>\dimen1                        
\rlap{\hbox to \dimen0{\hfil/\hfil}}      
#1                                        
\else                                        
\rlap{\hbox to \dimen1{\hfil$#1$\hfil}}   
/                                         
\fi}

\begin{document}

\font\cmss=cmss10 \font\cmsss=cmss10 at 7pt

\thispagestyle{empty}
\rightline{\scriptsize{IFT-UAM/CSIC-19-102}}
\vspace{0.5cm}
\begin{center}
\Huge{{{How many fluxes fit in an EFT?}
\\[15mm]}
\large{Stefano Lanza,$^1$ Fernando Marchesano,$^2$\\  Luca Martucci,$^1$ and Dmitri Sorokin$^1$ \\[10mm]}
\small{
${}^1$ Dipartimento di Fisica e Astronomia ``Galileo Galilei",  Universit\`a degli Studi di Padova \\ 
\& I.N.F.N. Sezione di Padova, Via F. Marzolo 8, 35131 Padova, Italy} \\[2mm] 
${}^2$ Instituto de F\'{\i}sica Te\'orica UAM-CSIC, Cantoblanco, 28049 Madrid, Spain 
\\[8mm]} 
\small{\bf Abstract} \\[5mm]
\end{center}
\begin{center}
\begin{minipage}[h]{15.0cm} 

We extend the recent construction of 4d $\caln=1$ three-form Lagrangians by including the most general three-form multiplets necessary to reproduce any F-term potential in string flux compactifications. In this context we find an obstruction to dualize all fluxes to three-forms in the effective field theory. This implies that, generically, a single EFT cannot capture all the membrane-mediated flux transitions expected from a string theory construction, but only a sublattice of them. The obstruction can be detected from the maximal number of three-forms per scalar in any supermultiplet, and from the gaugings involving three-forms that appear in the EFT. Some gaugings are related to the appearance of fluxes in the tadpole conditions, and give a general obstruction. Others are related to the anomalous axionic strings present in a specific compactification regime. We illustrate the structure of the three-form Lagrangian in type II and F/M-theory setups, where we argue that the above obstructions correlate with the different 4d membrane tensions with respect to the EFT energy scales.

\end{minipage}
\end{center}
\newpage


\vspace{1cm}


\thispagestyle{empty}

\newpage

\setcounter{footnote}{0}

\setcounter{page}{1}

\tableofcontents


\section{Introduction}

In the past few years, there has been a renewed interest in the conditions that quantum field theories need to satisfy in order to be embedded into a fully-fledged theory of quantum gravity, a line of research also known as the Swampland Program \cite{Vafa:2005ui} (see \cite{Brennan:2017rbf,Palti:2019pca} for reviews).  Progress on this front is oftentimes achieved by testing different conjectures on Effective Field Theories (EFTs) obtained from string theory compactifications, or by proposing new conjectures based on their general features. 

By definition, the Swampland Program is directly related to a good understanding of the String Landscape, seen as the set of meta-stable vacua that one can obtain from string compactifications. More importantly, it should determine the subset of vacua that can be seen as arising from the same EFT. In this respect, the Swampland Distance Conjecture \cite{Ooguri:2006in} and refinements \cite{Baume:2016psm,Klaewer:2016kiy} partially address this question, selecting a region of the space of solutions based on field space distances. Nevertheless, most of the results along these lines rely on constructions with at least eight supercharges, in which the vacua are continuously connected in a moduli space of solutions. The case of 4d with minimal or no supersymmetry, in which a potential is generated for the scalars of the compactification and the different vacua are typically isolated from each other, is on the other hand less understood. 

In this context, a particularly relevant class of vacua is the ensemble obtained from compactifications with background fluxes \cite{Douglas:2006es,Denef:2007pq}, from where we draw a great deal of our intuition about several corners of the String Landscape. In this case a large set of isolated vacua can be obtained from the same string theory construction, by simply scanning through a lattice of quantized fluxes. It is however not clear whether one can capture this whole ensemble of vacua in terms of a single EFT. In general, one would expect that this problem is easier to address for 4d compactifications with an underlying $\caln=1$ structure, due to the better control that we have over them. Nevertheless, the traditional formulations of 4d $\caln=1$ supergravities, where the flux quanta appear as fixed parameters of the superpotential and scalar potential, do not seem particularly suitable to answer this question. Indeed, a 4d EFT capturing a flux ensemble should describe a scalar potential with a multi-branched structure, with each  branch corresponding to a different set of flux quanta, and the possibility of jumping from one to another through membrane transitions.

Recently, it has been realized that such features are naturally incorporated in 4d EFTs that include three-form potentials, with each of them corresponding to a different internal flux. In particular, Lagrangians containing three-form potentials allow for a description of the multi-branched structure of flux-induced potentials in type II string compactifications (see e.g. \cite{Marchesano:2014mla,Bielleman:2015ina,Carta:2016ynn,Herraez:2018vae}) and of their relation to the discrete shift symmetries of the compactification. Triggered by this fact, substantial progress has been made in formulating EFTs that incorporate such non-propagating three-form potentials in $\caln=1$ supergravity multiplets \cite{Farakos:2017jme,Bandos:2018gjp}, unveiling a rich structure that allows for a `dynamical' description of flux quanta, and in particular membrane-mediated transitions between different flux sectors. 

One of the purposes of this paper is to generalize the analysis of \cite{Farakos:2017jme,Bandos:2018gjp} by including more general three-form multiplets, such that all flux-generated potentials known in the string theory literature can be captured by means of an EFT Lagrangian. In the standard formulation of 4d supergravity, such potentials arise from a superpotential that is given by a set of integers (the flux quanta) multiplying a set of holomorphic sections in field space (the periods), as in the archetypal example of \cite{Gukov:1999ya}. The more general class of multiplets, which we dub {\em master three-form multiplets}, are essentially defined in terms of the periods corresponding to each of the fluxes of the EFT, and provide a formalism overarching previous examples of supersymmetric three-form Lagrangians. 

As it turns out, within this more general scheme certain limitations of  supersymmetric three-form EFTs are exposed, more precisely the inability to treat as dynamical the whole set of fluxes that we typically associate to a given string theory compactification. Such an obstruction can be detected in a number of ways, like for instance from an upper bound on the amount of three-forms in the EFT in terms of the number of  fields that enter the flux-induced superpotential. 

\begin{figure}[!ht]
    \centering
	\includegraphics[width=12cm]{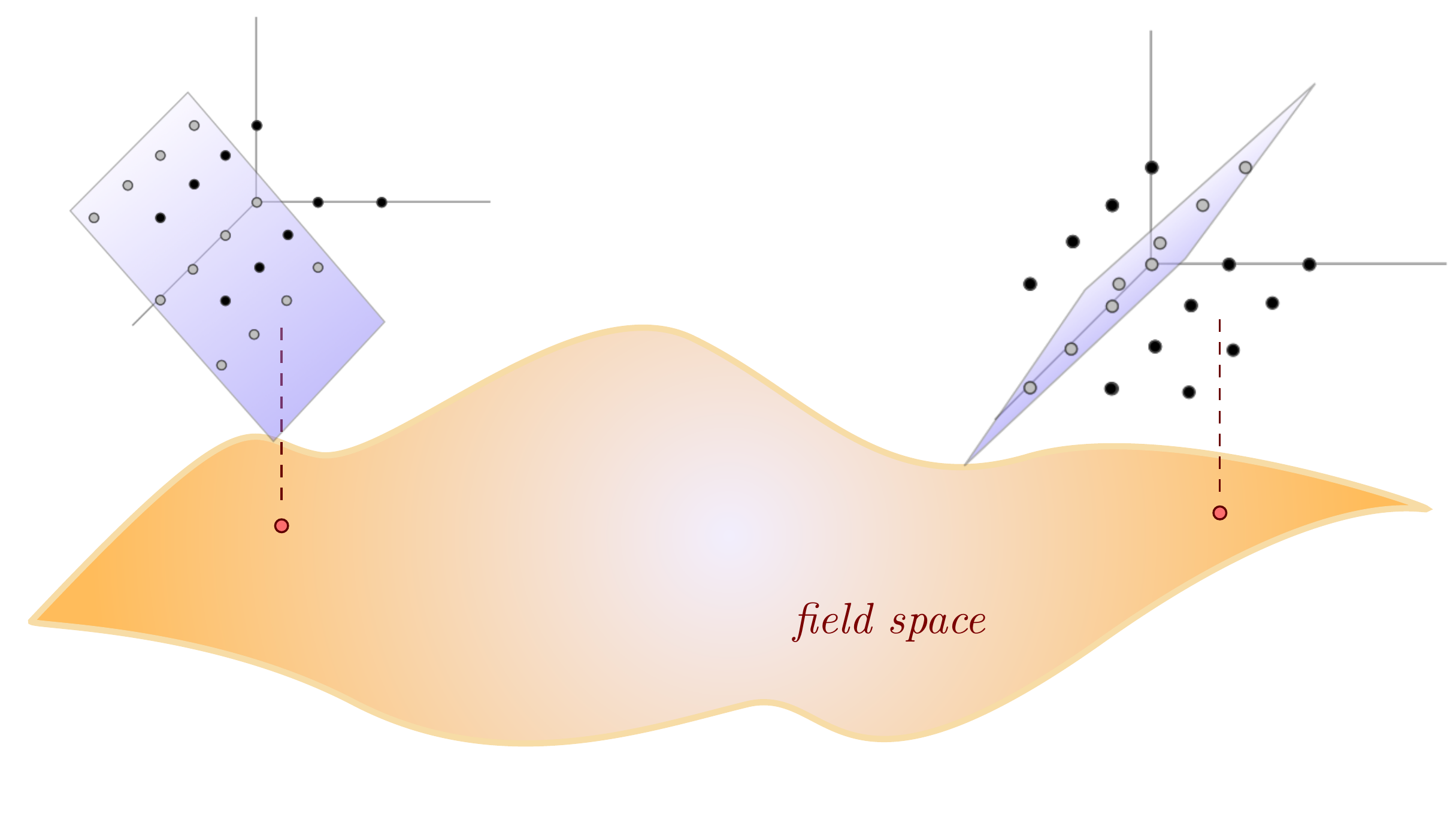}
	 \caption{\footnotesize{EFT flux lattices at different regions in the field space of a string theory construction.} }\label{fig:EFTLatt}
\end{figure}

A similar, a priori unrelated restriction is obtained by requiring the compatibility of the three-form Lagrangian with the gaugings of $p$-forms by $(p+1)$-forms that appear in the 4d EFT, and that generalize the well-known St\"uckelberg mechanism. In short, those fluxes that appear as gauging parameters cannot be treated as dynamical by the EFT. Each of these $p$-form gaugings have their counterpart in terms of 4d extended objects ending on each other, as 3-branes ending on membranes, membranes ending on axionic strings, and axionic strings ending on particles. Therefore, the set of gaugings that one may consider in a certain EFT will for instance depend on the spectrum of anomalous axionic strings that it contains, which in turn depends on the approximate continuous shift symmetries that are developed in certain regions of field space in a given string theory setup. Finally, one can see that one class of gaugings, that of three-form potentials by four-forms, is the 4d EFT manifestation of the well-known tadpole conditions that ensure the consistency of the compactification, to which fluxes typically contribute quadratically. Within our framework,  this particular gauging represents a generic obstruction to incorporate all the fluxes of a given string theory construction, represented by a lattice $\Gamma$, as dynamical fluxes in the 4d EFT. A given EFT will only be able to capture variations in a sublattice of fluxes $\Gamma_{\rm EFT} \subset \Gamma$, while the remaining fluxes will be seen as fixed EFT data, specifying the gauging parameters. More precisely, $\Gamma_{\rm EFT}$ must be such that the tadpole condition appears at most linearly on this sublattice. These differences between the string theory and the EFT perspective are summarized in table \ref{stEFT} and figure \ref{fig:EFTLatt}.

\bigskip

\begin{table}[!ht]
\begin{center}
\begin{tabular}{|c||c|c|}
\hline
& {\bf String Theory} & {\bf EFT}  \\
\hline\hline
Flux lattice & $\Gamma$ & $\Gamma_{\rm EFT}$\\
\hline
Tadpole condition & Quadratic & Linear\\
\hline
\end{tabular}
\caption{\footnotesize Main difference between the string theory and the 4d EFT perspectives. For each region of field space in the  string theory construction a different sublattice $\Gamma_{\rm EFT} \subset \Gamma$ may be selected, but the linear dependence of the tadpole condition on the EFT fluxes applies to all of them. \label{stEFT}}
\end{center}
\end{table}

\bigskip

\bigskip

We provide different examples of $\Gamma$ and $\Gamma_{\rm EFT}$ in type II, M-theory and F-theory compactifications, and argue that the EFT limitations to treat all fluxes as dynamical matches our expectations of which domain wall transitions are consistent with an EFT with a certain cut-off $\Lambda$.  This distinction is particularly clear for the case of the weakly-coupled type II setups, in which $\Gamma_{\rm EFT} = \{ \text{R-R fluxes} \}$ and the NS-NS fluxes are treated as fixed parameters of the theory. Indeed, in these cases the weak coupling regime selects 4d membranes arising from D-branes wrapping internal cycles of the compact manifold as they are much lighter than those coming from e.g.\  NS5-branes. Again, different perturbative regimes in field space may select different sublattices $\Gamma_{\rm EFT}$, in agreement with the different hierarchy of scales that holds on each of them, see e.g. the recent analysis in \cite{Font:2019cxq}. In fact, one may extend this analysis to understand why even different choices of $\Gamma$ are considered in opposite regimes of the same class of constructions, like it is illustrated by the weak and strong coupling regimes of type IIA compactifications.

Needless to say, this limitation of a single 4d EFT to capture the full spectrum of flux vacua and the possible transition between them could  be understood as a swampland criterium. In this sense, a natural direction would be to combine it with further criteria, in order to sharpen the set of conditions that EFTs describing flux compactifications must obey. We briefly comment on some possibilities, leaving a more detailed analysis of this direction for future work. 

 The paper is organized as follows: In section \ref{sec:IIBgauging} we illustrate some of the main ideas and results of this work by means of a familiar example: type IIB flux compactifications with D3-branes. The formalism required to describe more general compactifications is developed in section \ref{sec:genEFT}, where a three-form Lagrangians is constructed for a general 4d $\caln=1$ supergravity EFT, focusing for simplicity on its bosonic sector. There we discuss, in particular, the relation between the flux sublattice $\Gamma_{\rm EFT}$ and the 4d gaugings involving three-form potentials. The presence of the latter directly affects the physics of 4d extended objects, as we analyze in section \ref{sec:sm3}. Section \ref{sec:IIBstringy} revisits the IIB setup from the viewpoint of the general formalism, which we also apply to analyze more general F-theory constructions. It also discusses how to understand the lattice $\Gamma_{\rm EFT}$ in terms of the tension of the different 4d membranes, compared to the given EFT energy scales. The same applies to type IIA compactifications with D6-branes and fluxes, as shown in section \ref{sec:IIAstringy}. Sections \ref{sec:susy} provides, by means of the superspace formalism, the fully supersymmetric formulation of the three-form Lagrangian with gaugings, with the corresponding three- and two-form potentials properly embedded into multiplets. Section \ref{sec:susy3} uses $\kappa$-symmetry  to couple such multiplets to 4d extended objects, building supersymmetric actions for them that are fixed by their charge. We finally draw our conclusions in section \ref{sec:conclu}.
 
 Several technical details are given in the appendices. Appendix \ref{app:TAB} provides a geometrical interpretation of the kinetic matrix of the three-form potentials. Appendix \ref{app:SWL} discusses the super-Weyl invariant formalism and the associated bosonic Lagrangians used throughout the paper. Appendix \ref{app:bdcontr} describes the structure of the boundary terms that appear in the manifestly supersymmetric actions of the main text.


\section{Fluxes and \texorpdfstring{$p$}{p}-form gaugings in string compactifications}
\label{sec:IIBgauging}

In any string theory compactification to four dimensions, there are several $p$-forms that appear in the 4d effective theory. Most of them arise from direct dimensional reduction of the set of $p$-forms  in the higher dimensional, supergravity description of the theory. Typically, special attention is given to the 4d 0-forms $a^\Sigma$, seen as axions, and to the 1-forms $A_1^\sigma$, seen as vector bosons specifying the 4d EFT gauge sector. Depending on the discrete, topological data of the compactification, these two may be related via the St\"uckelberg mechanism, so that they always appear in the combination
\be
\d a^\Sigma - k^\Sigma_\sigma A_1^\sigma\, ,
\label{Stu}
\ee
with $k^\Sigma_\sigma \in \mathbb{Z}$.  This in turn implies the invariance under the combined transformations $A_1^\sigma \rightarrow A_1^\sigma + \d\Lambda^\sigma$, $a^\Sigma \rightarrow a^\Sigma + k^\Sigma_\sigma \Lambda^\sigma$, that can be understood as a gauging of the 4d EFT.

More recently, it has been realized that the same kind of structure applies to higher 4d $p$-forms, and that the corresponding gaugings contain quite relevant information for the 4d EFT. In the following we would like to discuss the relevance of those gaugings involving 4d three-forms. As already pointed out in \cite{Kaloper:2008fb,Kaloper:2011jz,Marchesano:2014mla}, some of these gaugings specify the discrete shift symmetries of the scalar potential generated by fluxes. As we will now see, a different kind of gaugings provide the 4d description of the consistency conditions known as tadpole cancellation conditions. Moreover, taking both of these gaugings into account constrains non-trivially the description of $\caln=1$ EFTs with three-form potentials. We will first illustrate these ideas in a simple class of type IIB compactifications, and then develop the general formalism in the next section.

\subsection{Type IIB with O3-planes and fluxes}

Let us consider a (warped) compactification of type IIB string theory on an orientifolded Calabi-Yau background $X={\rm CY}_3/\calr$, threaded by background three-form fluxes $F_3$, $H_3$. For simplicity, we assume the orientifold involution $\calr$ to be such that only O3-planes and possibly D3-branes are present, avoiding the presence of O7-planes and D7-branes for now. 

The presence of three-form fluxes generates a scalar potential for the axio-dilaton and complex structure moduli, which is encoded in the well-known GVW superpotential \cite{Gukov:1999ya} 
\be\label{IIBGVW}
W = \frac{\pi}{\ell^5_{\rm s}}\int_X \Omega\wedge (F_3-\tau H_3)\, ,
\ee
with $\ell_{\rm s}=2\pi\sqrt{\alpha'}$ and $\Omega$ the holomorphic $(3,0)$-form on $X$. One may rewrite the above expression in terms of the periods of the $(3,0)$-form, that one may define as
\be
M_P^3\, \Pi^A(\phi)=\frac{\pi}{\ell^3_{\rm s}}\int_X \Omega\wedge \varpi^A
\ee
with $M_{\rm P}$ the 4d Planck mass and $\varpi^A\in H^3_-(X;\mathbb{Z})$, $A=1,\ldots, b_3^-$ a cohomology basis of closed integral three-forms. Using the freedom to change the normalization of $\Omega$ to fix one of the periods to unity, say $\Pi^1 = 1$, allows one to work with periods $\Pi^A$ that depend on the standard complex structure moduli $\phi^i$, $i=1,\ldots, h^{2,1}_-$ \cite{Candelas:1990pi}. In addition, the quantization conditions for the three-form fluxes imply that in cohomology
\be
F_3= \ell^2_{\rm s}\,m_A\, \varpi^A\quad,\quad H_3=\ell^2_{\rm s}\,h_A\,\varpi^A\,.
\ee
with $m_A,h_A\in\mathbb{Z}$. As a result, one obtains a superpotential of the form
\be
 W (\phi)=M_{\rm P}^3\left( m_A - \tau h_A\right) \Pi^A(\phi)\, ,
\label{IIBsupo}
\ee
where $\phi^\alpha = (\tau, \phi^i)$, $\alpha=0,\ldots,h^{2,1}_-$ collectively denotes the axio-dilaton and the complex structure moduli. Finally, the three-form flux quanta are not unconstrained, as they need to satisfy the  R-R tadpole condition 
\be\label{D3tad2}
\calq=\tilde\calq^{\rm bg} +\mu_{\rm D3}+I^{AB} m_Ah_B=0 \,,
\ee
that guarantees the cancellation of D3-brane charge. Here $\tilde\calq^{\rm bg}$ accounts for the negative O3-plane charge,  $\mu_{\rm D3}$ is the number of space-time filling D3-branes and $I^{AB}=-I^{BA}$ are the intersection numbers 
\be
I^{AB}\equiv \int_X \varpi^A\wedge \varpi^B\,.
\label{Inter}
\ee

In this setup, the presence of three-form fluxes generate a scalar potential for the axio-dilaton and complex structure moduli. As in \cite{GKP}, one may compute such scalar potential from the superpotential \eqref{IIBsupo}, applying the standard $\caln=1$ Cremmer et al. formula. Alternatively, one may attempt to describe the potential in terms of the non-propagating three-form potentials $A_3^A$ present in the 4d EFT, that would arise from the dimensional reduction of the six-form potentials $C_6$, $B_6$ dual to the more familiar $C_2$, $B_2$ \cite{DallAgata:1998ahf,Bergshoeff:2001pv}. The advantage of this second strategy is that it allows, within a single EFT,  for a systematic description of all the possible choices of flux quanta, of the different membrane-mediated transitions between 4d vacua and of the discrete shift symmetries and multi-branched structure of the flux-induced scalar potential \cite{Farakos:2017jme,Bandos:2018gjp}. There are however certain limitations to fully implement this approach, that are already manifest in the present setup.

Indeed, to see such limitations let us follow the general strategy in \cite{Farakos:2017jme,Bandos:2018gjp} and interpret the flux quanta $\caln_\cala = (m_A, h_A)$ as expectation values of zero-form ``field-strengths", which are constant because of the Bianchi identities $\d \caln_\cala=0$. Dualizing the $\caln_\cala$'s to four-form field-strengths $F^\cala_4=\d A^\cala_3$ via the standard procedure, one can relax the constraint $\d \caln_\cala=0$ and promote $\caln_\cala$ to {\em real} scalar fields $y_\cala$. The Bianchi identities are then imposed at the level of the equations of motion of a parent effective action which includes a term $\int \d y_\cala\wedge A^\cala_3$, where the $A^\cala_3$'s play the role of Lagrange multipliers. More precisely, focusing on the $\phi^\alpha$ sector, one starts with a parent action of the form 
\be\label{IIBmaster}
S=- \int \left(\frac12 M^2_{\rm P}\, K_{\alpha\bar{\beta}}\,\d\phi^\alpha\wedge *\d\bar{\phi}^{\bar{\beta}}+\frac12T^{\cala\calb}y_\cala y_\calb\,*1 +\d y_\cala\wedge A^\cala_3\right)
\ee
with\footnote{For computing \eqref{IIBTAB} we have used the no-scale structure of the remaining chiral sector (paramerizing K\"ahler moduli, $C_4$, $C_2$ and $B_2$ axions) which we do not explicitly keep track of.\label{ft:noscale}}  
\be\label{IIBTAB}
T^{\cala\calb}(\phi)=2M^4_{\rm P}\, e^{K}\Re\left(K^{\alpha\bar \beta}D_\alpha\Pi^\cala\bar D_{\bar\beta}\bar\Pi^\calb\right)\ , 
\ee
being the matrix that specifies the scalar potential in terms of the flux quanta $N_\cala = (m_A, h_A)$, with $\Pi^\cala = (\Pi^A, - \tau \Pi^A)$ being the periods corresponding to  R-R and NS-NS fluxes, respectively. In the above formulas we use the notation $D_\alpha\Pi^A\equiv \Pi^A_\alpha+K_\alpha \Pi^A$, with  $\Pi^A_\alpha\equiv\partial_\alpha\Pi^A$ and  $K_\alpha\equiv\partial_\alpha K$, and $K^{\alpha\bar\beta}$ is the inverse of $K_{\alpha\bar{\beta}}\equiv \partial_\alpha\partial_{\bar\beta}K$. 

On the one hand, by integrating out the three-forms $A^\cala_3$, one obtains that $\d y_\cala=0\Rightarrow y_\cala=N_\cala$ and recovers an $\caln=1$ Lagrangian with potential $V = \frac12 T^{\cala\calb}N_\cala N_\calb$. Notice that $y_\cala$ can be interpreted as momenta conjugated to $A^\cala_3$ and then, in the quantum theory, get quantized values. On the other hand, one may integrate out the scalars $y_\cala$ and get
\be\label{*F}
*\!F^\cala_4= T^{\cala\calb}(\phi)y_\calb\, ,
\ee
hence, assuming that $T^{\cala\calb}(\phi)$ is invertible, one would obtain
\be\label{IIB3forml}
S=-\int_\calm\left(\frac12 M^2_{\rm P}\, K_{\alpha\bar{\beta}}\,\d\phi^\alpha\wedge *\d\bar{\phi}^{\bar{\beta}} +\frac12 T_{\cala\calb}\,F^\cala_4*\!F^\calb_4\right) -\int_{\del\calm}T_{\cala\calb} A^\cala_3 *\!F^\calb_4\, ,
\ee
namely, a dual Lagrangian in terms of complex scalar fields and three-form potentials. 

It however turns out that in the case at hand the matrix $T^{\cala\calb}(\phi)$ is not invertible for any value of the moduli, as can be seen from the fact that the matrix $D_\alpha\Pi^\cala$ has complex rank $\frac12b_3^-$ and so the real rank of $T^{\cala\calb}$ is $b_3^-$. Notice that $b_3^-$ is also the number of real scalars that are involved in the flux-induced scalar potential. Therefore, one may naively interpret this obstruction as having a compactification with too many fluxes, as compared to the number of scalars affected by them. As we will see in sections \ref{sec:genEFT} and \ref{sec:susy} this naive intuition is sustained in compactifications that correspond to supersymmetric EFTs like this one because, for a given number of  scalars,  the structure of supermultiplets implies a maximal  number of three-forms.

Based on this observation, one may attempt to solve the above problem by reducing the number of scalars $y^\cala$ in \eqref{IIBmaster} or, in other words, by reducing the number of fluxes whose vacuum expectation value can be understood dynamically. More precisely, the above discussion suggests that one should reduce the number of dynamical fluxes by half, and by inspection of the matrix $T^{\cala\calb}$ one deduces that one possibility would be to retain the R-R (or the NS-NS) three-form fluxes as dynamical. Rather than attempting either possibility, we will turn to discuss an independent set of constraints restricting the set of dynamical fluxes in the 4d effective field theory, namely the gaugings of the different $p$-forms present in it. We will pay special attention to those gaugings related to the implementation of the tadpole consistency condition from a 4d viewpoint, which we now turn to describe.

\subsection{Tadpole conditions as three-form gaugings}\label{tadpole1}

In any 4d EFT description in which background fluxes are allowed to vary, there must be a constraint implementing the consistency conditions that depend on them, such as the tadpole consistency conditions. In the class of type IIB compactifications described above, these amount to impose the condition \eqref{D3tad2}, which guarantees the cancellation of the total D3-brane charge along the compact internal manifold $X$. Given that such charge is measured by the  R-R four-form potential $C_4$ to which a space-time filling D3-brane couples, it is quite natural to consider the presence of such 4d four-form in the EFT and interpret it as a Lagrange multiplier implementing the said constraint. That is, we may regard \eqref{D3tad2} as the four-dimensional  equation of motion of $C_4$ arising as a result of the variation of the  following 4d coupling term in the action
\be
\calq \int   C_4
\label{IIBQC4}
\ee
which is clearly invariant under the gauge transformation 
\be\label{IIBC4gauge}
C_4 \rightarrow C_4+\d\Lambda_3\, . 
\ee

If we now promote our 4d EFT description to include dynamical fluxes via a parent Lagrangian of the form \eqref{IIBmaster}, we necessarily need to modify the above coupling to
\be\label{IIBC4new}
\int \Big(\calq^{\rm bg} +\frac12\cali^{\cala\calb}y_\cala y_\calb\Big)\, C_4
\ee
where for simplicity we have set $\calq^{\rm bg} = \tilde\calq^{\rm bg} +\mu_{\rm D3}$, and defined 
\be\label{IIBcali}
\cali\equiv \left(\begin{array}{cc}
0 & I \\
-I & 0
\end{array}
\right)\, .
\ee
Notice that now the combined Lagrangian \eqref{IIBmaster}+\eqref{IIBC4new} is no longer gauge invariant under \eqref{IIBC4gauge}, unless the three-form potentials simultaneously transform as
\be
A^\cala_3  \rightarrow A^\cala_3 - \cali^{\cala\calb}y_\calb\,\Lambda_3 \, ,
\label{A3gauge}
\ee
or in other words they are gauged. This gauging is however problematic, in the sense that compactness of the gauge symmetry would require $y_\calb$ to be an integer. Moreover, the obvious generalization for the four-form field strengths
\be
\label{bosgF42}
\hat F^\cala_4=\d A^\cala_3+ \cali^{\cala\calb}y_\calb\, C_4
\ee
is not gauge invariant under the gauge symmetry \eqref{IIBC4gauge}+\eqref{A3gauge}:  $\hat F^\cala_4\rightarrow \hat F^\cala_4-\cali^{\cala\calb}\d y_\calb\wedge \Lambda_3$. Therefore we find again an obstruction to describe the 4d EFT in terms of the naive three-form Lagrangian. 

While this new obstruction is independent of the rank of $T^{\cala\calb}$, it can be overcome by the same sort of prescription. One may reduce the number of dynamical fluxes, and take some of them to be fixed integers. Such non-dynamical fluxes will be the ones appearing as gauging coefficients of the dynamical three-forms, for which an integer will replace $y_B$ in \eqref{A3gauge}, and as a result their field strengths will be gauge invariant. By direct inspection of the matrix \eqref{IIBcali}, it is easy to see that one can achieve this by, e.g.,  setting the  R-R fluxes as dynamical and the NS-NS fluxes as non-dynamical, or the other way round. Let us for concreteness consider the former possibility. The four-form field strengths for the  R-R fluxes are 
\be\label{bosgF4}
\hat F^A_4=\d A^A_3 + I^{AB} h_B C_4\,.
\ee
They are clearly invariant under the gauge symmetry
\be
\begin{array}{rcl}
C_4 & \rightarrow& C_4+\d\Lambda_3\, ,\\
A^A_3  & \rightarrow& A^A_3-I^{AB}h_B\Lambda_3\, .
\end{array}
\label{gauging3}
\ee
Interestingly, the gauge invariant field strengths \eqref{bosgF4} can also be obtained by direct dimensional reduction of 10d type IIB supergravity, expressed in terms of the  R-R magnetic potentials $C_4$, $C_6$ and $C_8$. More precisely, they arise from expanding the gauge invariant field strength
\be
G_7 = \d C_6 - H_3 \wedge C_4
\label{G7}
\ee
over the basis of harmonic three-forms, as 
\be
C_6 = - I_{AB}\, A^A_3  \wedge \varpi^B
\ee
with $I_{AB}$ being the inverse of \eqref{Inter}. Needless to say, one could also obtain a gauging in which the roles of  R-R and NS-NS fluxes are interchanged or mixed-up, by first applying an S-duality or a more general $SL(2,\mathbb{Z})$-duality  to \eqref{G7} and then performing dimensional reduction.  Notice however that this operation takes us away from the perturbative type IIB regime in which we are working. As we will now discuss, staying in the weak-coupling regime creates an asymmetry between  R-R and NS-NS fluxes, indicating which ones should be dualized to three-forms.

\subsection{The 4d hierarchy of gaugings}
\label{sec:hierarchy}

Our discussion so far has indicated an obstruction to dualize all the three-form fluxes present in type IIB compactifications in terms of 4d three-forms. By direct inspection, we have indicated at least two possible simple choices of dualization. We may dualize the set of  R-R three-form fluxes and keep the NS-NS fluxes to a fixed value, or the other way round. There should be however a simple criterion to discriminate between these two choices, related by the $SL(2,\mathbb{Z})$ symmetry of the setup which may also be exploited to identify other choices. Indeed, let us consider the 4d membranes that couple to the  R-R and NS-NS three-form potentials. These are made of D5-branes and NS5-branes, respectively, wrapping the internal three-cycles of the internal compact geometry. In the weak coupling regime in which we are working, the 4d membranes that come from wrapped NS5-branes are much heavier than those coming from wrapped D5-branes. This suggests that the fluxes that should be treated as `dynamical' in this region of field space are the  R-R fluxes, while the NS-NS fluxes should be seen as constants unrelated to any three-form multiplet in the 4d EFT.

Rather than elaborating on this intuition, to be developed in section \ref{sec:IIBlowE}, let us turn into a different criterion to discriminate between these two choices of dynamical fluxes. Such criterion is based on the additional gaugings that emerge in different regions of field space, and that also involve three-form potentials. Indeed, at weak string coupling a shift symmetry is developed for the  R-R axion $C_0$, and so it can be dualized into a 4d two-form $\calb_2$. Such two-form is nothing but the dimensional reduction of the  R-R potential $C_8$ dual to $C_0$ in 10d. By dimensional reduction of the corresponding gauge invariant field strength in type II supergravity one obtains 
\be
G_9 = \d C_8 - H_3 \wedge C_6 \quad \rightarrow \quad \d\calb_2 +  h_A A_3^A\, ,
\ee
which will be the combination invariant under the gauge transformation
\be
\begin{array}{rcl}
A^A_3  & \rightarrow& A^A_3+\d\Lambda^A_2\, ,\\
\calb_2 & \rightarrow& \calb_2 - h_A \Lambda^A_2\, ,
\end{array}
\label{gauging2}
\ee
that appears in the 4d EFT. As pointed out in \cite{Marchesano:2014mla}, this is a clear example of flux-induced axion-four-form coupling $h_A C_0 \d A_3^A$, dual to the above two-form gauging. In the same spirit as in \cite{Kaloper:2008fb,Kaloper:2011jz}, this gauging signals a discrete shift symmetry in the flux-induced potential, that constrains its possible quantum corrections. 

Alternatively, one may detect the above two-form gauging in terms of the extended objects that appear in the 4d EFT. Indeed, let us consider a D7-brane wrapping the internal manifold $X$, and therefore coupled electrically to $\calb_2$. In the presence of internal NS-NS three-form fluxes such D7-brane develops a Freed-Witten anomaly, that is cured by D5-branes wrapping a three-cycle in the class  Poincar\'e dual to $h_A \varpi^A$, and ending on the D7-brane. From the 4d viewpoint, these are seen as a set of membranes ending on an axionic string, combined into a well-defined object under the gauge transformation  \eqref{gauging2}, see e.g. \cite{Evslin:2007ti,BerasaluceGonzalez:2012zn}. 

Clearly, this sector of the EFT treats differently  R-R and NS-NS background fluxes, as only the latter appear as gauging coefficients for $\calb_2$. As such, the $h_A$ should be treated as quantized constants. For instance, if one tried to promote these fluxes to dynamical variables, the modified field strength $\hat{\cal H}_3 = \d \calb_2 + h_A A_3^A$ would be no longer invariant under \eqref{gauging2}, similarly to the field strength \eqref{bosgF42}. In other words, compatibility with invariance under \eqref{gauging2} requires to treat the NS-NS fluxes as non-dynamical, while the  R-R fluxes can be dualized to 4d three-forms. 

Notice that this asymmetry between NS-NS and  R-R fluxes is a direct consequence of the field space regime under consideration. In the type IIB weak-coupling limit, $C_0$ develops an approximate continuous shift symmetry, and therefore a corresponding 4d axionic string is expected to exist. Such 4d string, which is nothing but the wrapped D7-brane above, has necessarily 4d membranes attached to it. Therefore it does not make sense to include the axionic string in the EFT without including the attached membranes as well, which in this case couple to the three-forms arising from the  R-R sector of the compactification. 

Having chosen the NS-NS fluxes as fixed parameters, it is important to verify that the different sets of gaugings in the theory are compatible with each other. In particular, the gaugings \eqref{gauging3} and \eqref{gauging2} must be such that  the three-forms participating in one must not participate in the other. In the case at hand this is guaranteed by the property $I^{AB} h_Ah_B \equiv 0$. Microscopically, it is a consequence of the fact that $\d_H^2 \equiv 0$, where $\d_H=\d-H\wedge$ is the twisted differential which appears in the democratic formulation of the R-R Bianchi identites. Macroscopically, it implies that domain walls ending on the axionic string do not affect the effective tadpole condition, as described below.

The two kinds of gaugings discussed above must be complemented with the more familiar ones like the St\"uckelber mechanism \eqref{Stu}, related to discrete gauge symmetries, or rather the dual version that gauges one-forms by two-forms. The whole set of gaugings that may be present in a 4d EFT and the compatibility conditions that they must satisfy is summarized in table \ref{hgauging}, together with familiar EFT quantities that they are associated to. Finally, as each of these gaugings can be described in terms of integers, their presence also contains more subtle, discrete data of the compactification. Microscopically, one may understand these data as torsional factors in the group classifying the charges of 4d strings, membranes and space-time filling branes \cite{BerasaluceGonzalez:2012zn}, which in the case at hand amounts to the twisted K-theory group \cite{Witten:1998cd}. In particular, the three-form gauging \eqref{bosgF4} determines the K-theory group $\mathbb{Z}_p$ for space-time filling D3-branes, with $p = {\rm g.c.d.} (I^{AB}h_B)$.

\begin{table}[!ht]
\begin{center}
\begin{tabular}{|c|c|c|}
\hline
{\bf Gauging} & {\bf EFT quantity} & {\bf discrete data} \\
\hline\hline
$\d A_1^\sigma - k_\Lambda^\sigma \calb_2^\Lambda$ & D-term potential & discrete gauge symmetries\\
\hline
$\d \calb_2^\Lambda - c^\Lambda_A A_3^A$ & F-term potential & discrete shift symmetries\\
\hline
$\d A_3^A - Q^A_I C_4^I$ & tadpole conditions & torsional 3-brane charges \\
\hline
\end{tabular}
\caption{Different gaugings in 4d and their related EFT quantities. Their compatibility requires that the constraints $k_\Lambda^\sigma c^\Lambda_A = c^\Lambda_A Q^A_I = 0$ are satisfied. \label{hgauging}}
\end{center}
\end{table}

In the case at hand, there are no gaugings regarding 4d one-forms, unless we include metric fluxes affecting the sector of orientifold-even three cycles or the presence of space-time filling D7-branes. In the latter case new non-trivial tadpole constraints will also appear, which can be taken into account by considering the complete set of 4d four-forms in the compactification. 

\subsection{The dual three-form Lagrangian}
\label{sec:IIBdual3}

Let us now proceed to construct the dual three-form Lagrangian taking account that we should distinguish between dynamical and non-dynamical fluxes. Namely, we have
\be
\begin{array}{rcl}
\Gamma & = & \left\{(m_A, h_A) | m_A, h_A \in \mathbb{Z}  \right\}=  \{ \caln_\cala \}  \,  , \\
\Gamma_{\rm EFT}  & = &  \left\{(m_A, 0) | m_A \in \mathbb{Z} \right\} \, ,
\end{array}
\ee
where $\Gamma_{\rm EFT}$ is the set of flux quanta that can be dualized to three-forms and that can be dynamically generated in the 4d EFT. In our case $\Gamma_{\rm EFT}$ can be identified with the  R-R fluxes $m_A$. Given an initial flux background $\caln_\cala^{\rm bg}=(m_A^{\rm bg},h_A^{\rm bg})$, the set of fluxes that is accessible in the 4d EFT description is 
\be
\Gamma_{\rm F}  \, \equiv \,  \Gamma_{\rm EFT}+\caln^{\rm bg}  \, = \, \left\{(m_A, h_A^{\rm bg}) |\  m_A \in \mathbb{Z},\  h_A^{\rm bg}\, {\rm fixed}  \right\} \, ,
\ee
which is an affine sublattice of the flux lattice $\Gamma$. These sublattices are parametrized by quotient elements $[\caln^{\rm bg}] \in \Gamma/\Gamma_{\rm EFT}$, which can be identified with the set of NS-NS fluxes $h_A$.

With these definitions in mind let us proceed to describe the dual Lagrangian. First notice that we may generalize \eqref{IIBsupo} to
\be
 W (\phi)=M_{\rm P}^3\, m_A  \Pi^A(\phi) + \hat{W}(\phi)
\ee
where 
now the chiral fields $\phi^\alpha$ comprise, in addition to $(\tau,\phi^i)$, also  K\"ahler moduli, two-form and four-form axions.  $\hat{W}$ contains the NS-NS flux dependence already present in \eqref{IIBsupo} and other possible contributions to the superpotential like e.g.\ those of non-perturbative origin: $\hat{W} = - M_{\rm P}^3\, \tau h_A  \Pi^A(\phi) + W_{\rm np}$. This piece of the superpotential can be treated as in \cite{Farakos:2017jme,Bandos:2018gjp} when dualizing the dynamical fluxes $m_A$. In practice, this implies that the last two terms in \eqref{IIBmaster} are rewritten as the first two terms of the following Lagrangian
\be\label{IIBmastersplit}
 -\Big(\frac12 T^{AB}y_A y_B +y_A \Upsilon^A + \hat V\Big)\,*\!1 - \d y_A\wedge A^A_3 + \Big(\calq^{\rm bg} +Q^A y_A\Big)C_4
\ee
while the last term implements the D3-brane tadpole condition. Here $Q^A = I^{AB} h_B$ and 
\begin{subequations}\label{IIBThV}
\begin{align}
T^{AB}& \equiv 2M^4_{\rm P}\, e^{K}\Re\left(K^{i\bar\jmath}D_i\Pi^A\bar D_{\bar\jmath}\bar\Pi^B+\Pi^A\bar\Pi^B\right)\,,\label{IIBThV1}\\
\Upsilon^A&=2M_{\rm P}\, e^{K}\Re\left(K^{\alpha\bar\beta}D_\alpha\hat W\bar D_{\bar\beta}\bar\Pi^A-3\hat W\bar\Pi^A\right)\,,\label{IIBThV2}\\
\hat V&=\frac{e^K}{M^2_{\rm P}}\left(K^{\alpha\bar\beta}D_\alpha \hat W\bar D_{\bar\beta}\overline{ \hat W}-3|\hat W|^2\right)\,,\label{IIBThV3}
 \end{align}
 \end{subequations}
where as usual $D_\alpha\hat W\equiv \hat W_\alpha+K_\alpha \hat W$, and we have made use of the same no-scale structure employed to obtain \eqref{IIBTAB} -- see footnote \ref{ft:noscale}. 
Since $T^{AB}$ is now invertible, one can integrate out the $y_A$ by solving their equations of motion as
\be
y_A = - T_{AB}\left(* \hat{F}_4^A + \Upsilon^A\right) \, ,
\ee
with $\hat{F}^A_4\equiv F^A_4+Q^A C_4$. By inserting this back into \eqref{IIBmastersplit} one obtains the following three-form action
\be\label{IIBdualF4lagr}
\begin{aligned}
S_{\text{three-forms}}=&-\int_\calm \Big[\frac12 T_{AB}\hat{F}^A_4*\!\hat{F}^B_4+ T_{AB}\Upsilon^A\hat{F}^B_4+\Big(
\hat V-\frac12 T_{AB}\Upsilon^A\Upsilon^B\Big)*\!1\Big] + \calq^{\rm bg} \int_\calm C_4\\
&+\int_{\del\calm}T_{AB}(* \hat{F}^A_4+\Upsilon^A)A^B_3\, ,
\end{aligned}
\ee
where recall that the three-forms and their field strengths are associated only with the  R-R fluxes of the compactification. Their equations of motion read
 \be\label{A3eom}
 \d[T_{AB}(*\hat{F}^B_4+\Upsilon^B)] =0\,,
 \ee
 and are solved by setting 
 \be\label{F4vev}
 T_{AB}(*\hat{F}^B_4+\Upsilon^B)= -m_A\, 
 \ee
 with $m_A \in \mathbb{Z}$ interpreted as the  R-R background fluxes. Finally, by inserting this solution into \eqref{IIBdualF4lagr}, one obtains the scalar potential
 \be\label{IIBpot}
V=\frac12 T^{AB}m_A m_B +m_A \Upsilon^A + \hat V
\ee
that reproduces the F-term scalar potential of type IIB flux compactifications. Notice however that, in the present approach, one is describing a multi-branched scalar potential within the same effective field theory. In addition one obtains a term of the form
\be
\int_{\cal M} \Big(\calq^{\rm bg} +Q^A m_A\Big)C_4
\ee
which, when integrating out $C_4$, gives the linear tadpole condition $\tilde\calq^{\rm bg} + \mu_{\rm D3} + Q^A m_A= 0$ to be imposed on the lattice of EFT fluxes.


\section{Three-form potentials and gaugings  in  EFTs}
\label{sec:genEFT}

In the previous section we have shown how higher order $p$-form potentials and their gaugings by higher $(p+1)$-form potentials naturally arise in string compactifications and encode relevant physical information. In general, it is important to understand how these ingredients can be consistently incorporated into the low-energy EFT from a purely four-dimensional perspective. In turn, this may allow one to identify the distinguishing patterns that characterise the EFT of string theory models and hence, possibly, of more general quantum gravity theories.  

Moreover, we have seen how the presence of these gaugings may affect the low-energy description of the set of fluxes in a given string compactification. 
Indeed, let us consider a 4d string theory model characterized by a lattice $\Gamma$ of quantized (ordinary, geometric or non-geometric) fluxes threading the internal compactification space. By expanding them in an appropriate basis, they can be identified by a set of quantized numbers $\caln_\cala$, which contribute linearly to the effective four-dimensional superpotential by terms of the form $\caln_\cala \Pi^\cala(\Phi)$, where $\Phi^i$ denotes a set of chiral multiplets.  From the discussion in the previous section and other string theory examples (see for instance \cite{Marchesano:2014mla,Dudas:2014pva,Bielleman:2015ina,Carta:2016ynn,Valenzuela:2016yny,Herraez:2018vae,Escobar:2018tiu,Escobar:2018rna}) it is expected that at least a sublattice $\Gamma_{\rm EFT}\subset \Gamma$  of these constants  can be promoted to expectation values of (appropriate combinations of) four-form field strengths. By introducing an appropriate basis of $v_\cala^A$ for such sublattice, we can split $\caln_\cala$  as follows
\be\label{Nsplit}
\caln_\cala=N_A v^A_\cala+\caln^{\rm bg}_\cala
\ee
and parametrize $\Gamma_{\rm EFT}$ by the constants $N_A$, while regarding the remaining $\caln^{\rm  bg}_\cala$ as background fluxes. We will then consider  $\mathcal N=1$ supersymmetric EFTs characterized by a superpotential of the form 
\be\label{gensup}
W(\Phi)= M^3_{\rm P}\, N_A\Pi^A(\Phi)+\hat W(\Phi)\, ,
\ee
where $\Pi^A\equiv v^A_\calb\Pi^\calb$ and  $\hat W(\Phi)$ contains the term $M^3_{\rm P}\, \caln^{\rm bg}_\cala \Pi^\cala(\Phi)$ as well as other (typically non-perturbative) contributions. 

One of the aims of this paper is to generalize the discussion of the previous section and show, from the purely four-dimensional perspective, how to select a flux sublattice $\Gamma_{\rm EFT}$ such that one can trade the constants $N_A$ for a set of dual field-strengths $F^A_4=\d A^A_3$. We will do it by streamlining and generalizing the recent constructions of four-dimensional $\caln=1$ EFTs involving three-form potentials $A_3^A$ \cite{Farakos:2017jme,Bandos:2018gjp}. In this way we will also identify the possible technical obstructions to this dualization, which should be taken into account to constrain the choice of $\Gamma_{\rm EFT}$. For instance, as discussed in the previous section, an obstruction will arise from implementing the tadpole constraints at the level of the EFT. In this section we will show in general how, if $\Gamma_{\rm EFT}$ is appropriately chosen, the tadpole conditions acquire a clean four-dimensional description in terms of a gauging of the potentials $A^A_3$. We will also consider the analogous gauging of two-form potentials (dual to axions) under gauge transformations $A^A_3\rightarrow A^A_3+\d \Lambda^A_2$. As seen in the last section and also discussed in sections \ref{sec:IIBstringy} and \ref{sec:IIAstringy}, in string models the charges specifying these gaugings are typically defined by the `non-dynamical' background fluxes $\caln_\cala^{\rm bg}$.\footnote{\label{foot:coslattice} Since the constants $N_A$ will be considered as dynamical variables, the physically independent choices of $\caln_\cala^{\rm bg}$ can be associated with the elements of the quotient $\Gamma/\Gamma_{\rm EFT}$. In particular, the gauging charges will depend only on the equivalence class defined by $\caln_\cala^{\rm bg}$.}

In order to emphasize some key points, in this section we will mostly focus on the relevant  bosonic sector of these EFTs.  The supersymmetric completion  of these models  will be discussed in section \ref{sec:susy} and the application to concrete  string models will be considered in sections \ref{sec:IIBstringy} and \ref{sec:IIAstringy}. As we will see, while the discussion in this and section \ref{sec:susy} is quite general, the different string theory examples of sections \ref{sec:IIBstringy} and \ref{sec:IIAstringy} will exhibit some interesting common features.

\subsection{Preliminaries on the Weyl-invariant formulation}
\label{sec:weyl}

As in \cite{Farakos:2017jme,Bandos:2018gjp}, in order to describe the general formulation of EFTs with three-form potentials,  it is convenient to start from a super-Weyl invariant EFT (see \cite{Buchbinder:1995uq} for an introduction and Appendix \ref{app:SWL} for details),   which is formally closer to the rigid supersymmetric case.\footnote{Alternatively, one may also adopt the  superconformal approach (see \cite{Freedman:2012zz} for an introduction) but the super-Weyl invariant formulation will allow use to couple branes in a manifestly supersymmetric way, see section \ref{sec:susy3}.}  Basically, one must extend the physical chiral multiplets  $\Phi^i$, $i=1,\ldots,n$, to a set of $n+1$ chiral multiplets $Z^a$ that transform with weight 3 under super-Weyl transformations. The connection with the ordinary formulation is then obtained by singling out a Weyl compensator $U$ by setting 
\be\label{Zsplit}
Z^a=U g^a(\Phi)
\ee 
for some set of functions $g^a$, and eventually gauge-fixing the Weyl compensator (see below). Note that $U$ has Weyl weight three, while $\Phi^i$ are inert under super-Weyl transformations. 
As anticipated above, in this section we will focus on the bosonic sector. Hence, for the time being we  restrict consideration to the lowest components $\phi^i$, $u$ and $z^a=ug^a(\phi)$ of the chiral multiplets $\Phi^i$, $U$ and $Z^a$, respectively.   

The (super-)Weyl invariant EFT is  specified by two  functions: a holomorphic one $\calw(z)$ and a real one $\calk(z,\bar z)$. These must satisfy the homogeneity conditions 
\be\label{superWeyl}
\calw(\lambda z)=\lambda \calw(z)\,, \qquad\calk(\lambda z,\bar\lambda\bar z)=|\lambda|^{\frac23}\calk( z,\bar z)\,. 
\ee
The ordinary 
superpotential and K\"ahler potential for the scalars $\phi^i$ are then obtained by setting  $W(\phi)\equiv M_{\rm P}^3\,\calw(g(\phi))$ and $K(\phi,\bar\phi)\equiv -3\log [-\frac13\calk(g(\phi),\bar g(\phi))]$.\footnote{The holomorphic  split $z^a=ug^a(\phi)$ introduced in  \eqref{Zsplit} is not unique, since one   may redefine 
$u\rightarrow e^{h(\phi)}u$ and $g^a(\phi)\rightarrow e^{-h(\phi)}g^a(\phi)$. 
In turn, under these redefinitions we have $K\rightarrow K+h+\bar h$ and $W\rightarrow e^{-h}W$. Hence this ambiguity corresponds to the K\"ahler invariance of the ordinary formulation. On the other hand notice that $\calk$ and $\calw$,  as well as $\calv^A$ and $\hat\calw$ introduced in \eqref{wvw} do not have this freedom and are uniquely defined.} In particular, in our class of models we have
 \be\label{wvw}
\calw(z)=N_A\calv^A(z)+\hat\calw(z) \, ,
\ee
so that $\Pi^A(\phi)\equiv \calv^A(g(\phi))$ and $\hat W(\phi)\equiv M_{\rm P}^3\,\hat\calw(g(\phi))$.
By  means of the homogeneity properties \eqref{superWeyl}, we can then make the following identifications
\be\label{WK}
\calw(z)= \frac{u}{M^3_{\rm P}} W(\phi)\quad,\quad \calk(z,\bar z)\equiv -3|u|^\frac23 e^{-\frac13 K(\phi,\bar\phi)}\, .
\ee
Analogously, we have $\hat\calw(z)= \frac{u}{M^3_{\rm P}} \hat W(\phi)$ and 
\be
\label{homVA}
\calv^A(z)=u\,\Pi^A(\phi)\, ,
\ee
so that we can trade the homogeneous scalars $z^a$ for the scalars $(u,\phi^i)$.

The formulas of the super-Weyl invariant EFT are  formally quite analogous to those of a rigid supersymmetric theory. One may interpret $\calk$ as a sort of  K\"ahler potential for the $z^a$ scalars with associated metric
\be
\calk_{a\bar b}\equiv \del_a\del_{\bar b}\calk\, .
\ee
However, one should keep in mind that this metric is not positive-definite, with a negative eigenvalue corresponding to the compensator $u$.  
 The EFT contains a scalar  potential which has the form of a standard rigid supersymmetric one 
\be\label{pot1}
V=\calk^{a\bar b}\,\calw_a\bar\calw_{\bar b}\,=\calk^{a\bar b}(N_A\calv^A_a+\hat\calw_a)(N_B\bar\calv^B_{\bar b}+\overline{\hat\calw}_{\bar b})\, ,
\ee
where $\calk^{a\bar b}$ is the inverse of $\calk_{a\bar b}$. 
 
We also recall that the Einstein-Hilbert term of the Weyl-invariant Lagrangian  takes the non-canonical form
\be\label{wEH}
-\frac16 \calk\, R\, ,
\ee
and that ordinary Poincar\'e supergravity can be recovered by imposing the (non-holomorphic) gauge-fixing condition  
\be\label{Efix}
u=M^3_{\rm P}\, e^{\frac12 K(\phi,\bar\phi)}\, ,
\ee
which sets $\calk=-3 M^2_{\rm P}$, see \eqref{WK}, so that \eqref{wEH} reduces to the canonical Einstein-Hilbert term $\frac12 M^2_{\rm P}R$.
The potential \eqref{pot1} becomes the usual $\caln=1$ potential
\be\label{Epot}
V=\frac{e^K}{M^2_{\rm P}}\left(K^{i\bar\jmath}D_i W\bar D_{\bar\jmath}\bar W-3|W|^2\right) \quad~~~\text{(Einstein frame)}
\ee
and the $\calv^A$'s can be identified with the standard normalized periods 
\be\label{homper}
\calv^A= M^3_{\rm P} e^{\frac12 K}\Pi^A\quad~~~\text{(Einstein frame)}\, .
\ee
Finally, the  potential \eqref{Epot} combines with the usual Einstein frame kinetic terms
\be\label{boskin}
 M^2_{\rm P}\int \left(\frac12R*\!1- K_{i\bar\jmath}\,\d\phi^i\wedge *\d\bar\phi^{\bar\jmath}\right)\, .
\ee

Even though we may work directly with the Weyl-fixed formulation, in the rest of this section we will mostly use the Weyl-invariant formulation and impose the Weyl gauge fixing only at the very end. The reason is that we want to make clear the connection with the  supersymmetric extension of the following arguments,  which will be discussed in sections \ref{sec:susy} and \ref{sec:susy3} and are naturally formulated in the (super) Weyl-invariant framework. Furthermore, the Weyl-invariant formulation has also a more superficial advantage of simplifying the formulas and to be immediately adaptable to a rigid theory, in which $\calw$ and $\calk$  are the ordinary (non necessary homogeneous) superpotential and K\"ahler potential.     


\subsection{Dual formulation with three-form potentials}
\label{sec:bosduality}

Using the Weyl-invariant formulation, one may easily generalize the discussion of section \ref{sec:IIBdual3} and outline, in bosonic terms, how to derive an EFT in which the constants $N_A$ are substituted by dynamical (although non-propagating) three-form potentials $A^A_3$. In section \ref{sec:susy} we will see how this procedure can be made manifestly supersymmetric, extending and improving  the strategy adopted in \cite{Farakos:2017jme,Bandos:2018gjp}.  

As in section \ref{sec:IIBgauging}, the basic trick is to consider the constants $N_A$ as expectation values of zero-form field-strengths, and then promote them to real scalar fields $y_A$. Adding the term $\int \d y_A\wedge A^A_3$ to the parent effective action, one may dualize the fluxes to four-form field-strengths $F^A_4=\d A^A_3$.  Alternatively, treating the $A^A_3$'s as Lagrange multipliers, allows one to impose the Bianchi identities $\d y_A=0$ at the level of the equations of motion (a similar trick was used e.g. in \cite{Kaloper:2008qs}).

In our (Weyl-invariant) bosonic EFTs the $N_A$'s only appear in the  potential \eqref{pot1}, which can be rewritten in the form
\be\label{pot3}
V=\frac12 T^{AB}N_A N_B +N_A \Upsilon^A + \hat V\, ,
\ee
where 
\begin{subequations}\label{ThVdef}
\begin{align}
T^{AB}(z,\bar z)&\equiv 2\,\Re\left(\calk^{a\bar b}\,\calv_a^A\overline\calv_{\bar b}^B\right)\,,\label{ThVdef1}\\
\Upsilon^A(z,\bar z)&\equiv 2\,\Re \left(\calk^{a\bar b}\hat\calw_a\overline\calv_{\bar b}^A \right)\,,\label{ThVdef2}\\
\hat V(z,\bar z)&\equiv \calk^{a\bar b}\,\hat\calw_a\overline{\hat\calw}_{\bar b}\,.\label{ThVdef3}
 \end{align}
 \end{subequations}
Hence, the relevant terms of  the bosonic parent effective Lagrangian are
\be\label{bosmaster}
 -\Big(\frac12 T^{AB}y_A y_B +y_A \Upsilon^A + \hat V\Big)\,*\!1 - \d y_A\wedge A^A_3 \, .
\ee
As mentioned above, we can integrate $A^A_3$ out, getting the equation of motion/constraint $\d y_A=0$, which is solved by setting $y_A=N_A$. Then, plugging $y_A=N_A$ back into \eqref{bosmaster} one gets back the original potential \eqref{pot3}. 

Before proceeding, let us comment on our choice of boundary conditions for the fields. First of all, observe that one can rewrite the contribution to the effective action of the last term in \eqref{bosmaster} in the form 
\be\label{Ayterms}
\int_\calm y_A F^A_4+\int_{\del\calm}y_A\, A^A_3
\ee
where $\calm$ is the four-dimensional spacetime.
We choose the scalars $y_A$  to take (possibly different) constant values $y_A|_{\rm bd}=N^{\rm bd}_A$ on each connected component of the spacetime boundary $\del\calm$, while the three-form potentials $A^A_3$ are unconstrained. With this boundary conditions, the terms appearing in  \eqref{Ayterms} are gauge invariant. Furthermore, the form of \eqref{Ayterms} makes it clear that: first, the scalars $y_A$ can be considered as momenta canonically conjugated to the gauge fields; second, in a path-integral formulation the boundary term in \eqref{Ayterms}  fixes the asymptotic states to have definite momenta $N^{\rm bd}_A$ -- see for instance \cite{Duncan:1989ug}. Once we have fixed these boundary conditions, \eqref{Ayterms} becomes invariant under unrestricted two-form gauge transformations 
\be\label{2gaugeb}
A^A_3\rightarrow A^A_3+\d\Lambda^A_2\, .
\ee
Notice that for fixed $N_A^{\rm bd}$,  in order to have (step-wise) varying values of $y_A$ within a connected space-time component, one needs the presence of membranes charged under the three-form potentials $A^A_3$. These will be introduced in section \ref{sec:sm3}.

As a related specification, we will also assume the compactness of the two-form gauge symmetries associated with the  three-form potentials $A^A_3$, which is expected in consistent quantum gravity theories \cite{Banks:2010zn}. Quantum mechanically, this  implies that the conjugate momenta $y_A$, and hence $N_A$,  can only take appropriately quantized values. This is indeed what happens in  string compactifications, where the constants $N_A$ correspond to flux quanta and can be considered as components in an appropriate  basis  of an element of the flux lattice $\Gamma_{\rm EFT}$. Correspondingly, the Wilson lines $\frac{1}{2\pi}\int A^A_3 $ are defined modulo elements of the dual lattice $\Gamma^*_{\rm EFT}$. 
We will  work with an integral basis, in which $N_A\in\mathbb{Z}$ and $\frac{1}{2\pi}\int A^A_3 \simeq \frac{1}{2\pi}\int A^A_3+1$, although other choices would also be  possible. Notice that this viewpoint motivates, from a purely four-dimensional perspective, the quantization of the constants $N_A$ -- see also \cite{Kaloper:2011jz,Bandos:2018gjp}.  

We can now come back to 
\eqref{bosmaster} and proceed with the dualization. By integrating out $y_A$, we get the identification 
\be\label{Fdual}
F^A_4=(T^{AB}y_B+\Upsilon^A)*\!1\, .
\ee
If $T^{AB}$ is invertible, one can see this as fixing the $y_A$ as ($z$-dependent) functions of the $F^A_4$'s. The invertibility of $T^{AB}$ is discussed in detail in appendix \ref{app:TAB} and in particular it requires that the number of three-forms is not bigger than twice the number of scalars $z^a$ appearing in the periods $\calv^A(z)$.  

As follows from the discussion in appendix \ref{app:TAB}, one may avoid invertibility issues by adjusting the choice of the lattice $\Gamma_{\rm EFT}$ of dualizable fluxes.  Let us in particular  assume that we have chosen $\Gamma_{\rm EFT}$ so that $T^{AB}$ admits an inverse $T_{AB}$, as can be done in all the concrete  examples that will be discussed later on. Then \eqref{Fdual} can be solved for $y_B=-T_{AB}(*F_4^A+\Upsilon^A)$, which inserted back into \eqref{bosmaster} gives the action  
\be\label{dualF4lagr}
\begin{aligned}
S_{\text{three-forms}}=&-\int_\calm \Big[\frac12 T_{AB}F^A_4*\!F^B_4+ T_{AB}\Upsilon^AF^B_4+\Big(
\hat V-\frac12 T_{AB}\Upsilon^A \Upsilon^B\Big)*\!1\Big] \\
&+\int_{\del\calm}T_{AB}(* F^A_4+\Upsilon^A)A^B_3\, .
\end{aligned}
\ee
 One can then easily check that the variational principle  for an unconstrained $A^A_3$ is well defined thanks to the presence of the boundary term  in \eqref{dualF4lagr} and that the corresponding equations of motion are 
 \be\label{A3eomb}
 \d[T_{AB}(*F^A_4+\Upsilon^A)] =0\,,
 \ee
 which are solved by setting 
 \be\label{F4vevb}
 T_{AB}(*F^A_4+\Upsilon^A)= -N_B\,.
 \ee
Upon inserting this solution into \eqref{dualF4lagr}, one gets back the potential \eqref{pot3}. Again, the presence of the boundary term in \eqref{dualF4lagr} is crucial, since it gives a non-vanishing contribution to the action. Notice also that the above arguments justifying the quantization of $N_A$ still hold, up to replacing the role of $y_A$ with $-T_{AB}(*F^B_4+\Upsilon^B)$. 
 
The terms \eqref{dualF4lagr} provide the contribution of the gauge three-forms to the bosonic dual Lagrangian. In particular, from \eqref{ThVdef1} we see that the kinetic matrix $T_{AB}$ is completely determined by the data defining the scalar sector of the theory.  In order to highlight the physical content of this action, let us gauge fix the Weyl symmetry as explained in section \ref{sec:weyl}. Then the inverse of the kinetic matrix $T_{AB}$ takes the following form
\be\label{wfTAB}
T^{AB}=2 M^4_{\rm P}\,e^{K} \Re\left(K^{i\bar\jmath}D_i\Pi^A\bar D_{\bar\jmath}\bar\Pi^B-3\Pi^A\bar\Pi^B\right)\,.
\ee
We then see that the kinetic matrix $T_{AB}$ is completely determined by the K\"ahler potential $K(\phi,\bar\phi)$ and the periods $\Pi^A(\phi)$.
This fact is essentially due to supersymmetry, as it will be clearer later on. 

After Weyl-symmetry fixing, $\Upsilon^A$ and $\hat V$ defined in \eqref{ThVdef} become
\begin{subequations}\label{hatV}
\begin{align}
\Upsilon^A&=2 M_{\rm P}\,e^{K}\Re\left(K^{i\bar\jmath}D_i\hat W\bar D_{\bar\jmath}\bar\Pi^A-3\hat W\bar\Pi^A\right)\, ,\label{hatV1}\\
\hat V&=\frac{e^K}{M^2_{\rm P}}\left(K^{i\bar\jmath}D_i \hat W\bar D_{\bar\jmath}\overline{ \hat W}-3|\hat W|^2\right)\, .\label{hatV2}
\end{align}
\end{subequations}
The complete bosonic Einstein frame effective action  is obtained by summing  \eqref{boskin} and \eqref{dualF4lagr}, and taking into account \eqref{wfTAB} and \eqref{hatV}. We emphasize that, assuming a non-degenerate $T^{AB}$, the above formulas are completely general and hold for any K\"ahler potential $K$, periods $\calv^A$ and superpotential term $\hat W$. For instance, they reproduce as particular sub-cases the bosonic sectors of the supersymmetric EFTs constructed in \cite{Farakos:2017jme,Bandos:2018gjp}, as we now discuss by looking at two special cases. In the following formulas we could easily include  a non-vanishing $\hat W$, but for clarity we set  $\hat W=0$.

\subsubsection*{Linear superpotential}
\label{sec:minbos}

The simplest non-trivial possibility is realized if the number  $k$ of three-forms $A^A_3$, $A=1,\ldots,k$, is at most equal to the number  $n+1$ of complex scalars $z^a$, $a=0,\ldots,n$ (which include the Weyl compensator) and  the  matrix $\calv^A_a(z)$ have maximal rank $k$. We can then  choose `adapted' coordinates $z^a=(z^A,\tilde z^\alpha)$ such that $\calv^A(z)\equiv  z^A$ and the homogeneous superpotential takes the form $\calw=N_A z^A$.
 It is clear that the scalars $\tilde z^\alpha$ play the role of spectators in the three-form description. (A similar comment holds, more generically, whenever $\calv^A$ does not depend on some set of scalars $\tilde z^a$.) Hence, we can restrict to the case $k=n+1$, the generalization to $k< n+1$ being obvious.    In the terminology of \cite{Farakos:2017jme},  this case corresponds to the single three-form multiplets -- see section \ref{sec:susy1} below. 

One may isolate the compensator $u$ by setting $z^0=u$ and $z^i=u\phi^i$, with $i=1,\ldots,n$, so that $\Pi^0=1$ and $\Pi^i=\phi^i$.  Then, before dualization, the superpotential takes the form $W=M^3_{\rm P}(N_0+N_i \phi^i)$.  For instance, we will encounter this kind of superpotential when we will discuss M-theory compactifications on $G_2$-holonomy spaces in section  \ref{sec:IIAscales}.

Having set $\hat W=0$ for simplicity, the corresponding  three-form action \eqref{dualF4lagr} reduces to 
\be\label{dualF4lagr2}
\begin{aligned}
S_{\text{three-forms}}=&-\frac12\int_\calm T_{AB}\, F^A_4*F^B_4 +\int_{\del\calm}T_{AB}A^A_3* F^B_4 \, ,
\end{aligned}
\ee
where the kinetic matrix $T_{AB}$ is the inverse of $T^{AB}$ as defined in \eqref{wfTAB}. This can be most readily computed by observing that $D_i\Pi^0=K_i$ and $D_i\Pi^j=\delta_i^j+\phi^i K_j$.

\subsubsection*{Maximally non-linear case}
\label{sec:maxbos}

We now consider the opposite case $k=2n+2$,  in which 
the number of three-forms $A^A_3$ is twice the number of complex scalars $z^a$, and we are still assuming $\calv^A_a$ be of maximal rank. Locally in field space, we can make a field-redefinition such that 
\be\label{maxrank}
\calv^A(z)\equiv\left(\begin{array}{c} 
z^a \\
\calg_a(z)
\end{array}\right)\,.
\ee
By the homogeneity of $\calg_a(z)$, we can also write $\calg_{a}=\calg_{ab} z^b$. This case corresponds to the double three-form multiplets of \cite{Farakos:2017jme} -- see also section \ref{sec:susy1} below. As we will see, it appears in the description of weakly-coupled type II compactifications.

Having set $\hat W=0$, the three-form action takes again the form \eqref{dualF4lagr2} with $T_{AB}$ the inverse of  \eqref{wfTAB}. Again, for any given K\"ahler potential $K$, $T^{AB}$ can be straightforwardly computed. Alternatively, one may start from the Weyl-invariant formulation, in which $T^{AB}$ takes the simpler-looking form \eqref{ThVdef1} with
\be\label{VAa}
\calv^A_b(z)\equiv\left(\begin{array}{c} 
\delta^a_b \\
\calg_{ab}(z)
\end{array}\right)\,.
\ee
This case was also examined in \cite{Bandos:2018gjp}, where the kinetic matrix for gauge three-forms was computed, assuming  the chiral fields $z^a$ to be homogeneous coordinates describing a special K\"ahler manifold.

\subsection{Tadpoles and three-form gaugings}
\label{sec:tadpole}

As we discussed in Section \ref{tadpole1} with the type IIB string example, in general, the choice of a sublattice $\Gamma_{\rm EFT}$ is also forced by incorporating into the effective theory the non-trivial tadpole conditions present in string theory compactifications. Indeed, there one often encounters linear or quadratic tadpole conditions for the full lattice $\Gamma$ of internal fluxes $\caln_\cala$. These may be given by a set of constraints of the form
\be\label{4charge}
\calq_I\equiv \frac12\cali^{\cala\calb}_I\caln_\cala \caln_\calb+ \tilde Q_I^\cala \caln_\cala+\tilde\calq_I^{\rm bg}=0\,,
\ee
where the index $I$ labels the different tadpole conditions, $\cali^{\cala\calb}_I=\cali^{\calb\cala}_I$ defines a symmetric pairing between the fluxes $\caln_A$, ${\tilde Q_I^\cala}\caln_\cala$ stands for a possible linear contribution of fluxes to the tadpoles, and $\tilde\calq_I^{\rm bg}$ denote some background `charge' that needs to be cancelled by the flux contribution. In string theory the last contribution is typically generated by orientifolds or curvature corrections. The contribution to $\calq_I$ of possible dynamical space-filling branes will be more explicitly discussed in section \ref{sec:sm3}.

Since we are interested in dualizing only the flux sublattice $\Gamma_{\rm EFT}$ labelled by $N_A$, it is convenient to use the splitting \eqref{Nsplit} and define 
\begin{subequations}\label{3Q}
 \begin{align}
 \cali^{AB}_I&\equiv v^A_\cala v^B_\calb \cali^{\cala\calb}_I\,,\label{gfd1}\\
 \calq^{\rm bg}_I&\equiv \tilde\calq_I^{\rm bg}+\frac12\cali_I^{\cala\calb}\caln_\cala^{\rm bg} \caln_\calb^{\rm bg}\,,\label{3Q2}\\
 Q_I^A&\equiv v^A_\cala \tilde Q^\cala_I+ \cali^{\cala\calb}_I v^A_\cala \caln^{\rm bg}_\calb\,.\label{3Q3}
 \end{align}
 \end{subequations}
We can then rewrite the condition \eqref{4charge}  as follows
\be\label{4charge2}
\calq_I\equiv \calq_I^{\rm bg}+\frac12\cali^{AB}_IN_AN_B+ Q_I^A N_A=0\,.
\ee

We would now like to understand how to dualize the constants $N_A$ to three-forms by  taking into account \eqref{4charge2}. 
As a first step, it is convenient to impose \eqref{4charge} at the level of the four-dimensional  equations of motion, by adding the following coupling to a set of four-form potentials $C^I_4$
\be\label{minimalC4}
\calq_I\int C^I_4\, ,
\ee
which is clearly invariant under the gauge transformations
 \be\label{C4gauge}
 C_4^I\rightarrow C_4^I+\d\Lambda_3^I\, .
 \ee

Let us now try to run the dualization prescription described in the previous subsection. We should promote the constants $N_A$ to real scalar fields $y_A$ and to add the last term in \eqref{bosmaster} to the Lagrangian.  The coupling \eqref{minimalC4} should then be replaced by
\be\label{minC4new}
\int \Big(\calq^{\rm bg}_I+\frac12\cali_I^{AB}y_A y_B+Q^A_I y_A\Big)C_4^I
\ee
Clearly, \eqref{minC4new} breaks the gauge invariance under \eqref{C4gauge}, which should be re-installed. A general way to restore a symmetry is the St\"uckelberg trick, which in the considered case amounts to introducing the following St\"uckelberg gauge transformations of the three-form potentials
\be\label{A3gauge2}
 A^A_3\rightarrow A^A_3-(\cali_{I}^{AB}y_B+Q^A_I)\Lambda_3^I\, .
 \ee
Indeed, it is easy to see that in this way the variations of \eqref{minC4new} and of the last term in \eqref{bosmaster} under \eqref{C4gauge}  precisely cancel each other. 

This is however not the end of the story, because the contribution $\cali_I^{AB}y_B$ to the charge that defines the three-form gauging \eqref{A3gauge2} is not a constant. This not only introduces consistency issues regarding the compactness of the three-form gauge symmetry, but actually results in an obstruction to the dualization procedure, basically because the $y_A$ should eventually be expressed in terms of gauge invariant field-strengths of $A^A_3$, which should in turn define their own charges under \eqref{C4gauge}.       
 
The only apparent way to get out of this impasse is to  choose the lattice $\Gamma_{\rm EFT}$ of fluxes to  be dualized
so that the induced pairings defined in \eqref{gfd1} vanish
\be
\cali^{AB}_I=0\,.
\ee
In other words, $\Gamma_{\rm EFT}$ must be {\em isotropic}  with respect to all the pairings  $\cali_I^{\cala\calb}$  entering the tadpole conditions \eqref{4charge}. 
With such a choice, the tadpole conditions  \eqref{4charge2} become linear 
\be\label{lineartad}
\calq_I=\calq^{\rm bg}_I+Q^A_IN_A =0\, , 
\ee
and one no longer encounters any obstruction  to dualize  the constants $N_A$. Indeed, the term \eqref{minC4new} reduces to  
\be\label{minC4newb}
\int \Big(\calq^{\rm bg}_I+Q^A_I y_A\Big)C_4^I\, ,
\ee
and \eqref{A3gauge2} becomes a well-defined gauge symmetry
\be\label{lin3gauge}
A^A_3\rightarrow A^A_3-Q^A_I \Lambda_3^I\,.
\ee
Upon integrating out $y_A$ from the resulting parent Lagrangian one gets the dual action.  This can be obtained from the three-form action \eqref{dualF4lagr} by  adding the term
\be\label{C4cc}
\calq^{\rm bg}_I\int C^I_4\, ,
\ee 
and replacing   $F_4^A=\d A^A_3$ with the  field-strengths 
\be\label{tildeF4}
\hat F^A_4\equiv F^A_4+Q^A_I C^I_4
\ee
which are gauge invariant under \eqref{lin3gauge}.\footnote{As a check, one can rederive the tadpole condition in the dual formulation, by extremizing the action with respect to $C_4^I$.  One gets the equations $\calq^{\rm bg}_I- T_{AB}Q^A_I(*\hat F^B_4+h^B)=0$, which indeed reduce to  \eqref{lineartad} after having solved the $A^A_3$ equations of motion by setting $T_{AB}(*\hat F^B_4+h^B)=-N_A$.} 

Finally, we observe that one may further restrict $\Gamma_{\rm EFT}$ to an affine sublattice which identically solves \eqref{lineartad}. Then, the corresponding three-form description would not require any four-form potential. However,  as we will discuss in section \ref{sec:sm3}, the formulation with potentials $C_4^I$ allows one to discuss tadpole-changing configurations with 3-branes ending on membranes.


\subsection{Axions and two-form gaugings}
\label{sec:2gauging}
  
Suppose now that we are in (typically asymptotic) region of the field space in which the theory develops a set of approximate axionic symmetries. We may single out a corresponding set of complex fields $t_\Lambda$ with periodicity $t_\Lambda\simeq t_\Lambda+1$ such that the approximate shift symmetry is described by a constant shift of the axions $a_\Lambda\equiv \Re t_\Lambda$. The EFT can be approximated by an EFT in which the axions are dualized to two-form potentials $\calb^\Lambda_2$, with field-strengths $\calh^\Lambda_3=\d\calb^\Lambda_2$. Manifest supersymmetry then requires that $\Im t_\Lambda$ are traded for their `conjugated' real fields 
\be
l^\Lambda\equiv -\frac{1}{2}\frac{\del \calk}{\del \Im t_\Lambda}\, .
\ee
Here we are adopting the Weyl-invariant formulation, in which $l^\Lambda$  have Weyl dimension $2$. The EFT is then specified by a kinetic function $\calf(z,\bar z,l)$ which is obtained by Legendre-transforming  $\calk$:
\be
\calf=\calk+2l^\Lambda\Im t_\Lambda\,.
\ee


It is natural to consider a St\"uckelberg gauging of   $\calb^\Lambda_2$ under the two-form gauge symmetries $A^A_3\rightarrow A^A_3+\d\Lambda_2^A$:
\be\label{2gauge}
\calb^\Lambda_2\rightarrow \calb^\Lambda_2-c^\Lambda_A\Lambda^A_2\,,
\ee
where $c^\Lambda_A$ should be appropriately quantized constants. 
For instance, by normalizing the two-form potentials so that the Wilson lines $\frac{1}{2\pi}\int\calb^\Lambda_2$ have periodicity one, we must require that  $c^\Lambda_A\in\mathbb{Z}$. 
One then constructs the modified field-strengths 
\be\label{tildeH3}
\hat \calh^\Lambda_3\equiv \calh^\Lambda_3+c^\Lambda_AA^A_3\, ,
\ee
which substitute $\calh^\Lambda_3=\d\calb^\Lambda_2$ in the EFT.
Note that the gaugings \eqref{2gauge} and \eqref{lin3gauge} are mutually consistent only if 
\be\label{23cond}
c^\Lambda_AQ^A_I=0\, .
\ee
For instance, this condition guarantees  that the  modified field-strengths \eqref{tildeH3} are   invariant under both \eqref{2gauge} and \eqref{lin3gauge}. 

As  will be explained in section \ref{sec:susy2}, one can perform this gauging in supersymmetric way, in which the two-forms $\calb^\Lambda_2$ are embedded in linear multiplets. The corresponding Weyl-invariant bosonic action is discussed in detail in appendix \ref{app:SWLc} -- see equations \eqref{SW_CLgaugedb} and \eqref{ThVdefgl}. In the same appendix, it is also explained how to pass to the Einstein frame EFT, which is better described in terms of the Legendre transform
\be\label{elldef}
F(\phi,\bar\phi,\ell)=K+2\ell^\Lambda\Im t_\Lambda\quad  \ \text{with} \ \quad \ell^\Lambda\equiv -\frac{1}{2}\frac{\del K}{\del \Im t_\Lambda}\, ,
\ee
of the ordinary K\"ahler potential $K(\phi,\bar\phi,\Im t)$. In the Einstein frame, we can make the identification
\be\label{weylfixL}
l^\Lambda =M^2_{\rm P}\, \ell^\Lambda\quad~~~~~~~~~\text{(Einstein frame)}\, .
\ee
Notice that the new scalars $\ell^\Lambda$ are dimensionless. 
In the Einstein frame, the bosonic part of this EFT action has the following form (see appendix \ref{app:SWLc})
	\be
	\label{SW_GF_Lag_LCh}
	\begin{split}
		S_{\rm bos} &= M^2_{\rm P} \int \left( \frac12\, R *1   -F_{i\bar \jmath}\,  \d \phi^i \wedge *\d \bar \phi^{\bar\jmath} + \frac14 F_{\Lambda\Sigma}  \d \ell^\Lambda \wedge * \d \ell^\Sigma \right)
		\\
		&\quad\,+\frac{1}{4 M_{\rm P}^2} \int F_{\Lambda\Sigma}\,  \hat\calh_3^\Lambda \wedge * \hat \calh^{\Sigma}_3 +  \int \left(\frac{\ii}{2} F_{\bar \imath \Sigma}\, \d \bar \phi^{\bar\imath} \wedge \hat\calh^{\Sigma}_3  + {\rm c.c.}\right)+ S_{\text{three-forms}}\,,
	\end{split}
	\ee
where the three-form action has the same form as \eqref{dualF4lagr}, with
	\begin{subequations}\label{ThVdefgl_wf2}
		\begin{align}
		T^{AB}&\equiv 2M^4_{\rm P}\, e^{\tilde F}\,\Re\left({F}^{i \bar \jmath}\,D_i  \Pi^A \bar D_{\bar \jmath} \bar\Pi^B-(3 - \ell^\Lambda \tilde F_\Lambda) \Pi^A\bar\Pi^B\right)\,,\label{ThVdef1gl_wf2}
		\\
		\begin{split}
		\Upsilon^A&\equiv 2M_{\rm P}\, e^{\tilde F}\Re \Bigg[{F}^{i \bar \jmath}\,\left( D_{\bar \jmath}  \bar{\hat W}  + \frac{\ii}{2} M_{\rm{P}}^3 F_\Lambda c_{B}^\Lambda D_{\bar \jmath} \bar \Pi^B + \ii M_{\rm{P}}^3 F_{\Lambda \bar \jmath} c_B^\Lambda \bar \Pi^{\bar B}\right) D_i\Pi^A 
		\\
		&\qquad\qquad\, - (3 - \ell^\Lambda \tilde F_\Lambda) \left(\bar{\hat W}+ \frac\ii2 M_{\rm{P}}^3 F_\Lambda c_D^\Lambda \bar\Pi^D\right) \Pi^A - \ii M_{\rm{P}}^3 \tilde F_{\Lambda} c_B^\Lambda \bar \Pi^{\bar B} \Pi^A\Bigg]\,,\label{ThVdef2gl_wf2}
		\end{split}
		\\
		\begin{split}
		\hat V&\equiv \frac{e^{\tilde F}}{M_{\rm P}^2} {F}^{i \bar \jmath} \left( D_{i}  {\hat W} - \frac{\ii}{2} M_{\rm{P}}^3 F_\Lambda c_{A}^\Lambda D_{i} \Pi^A - \ii M_{\rm{P}}^3 F_{\Lambda i} c_A^\Lambda \Pi^{A} \right) \times
		\\
		&\quad\,\qquad\qquad  \times \left(D_{\bar \jmath}  \bar{\hat W} + \frac{\ii}{2} M_{\rm{P}}^3 F_\Lambda c_{B}^\Lambda D_{\bar \jmath} \bar \Pi^B + \ii M_{\rm{P}}^3 F_{\Lambda \bar \jmath} c_B^\Lambda \bar \Pi^{\bar B} \right)
		\\
		&\quad\,- (3 - \ell^\Lambda \tilde F_\Lambda) \frac{e^{\tilde F}}{M_{\rm P}^2} \left|\hat W - \frac\ii2 M_{\rm{P}}^3 F_\Lambda c_A^\Lambda \Pi^A \right|^2 - M_{\rm{P}}^4 e^{\tilde F} F_{\Lambda \Sigma} c_A^\Lambda c_B^\Sigma \Pi^A \bar \Pi^B
		\\
		&\quad\,- M_{\rm{P}} e^{\tilde F}  \left[ - \ii c_B^\Sigma \tilde F_\Sigma \Pi^B \left( \bar{\hat W} + \frac\ii2 M_{\rm{P}}^3 F_\Lambda c_A^\Lambda \bar\Pi^A \right)+{\rm c.c.}\right] .\label{ThVdef3gl_wf2}
		\end{split}
		\end{align}
	\end{subequations}

It is instructive to anticipate also what happens if one dualizes the linear multiplets back to the chiral multiplets associated with the complex scalars $t_\Lambda$. In practice, it is easier to also simultaneously dualize the field strengths $F^A_4$ back to the constants $N_A$.  Using $z^a$ to denote the other complex fields including the Weyl compensator, one gets  a supersymmetric EFT with  homogeneous superpotential of the form \eqref{wvw}, with $\hat\calw$ replaced by
\be\label{gpsup0}
\begin{aligned}
\hat\calw'(t,z) = -c^\Lambda_A\, t_\Lambda\calv^A(z)+\hat
\calw(z)\, .
\end{aligned}
\ee
Correspondingly, the Einstein-frame superpotential  takes the form 
\be\label{gpsup}
\begin{aligned}
W(\phi,t)=&M^3_{\rm P}\,N_A\Pi^A(\phi)+\hat W'(\phi,t)\\
&\quad~~~~\text{with}\quad\hat W'(\phi,t)\equiv-M^3_{\rm P}\, c^\Lambda_A\, t_\Lambda\Pi^A(\phi)+\hat W(\phi)\, .
\end{aligned}
\ee
One may then  extend in an obvious way the formulas of section \ref{sec:bosduality} in order to include also the chiral fields $t_\Lambda$ and dualize the constants $N_A$ back to three-form potentials $A^A_3$, and with $\hat{W}$ substituted by $\hat W'$.  As we will see, most of  the perturbative superpotentials of string/M-theory compactifications can be described by the first two terms of the superpotential \eqref{gpsup} and can  be  interpreted as generated by the supersymmetrization of the gaugings \eqref{2gauge} of the dual two-form potentials $\calb^\Lambda_2$. 

As in section \ref{sec:hierarchy}, the gauging of three-forms by four-forms and the gauging of two-forms by three-forms will coexist with other gaugings, that appear in the presence of vector multiplets. These can be either expressed as the familiar St\"uckelberg gauging \eqref{Stu} or as its magnetic dual gauging of a one-form by a two-form. If one expresses all of them in terms of the latter, the complete set of gaugings of the EFT will be similar to the content of table \ref{hgauging}. The gaugings involving 4d gauge bosons as well as their supersymmetrization have been largely studied in the literature. Although generically present in string compactifications,  they will not play any particular role in our discussion and we will not further consider them.


\subsection{The EFT duality group}
\label{sec:dualities}

In general, there may exist a duality group of transformations which acts on   the flux   lattice $\Gamma$ as well as on the EFT fields. 
Hence, in a traditional  EFT depending on fixed non-vanishing fluxes $\caln_A$, part or all of the duality group is explicitly  broken. 

Suppose now that $\caln_\cala$ is split as in \eqref{Nsplit} and consider a duality subgroup $G_{\rm dual}$ that leaves $\caln_{\cala}^{\rm bg}$ unchanged, hence acting only on the constants $N_A$. By using our  EFT formulation in terms of three-forms, the constants $N_A$ are traded for dynamical three-forms $A^A_3$. In such a formulation $G_{\rm dual}$ acts as an actual symmetry group of the action, which is  only {\em spontaneously} broken.

More concretely, given a homogeneous K\"ahler potential $\calk(z,\bar z)$ and a superpotential \eqref{wvw}, a class of  dualities are given by isometries $z^a\rightarrow z'^a$ which leave $\calk(z,\bar z)$ and $\hat\calw(z)$ invariant and act linearly on the periods 
\be\label{perduality}
\calv^A(z)\rightarrow \calv^A(z')=R^A{}_B\calv^B(z)\, .
\ee 
In the presence of non-vanishing  flux quanta $N_A$, the superpotential   \eqref{wvw} is clearly not invariant under such a transformation, which may be regarded as a `spurionic' symmetry if $N_A$ transform in an opposite way: 
$N_A\rightarrow N_A'=(R^{-1})^B{}_AN_B$.\footnote{Notice that, by imposing the Weyl-fixing and splitting the homogeneous periods $\calv^A$ as in \eqref{homper}, the periods $\Pi^A$ may transform linearly only up to a pre-factor, which should then be compensated by a K\"ahler tranformation of $K$.} Recalling the procedure followed for constructing the three-form formulation, or more directly from  explicit formulas like \eqref{dualF4lagr}, with \eqref{wfTAB} and \eqref{hatV}, or \eqref{SW_GF_Lag_LCh}, with \eqref{ThVdefgl_wf2}, it is clear that the three-form theory is exactly invariant under the duality transformation provided that the three-form potentials transform as follows, 
\be
A^A_3\rightarrow A'^A_3=R^A{}_BA^B_3\, .
\ee
Such a symmetry is  spontaneously broken once a certain vacuum sector specified by  \eqref{F4vevb} is selected. 

Another class of possible duality transformations appear in the models with superpotentials of the forms \eqref{gpsup}. These are associated with the integral shifts $t_\Lambda\rightarrow t_\Lambda +n_\Lambda$, with $n_\Lambda\in\mathbb{Z}$, and are explicitly broken, even though the explicit breaking may be compensated by shifting $N_A$  to $N_A'=N_A+c_A^\Lambda n_\Lambda$. 
To analyze this case, let us consider the corresponding Weyl-invariant EFT  with three-forms $A^A$  and chiral fields $(z^a,t_\Lambda)$. By extending the formulas \eqref{ThVdef} in order to include the chiral fields $t_\Lambda$ in an obvious way and to replace $\hat\calw$ with  $\hat\calw'$ as defined in \eqref{gpsup0}, it is immediate to check that a  shift $t_\Lambda\rightarrow t_\Lambda+n_\Lambda$ induces the shifts
$\Upsilon^A\rightarrow \Upsilon^A-n_\Lambda c^\Lambda_B T^{AB}$ and $\hat V\rightarrow \hat V-n_\Lambda c^\Lambda_B h^B+\frac12 n_\Lambda n_\Sigma   c^\Lambda_Ac^\Sigma_BT^{AB}$. It  follows that the three-form action \eqref{dualF4lagr} is exactly invariant under such transformations. Hence, also in this case, the duality transformations are proper symmetries of the three-form action, which are only spontaneously broken by the choice of a vacuum sector. Notice that, on the one hand, the same conclusion would hold also in the presence of corrections depending on $e^{2\pi\ii k^\Lambda t_\Lambda}$, with $k^\Lambda\in\mathbb{Z}$, which break the continuous shift symmetries, but preserve the discrete ones.  On the other hand, in absence of such corrections, one may make a further step and dualize the chiral fields $t_\Lambda$ to the linear multiplet bosons  $(l^\Lambda,\calb^\Lambda_2)$. In the resulting formulation the original shift symmetries are  traded for the compact gauge symmetry of the two-form potentials $\calb^\Lambda_2$.


\section{Effective strings, membranes and 3-branes}
\label{sec:sm3}

Having at hand a general EFT with two-, three- and four-form potentials, it is natural to introduce  strings, membranes and 3-branes which minimally couple to them.  In fact, the presence of strings and membranes of any charge would be compatible with the completeness conjecture  \cite{Polchinski:2003bq}   for these extended objects. The extension of the completeness conjecture to  3-branes is less obvious since, as we will see, they have somewhat peculiar properties.

Membranes and strings will be treated as effectively fundamental, in the sense that one cannot `resolve' their microscopic structure at the low-energy EFT level.  It will be assumed that the EFT admits a parametrically controlled regime in which their tensions are high enough, so that they can be treated  semiclassically. This indeed happens in our examples of sections \ref{sec:IIBstringy} and \ref{sec:IIAstringy}. 

On the other hand 3-branes are more peculiar, since they have trivial dynamics in four dimensions. However, they may end on membranes and, since they contribute to the tadpole conditions, they can introduce interesting changes thereof. Furthermore, in stringy motivated EFTs, 3-branes may support non-trivial (typically gauge) sectors, which can change as one crosses a membrane where 3-branes end. 

Let us start from the strings. Their inclusion  as `fundamental' objects of the EFT assumes the existence of a regime 
with an (approximate) symmetry under constant shifts of the axions $a_\Lambda\equiv \Re t_\Lambda$. A string carrying a set of charges $e_\Lambda$ couples to the dual two-forms $\calb^\Lambda_2$ as usual through a WZ-term of the form
$e_\Lambda\int_\calc \calb^\Lambda_2$, where $\calc$ denotes the string world-sheet, which induces a non-trivial monodromy $t_\Lambda\rightarrow t_\Lambda+e_\Lambda$ around $\calc$. Of course, the string will couple to the metric through a (generically field-dependent) tension.   In a general non supersymmetric EFT,  the tension of the string could be completely unrelated to its charges $e_\Lambda$. In the supersymmetric cases, instead, as we will discuss in detail in section \ref{sec:susy3}, a local fermionic worldvolume kappa-symmetry  completely fixes the relation between them, and makes the strings BPS objects that preserve half supersymmetry of the bulk. As a result, the complete bosonic string action is given by
\be\label{bostring}
\begin{aligned}
S_{\rm string}&=-\int_\calc\d^2\zeta\, \left|e_\Lambda l^\Lambda \right|\sqrt{-\det h}+e_\Lambda\int_\calc \calb^\Lambda_2\quad~~~~~~~~~ \text{(Weyl invariant)},\\
&=-M^2_{\rm P}\int_\calc\d^2\zeta\, \left|e_\Lambda \ell^\Lambda \right|\sqrt{-\det h}+e_\Lambda\int_\calc \calb^\Lambda_2\quad~~~~~~\text{(Einsten frame)}.
\end{aligned}
\ee
where $\zeta^i$ are world-sheet coordinates and $h_{ij}$ denotes the induced metric (the world-volume indices $i,j=0,1$ should not be confused with the indices of the physical bulk chiral fields $\phi^i$). 
Hence, in the Einstein frame, the field-dependent string tension is 
\be\label{stringT}
\calt_{\rm string}=M^2_{\rm P}\,\left|e_\Lambda \ell^\Lambda\right|\, .
\ee

Let us now consider membranes. They are characterized by a set of charges $q_A$ which define their minimal coupling to the three-form potentials $q_A\int_\Sigma A^A_3 $, where $\Sigma$ is the membrane world-volume. This coupling modifies the equation of motion \eqref{A3eomb} by a delta-function localized on $\Sigma$,  hence sourcing a jump $N_A\rightarrow N_A+q_A$. As the flux quanta $N_A$, the membrane charges $q_A$ must be appropriately quantized. 

By generalizing the results of \cite{Bandos:2018gjp, Bandos:2019wgy}, in section \ref{sec:susy3} we will show that also the (field dependent) tension of the membranes is completely fixed by kappa-symmetry in relation to the three-form coupling. As a result,  the bosonic part of the membrane effective action reads
\be\label{bosmem}
\begin{aligned}
S_{\rm mem}&=-2\int_\Sigma\d^3\zeta\, \left|q_A \calv^A \right|\sqrt{-\det h}+q_A\int_\Sigma A^A_3\quad~~~~~~~~~~~~~~~~~~ \text{(Weyl invariant),}\\
&=-2 M^3_{\rm P}\int_\Sigma\d^3\zeta\, e^{\frac{1}{2}K}\left|q_A \Pi^A \right|\sqrt{-\det h}+q_A\int_\Sigma A^A_3\quad~~~~~~~~~ \text{(Einstein frame).}
\end{aligned}
\ee
From the Nambu-Goto term of this action one can deduce that in the Einstein-frame the membrane tension is
\be\label{memT}
\calt_{\rm mem}=2 M^3_{\rm P}\,e^{\frac12K}\left|q_A \Pi^A \right| \, .
\ee

Notice now that, in the presence of a two-form gauging \eqref{2gauge}, if $c^\Lambda_A e_\Lambda \neq 0$ the string WZ-term is not invariant under the gauge transformation $A^A_3\rightarrow A^A_3+\d\Lambda^A_2$.  However, one can cure this anomaly by attaching to the string one or more open membrane(s) (such that $\del\Sigma=\calc$) with  charges
\be\label{SMcond}
q_A= c^\Lambda_Ae_\Lambda\,.
\ee
As it will be clearer from our examples, in  string theory this effect is associated with the  Freed-Witten anomaly -- see for instance  \cite{BerasaluceGonzalez:2012zn,Herraez:2018vae}.

Analogously, the gaugings \eqref{lin3gauge} make anomalous the membrane WZ-terms if $Q^A_I q_A\neq0$. In turn, these anomalies can be cancelled by introducing open 3-branes. A 3-brane  contributes to the effective action by a WZ-term 
\be\label{3braneWZ}
\mu_I\int_\cals C^I_4\,.
\ee
One can cancel the anomaly by choosing 
\be\label{M3cond}
\mu_I=Q^A_I q_A\, ,
\ee 
and a 3-brane world-volume $\cals$ with boundary $\del\cals=\Sigma$. Notice that the 3-brane charges $\mu_I$ contribute to the background charges $\calq^{\rm bg}_I$, and hence to the tadpole conditions, which can then vary (stepwise) in space in the presence of open 3-branes. Differently from the string and membrane WZ-terms, we will see that the extension of \eqref{3braneWZ} to the supersymmetric case can be made kappa-symmetric without the need of the contribution of a  tension-dependent Nambu-Goto term.\footnote{Actually, when the Nambu-Goto term is included, the 3-brane becomes a `Goldstino brane' \cite{Bandos:2015xnf,Bandos:2016xyu} on which the local supersymmetry is only non-linearly realized.}

One can consider more general networks of  3-branes, membranes and strings.\footnote{See e.g. \cite{Evslin:2007ti} for string theory realizations of such configurations in terms of D-brane networks.}    
By adapting the results of \cite{Bandos:2019qok,Bandos:2019lps}, in section \ref{sec:susy3} we will show that the combined effective actions for these  brane networks can be made manifestly supersymmetric and kappa-symmetry invariant. Furthermore, the above branes can support some world-volume fields, in addition to the embedding ones (and their supersymmetric partners). It would be interesting to understand how to incorporate them in a supersymmetrically controlled way, but this is beyond the scope of present work.

\section{Type IIB models}
\label{sec:IIBstringy}

The above class EFTs can be immediately applied to describe all known classes of string/M-theory flux compactifications to four dimensions, which  share a GVW-like \cite{Gukov:1999ya} superpotential  of the form $\caln_\cala \Pi^\cala(\phi)$. As shown in section \ref{sec:IIBgauging}, many general features  of these EFTs are well illustrated in type IIB flux compactifications with O3-planes, which we now revisit from the vantage point of the general formalism developed in section \ref{sec:genEFT}.

\subsection{Weakly coupled IIB models}
\label{sec:weakIIB}

Let us again consider the simplest IIB (warped) compactifications on an orientifolded Calabi-Yau 3-fold $X={\rm CY}_3/\calr$. As in section \ref{sec:IIBgauging} we only consider the presence of O3-planes and D3-branes, while more general F-theory compactifications will be considered in subsection \ref{sec:Ftheory}. The effective superpotential takes a similar form as in \eqref{IIBGVW}
\be\label{IIBsup}
\calw= \frac{\pi}{\ell^5_{\rm s}}\int_X \Omega\wedge (F_3-\tau H_3)\, ,
\ee
with $\Omega$ the holomorphic $(3,0)$-form on $X$. However, we are now adopting the conformally invariant formulation used in \cite{eff1,Martucci:2014ska} in which $\Omega$ is dimensionless and with a fixed normalization,\footnote{Fixed such that $\frac{2\pi}{\ell^3_{\rm s}}\int_\Sigma\Omega$ gives the tension of an effective membrane obtained by wrapping a calibrated D5-brane \cite{luca1} on a 3-cycle $\Sigma$.} which is compatible with the discussion of section \ref{sec:genEFT}. 

Again, the fluxes parametrizing the $2b_3^-$-dimensional lattice $\Gamma$ are given by $\caln_A=(m_A,h_B)$. The sublattice
\be\label{IIBGamma}
\Gamma_{\rm EFT}=\{\text{R-R three-form fluxes}\}\,,
\ee
satisfies the conditions of {\it i)} being maximally isotropic\footnote{One can see that any isotropic sublattice $\Gamma_{\rm EFT}$ must have at most dimension $b_3^-$. Indeed, by  writing $I^{AB}$ in a symplectic basis of three-forms $\varpi^A$,  $\cali^{AB}$ can be  recognized  as a  metric of signature $(b_3^-,b_3^-)$.} with respect to the pairing \eqref{IIBcali} and {\it ii)} have dimension smaller than twice the number of complex fields (including the Weyl compensator) entering the superpotential \eqref{IIBsup}.
At a purely technical level, other choices would also be possible. However, as we will discuss in section \ref{sec:IIBlowE}, the choice \eqref{IIBGamma} is  the  sensible one in an EFT perturbative regime defined at large volume and weak string coupling.  

In the Weyl-invariant formulation the relevant periods are
\be
\calv^A(z)=\frac{\pi}{\ell^3_{\rm s}}\int_X \Omega\wedge \varpi^A\,,
\ee
where $z^a$, $a=0,\ldots,h^{2,1}_-$ are  homogeneous coordinates parametrizing the complex structure moduli and the Weyl compensator.  We may choose a symplectic basis of internal three-forms $\varpi^A$ to write $\calv^A(z)$ in the form \eqref{maxrank} and identify the superconformal chiral fields $z^a$ with the standard projective coordinates for the complex structure moduli \cite{Candelas:1990pi}. We may then go to the Einstein frame as described in section \ref{sec:weyl}, and write the effective superpotential in the form \eqref{wvw}
\be\label{IIBsupE}
W(\phi)=M_{\rm P}^3\left[m_A \Pi^A(\phi) -\tau \,h_A\Pi^A(\phi)\right]\, ,
\ee
recovering the expression \eqref{IIBsupo}. Notice that \eqref{IIBsupE} has precisely the structure \eqref{gpsup} (with $t_\Lambda\rightarrow \tau$, $c_A^\Lambda\rightarrow h_A$ and $\hat W=0$) of an effective superpotential generated by a two-form gauging. 

Applying the general discussion of section \ref{sec:genEFT} one recovers that this class of type IIB compactifications admits an EFT in which the internal  R-R quanta $m_A$ are not frozen, but are traded for a set of three-form potentials $A^A_3$. One must also include one four-form potential $C_4$ in the EFT, which gauges the three-forms $A^A_3$ with charges $Q^A\equiv I^{AB}h_B$, cf.~\eqref{3Q3}. 
We then have a three-form gauge symmetry acting as \eqref{gauging3}. Finally, the axio-dilaton $\tau$ can be dualized to a real scalar $\ell$ and  two-form potential $\calb_2$, which is gauged under  $A^A_3\rightarrow A^A_3+\d\Lambda^A_2$, with charges $c_A= h_A$, that is as in \eqref{gauging2}.

The resulting EFT is completely specified by the K\"ahler potential, the periods $\Pi^A$ and the charges $Q^A\equiv I^{AB}h_B$ and $c_A\equiv h_A$. These charges are determined by the NS-NS fluxes $h_A$, consistently with choosing $h_A$ as non-dynamical, and satisfy the consistency condition $Q^Ac_A=0$. To sum up, we arrive at an EFT in which one half of the original fluxes have been  traded for dynamical three-form potentials, while the other half have become the charges of two- and three-form gaugings.  

Let us now discuss  the microscopic origin of the possible effective branes which couple to such $p$-form potentials. First, there can be 3-branes, which are nothing but D3-branes. By neglecting the matter that they can support, they reduce to the effective 3-branes of section \ref{sec:sm3}.  They couple to the  R-R four-form $C_4^{\rm RR}$,  which is known to have a non-closed contribution which precisely cancels the contribution of the DBI part of the D3-brane action. A remaining closed $\Delta C_4^{\rm RR}$ ($\d_{10}\Delta C_4^{\rm RR}=0$) with four external legs  can be identified with the four-form potential $C_4$ appearing in the EFT by $\Delta C_4^{\rm RR}=\frac{\ell_s^4}{2\pi}C_4$. Hence, a D3-brane  produces the effective topological coupling $\int_\cals C_4$, which is  of the form \eqref{3braneWZ}. Notice that the sign of the charge of the effective 3-brane is assumed 
to be positive. The negative sign would correspond microscopically to an anti-D3-brane. In that case there would be no cancellation between the DBI and WZ parts, which would correspond to a goldstino brane contribution \cite{Bandos:2015xnf,Bandos:2016xyu}. 

A membrane coupling to the three-form potentials $A^A_3$  with charges $q_A$ corresponds to  a D5-brane wrapping an internal 3-cycle $\Sigma$ Poincar\'e dual to $q_A\,\omega^A\in H^3_-(X,\mathbb{Z})$. In order not to break supersymmetry, $\Sigma$ must be a special Lagrangian cycle \cite{Becker:1995kb,luca1}, giving a corresponding effective tension $\frac{2\pi}{\ell^3_{\rm s}}\int_\Sigma\Omega=2\, q_A\calv^A(z)=2\,  e^{\frac{K}{2}}q_A\Pi^A(\phi)$, in agreement with \eqref{bosmem}. These D5-branes suffer from  a Freed-Witten anomaly  if $\frac{1}{\ell^2_{\rm s}}\int_\Sigma H_3= h_Aq_B\cali^{AB}\neq 0$, but this can be cured by allowing $\mu_{\rm D3}=q_Ah_B\cali^{AB}=q_AQ^A$ D3-branes to end on them. This indeed coincides with the effective anomaly cancellation mechanism described in section \ref{sec:sm3}.
   

\subsection{Gauged linear multiplet formulation}

In the above picture, we have considered the  R-R flux quanta $m^A$ in \eqref{IIBsupE} as dynamical and generated by gauge three-forms, leaving the NS-NS quanta $h^A$ frozen, which then contribute to the superpotential of the three-form EFT by a term $-\tau h_A\Pi^A(\phi)$. However, following section \ref{sec:2gauging}, we can take a step forward and understand this contribution as the gauging of a single linear multiplet.  This will also provide an explicit example of the procedure outlined in section \ref{sec:2gauging}.
	
For the sake of clarity, consider the unwarped K\"ahler potential \cite{GKP,Grimm:2004uq}
\be\label{KIIB}
K =  -\log(2\Im\tau) -2\log V_{\rm E}+K_{\rm cs}\,,
\ee
where $K_{\rm cs}$ is the K\"ahler potential of the complex structure moduli and $V_{\rm E}$ the internal Einstein-frame volume, measured in string units. One can explicitly check that,  in the large $\Im\tau$ and large $V_{\rm E}$ limit, the  warped K\"ahler potentials of \cite{Martucci:2014ska,Martucci:2016pzt} are well approximated by \eqref{KIIB}. Furthermore, for simplicity, we assume that $h^{1,1}_- = 0$, so that $V_{\rm E}$ does not carry any implicit dependence on the axio-dilaton.
	
First, we dualize the axio-dilaton $\tau$ to the scalar component $\ell$ of a linear multiplet  by using \eqref{elldef}
\be\label{lImtau}
\Im \tau = \frac{1}{2 \ell}\,.
\ee
The field metric is then determined by the Legendre transform of \eqref{KIIB}, that is
\be\label{FIIB}
F =  \log \ell + 1 -2\log V_{\rm E}+K_{\rm cs}\,.
\ee
Owing to the block  diagonal structure of the metric, the Lagrangian \eqref{SW_GF_Lag_LCh} gives
\be
\label{IIBLLeg}
\begin{split}
		S_{\rm bos} &= M^2_{\rm P} \int \left( \frac12\, R *1   -F_{M\bar N}\,  \d \varphi^M \wedge *\d \bar \varphi^{\bar N} - \frac1{4\ell^2}   \d \ell \wedge * \d \ell \right)
		\\
		&\quad - \frac{1}{M_{\rm P}^2} \int \left( \frac1{4\ell^2} \hat\calh_3 \wedge * \hat \calh_3 \right)  + S_{\text{three-forms}}+ \calq^{\rm bg} \int C_4\,.
\end{split}
\ee
Here we have collectively denoted the complex structure and (complexified) K\"ahler moduli by $\varphi^M = (\phi^i, t^{\hat\alpha})$, with $i=1,\ldots,h^{2,1}_-$ and $\hat\alpha = 1,\ldots, h_+^{1,1}$. The field strength $\hat\calh_3$ is gauged as in \eqref{tildeH3}, with the charges $c_A^\Lambda \to h_A$, that is
	\be\label{IIBHtildeH3}
	\hat \calh_3\equiv \calh_3 + h_A A^A_3\, .
	\ee
Furthermore, as stated above, the tadpole condition is dynamically implemented  by including the last term in the Lagrangian \eqref{IIBLLeg}, with $\calq^{\rm bg}=\tilde\calq^{\rm bg}+\mu_{\rm D3}$ and by replacing  the four-form field strengths $F_4^A$ by their gauged versions: 
\be
\hat F_4^A = \d A_3^A + Q^A C_4 = \d A_3^A + I^{AB} h_B C_4\,.
\ee

The potential is fully encoded in the three-form part of the Lagrangian, which is still in the form \eqref{dualF4lagr}, with 
	\be
	\begin{aligned}
		T^{AB}& = 2M_{\rm P}^4 e^{\tilde F}\,\Re\left(K^{i \bar \jmath}_{\rm cs}\,D_i  \Pi^A \bar D_{\bar \jmath} \bar\Pi^B+   \Pi^A\bar\Pi^B\right)\,,
		\\
		\Upsilon^A& = -\frac{M_{\rm P}^4 e^{\tilde F}}{\ell}\Im \left(  h_B K^{i \bar \jmath}_{\rm cs} \bar D_{\bar \jmath} \bar \Pi^B  D_i \Pi^A -  h_B \bar \Pi^{\bar B} \Pi^A\right)\,,
		\\
		\hat V &=  \frac{M_{\rm P}^4 e^{\tilde F}}{4  \ell^2} \left( h_A h_B K^{i \bar \jmath}_{\rm cs} D_i \Pi^A \bar D_{\bar \jmath} \bar \Pi^B +    h_A h_B \Pi^A \bar \Pi^B \right)\,,
	\end{aligned}
	\ee
as it can be easily computed from the general formulas \eqref{ThVdefgl_wf2}.

Finally, we may consider a string coupled to the linear multiplet $\ell$ dual to $\tau$, say of charge $e=1$. This has tension $\calt_{\rm string}=M^2_{\rm P}\ell$ and corresponds to a D7-brane wrapping the entire compactification space $X$. As discussed in section \ref{sec:hierarchy}, this D7-brane is FW-anomalous whenever $H_3$ is non-trivial. However, the anomaly can be corrected by attaching to it a D5-brane wrapping an internal 3-cycle $\Sigma$ dual to $H_3$. The corresponding effective membrane has charges $q_A=h_A$, which, in turn, implies the identification of the gauge invariant field strength \eqref{IIBHtildeH3}, again in agreement with the expected four-dimensional condition derived in section \ref{sec:sm3}.


\subsection{Compatibility with EFT cut-offs}
\label{sec:IIBlowE}

In the above description, we have observed how the tadpole conditions and low-energy supersymmetry force us to treat the  internal fluxes in a nondemocratic way. In particular, the main feature of the above EFT with three-form potentials $A^A_3$ is that the internal R-R fluxes can actively participate in the dynamics, jumping among different values through membranes, while the NS-NS fluxes are kept fixed. At the same time, notice that the perturbative type IIB regime $\Im\tau\gg 1$ at which we are working generates a hierarchy between the energy scales associated with R-R and NS-NS fluxes. We would now like to argue that these two features are correlated with each other, and the combined picture is compatible with the standard features of a low-energy EFT.

\subsubsection*{Setting an EFT cut-off}

\begin{figure}[!th]
    \centering
	\includegraphics[width=7cm]{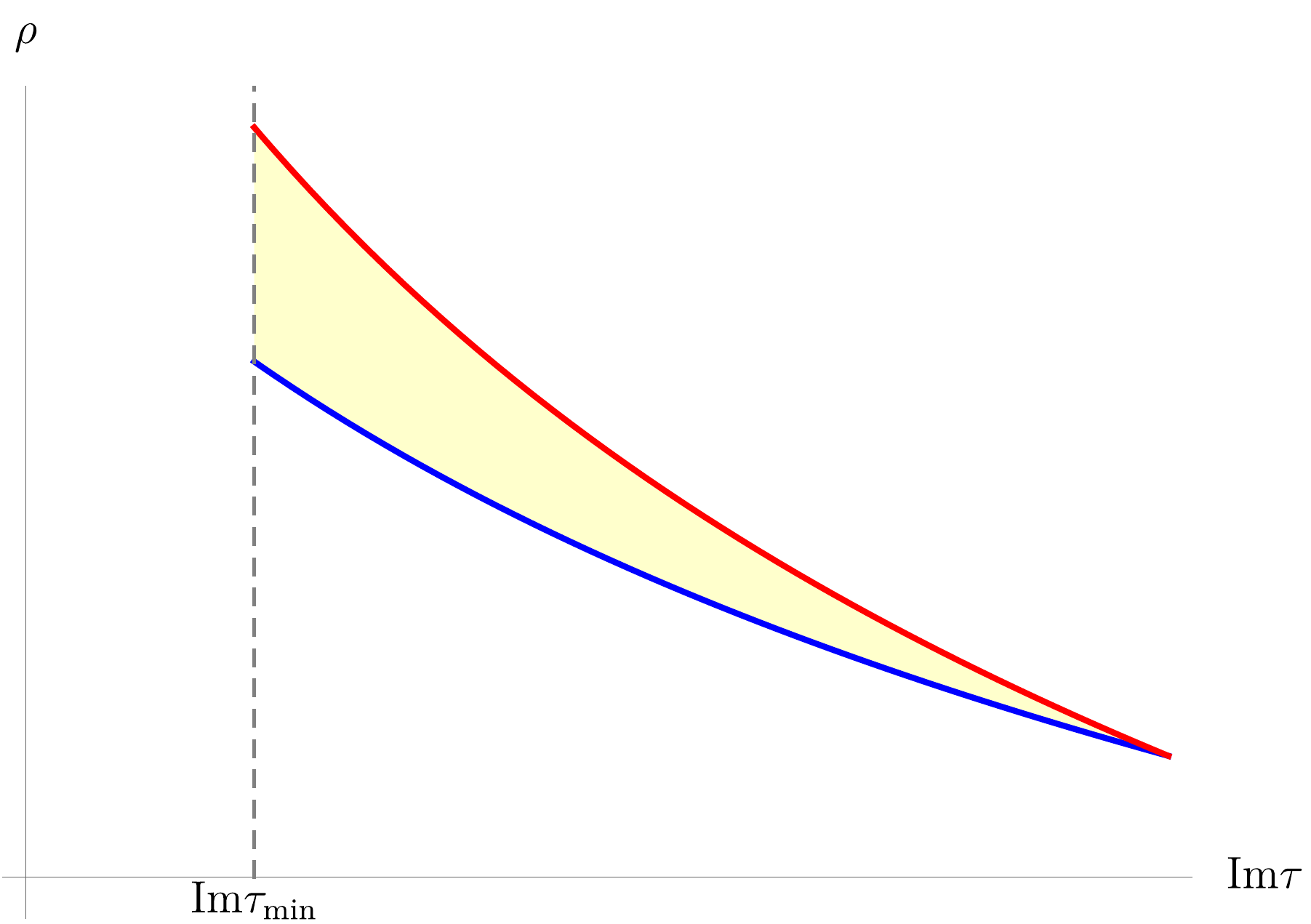}
	 \caption{\footnotesize{ Region of validity of the EFT specified by \eqref{taurange}.} }\label{fig:RegVt}
\end{figure}

In general, to characterize an EFT we may fix a (moduli-independent) UV cut-off scale $\Lambda_{\rm UV}$. In the model at hand, it is natural to set an upper bound on $\Lambda_{\rm UV}$ with the KK-scale $m_{\rm KK}$: $\Lambda_{\rm UV}\leq m_{\rm KK}$. By approximating the internal string-frame metric in a simple factorized form, one can obtain the following estimate of $m_{\rm KK}$ in Planck units  (dropping $2\pi$'s and $\calo(1)$ numerical factors for simplicity):
\be\label{LKK}
m_{\rm KK}\simeq \frac{1}{\rho^2\,\Im \tau }\,M_{\rm P}\, .
\ee 
Here we have introduced the volume modulus
\be\label{IIBrho}
\rho\equiv V^{\frac13}_{\rm s}\, ,
\ee 
where $V_{\rm s}$ is the string-frame volume of the internal space, measured in string units.
Furthermore, we require $\Lambda_{\rm UV}$ to be somewhat bigger than the mass scale $m_\phi$ induced on the  moduli by the flux potential. This mass scale is of the same order as the  flux-induced mass for the axio-dilaton, or equivalently of the three-forms involved in the corresponding $h_A$-induced St\"uckelberg gauging. A direct computation shows that
\be\label{Lflux}
m_\phi \simeq \frac{|h|\|\Pi(\phi)\|}{\rho^3\,\Im\tau}\,M_{\rm P}\,,
\ee 
where $|h|$ represents the typical $h_A$ flux quanta and  $\left\| \Pi(\phi)\right\|^2$ schematically  denotes  a contribution of the form $e^{K_{\rm cs}}K_{\rm cs}^{i\bar\jmath}\Pi^A_i\Pi^B_j$, $e^{K_{\rm cs}}\Pi^A\Pi^B$, \ldots We  assume that these terms are all of the same finite order. Therefore the above requirements
\be \label{mphiKK}
m_\phi\lesssim \Lambda_{\rm UV}\lesssim m_{\rm KK}\,, 
\ee 
translate into the following EFT condition on $\Im\tau$ and $\rho$: 
\be\label{taurange}
 \rho^2\,\Im \tau\ \lesssim\  \frac{M_{\rm P}}{\Lambda_{\rm UV}}\  \lesssim \ \frac{ \rho^3\,\Im \tau}{|h|\|\Pi\|}    \, ,
\ee
which specifies the region depicted in Fig.~\ref{fig:RegVt}. 

From these conditions we obtain the following minimal and maximal values of $\rho$ and $\Im\tau$ respectively, allowed by the EFT bounds \eqref{LKK}
\be
\rho_{\rm min}\simeq |h|\|\Pi\|\ ,\ \quad \Im\tau_{\rm max}\simeq \frac{1}{|h|\|\Pi\|}\left(\frac{M_{\rm P}}{\Lambda_{\rm UV}}\right)\, .
\ee
Hence, if the combination $|h|\|\Pi\|$ entering the estimate of $m_\phi$ in \eqref{LKK} is moderately large, the conditions \eqref{taurange} guarantee the geometric regime. Furthermore, by taking $\Lambda_{\rm UV}\ll M_{\rm P}$,  $\Im\tau$ can reach very large values. On the other hand, we will take 
\be\label{taurange2}
\Im\tau\geq \Im\tau_{\rm min}\,,
\ee
for any $\Im\tau_{\rm min}$ which should be large but much smaller than $M_{\rm P}/\Lambda_{\rm UV}$, so that 
\be
\rho^2_{\rm max}=\frac{1}{\Im\tau_{\rm min}} \left(\frac{M_{\rm P}}{\Lambda_{\rm UV}}\right)
\ee 
is large too. 

\subsubsection*{Hierarchies of membranes}

Let us now consider the formulation of this system in terms of three-form potentials and ask whether it is compatible with the above EFT picture. As discussed above the non-trivial dynamics of the three-form description corresponds to discrete flux transitions $\Delta m_A$ mediated by membranes of charges $q_A=\Delta m_A$. An estimate of the gravitational energy scales involved in such transitions is provided by  $\calt_{\rm mem}M^{-2}_{\rm P}$, where  $\calt_{\rm mem}$ is the tension of the corresponding membranes. It then follows that the set of membranes included in an EFT with a given cut-off $\Lambda_{\rm UV}$ must satisfy
\be\label{caltcond}
\frac{\calt_{\rm mem}}{M^2_{\rm P}}\lesssim \Lambda_{\rm UV} \,.
\ee
From this viewpoint, a sensible choice of EFT flux lattice $\Gamma_{\rm EFT} \subset \Gamma$ should correspond to a set of membranes that satisfy \eqref{caltcond}, for some choice of $\Lambda_{\rm UV}$ within the range \eqref{mphiKK}.

Let us consider the set of type IIB membranes made up from D5-branes and NS5-branes wrapping special Lagrangian three-cycles on the compact manifold $X$, which correspond to the R-R and NS-NS fluxes of the full flux lattice $\Gamma$, respectively. One can use \eqref{memT} to evaluate our R-R membrane tension, and that the tension of the NS-NS membrane on the same three-cycle is $\Im \tau$ times larger. Assuming small internal warping and using the effective K\"ahler potential \eqref{KIIB} we obtain
\be\label{caltcond0}
\frac{\calt_{\rm mem}^{\rm RR}}{M^2_{\rm P}}\simeq  \frac{|q|\|\Pi\|}{\rho^3(\Im\tau)^2}\, M_{\rm P}\,, \quad \quad \frac{\calt_{\rm mem}^\vee}{M^2_{\rm P}}\simeq  \frac{|q^\vee|\|\Pi\|}{\rho^3\Im\tau}\, M_{\rm P}\,,
\ee
where $\calt_{\rm mem}^{\rm RR}$ and $\calt_{\rm mem}^\vee$ stand for the tensions of R-R and NS-NS membranes, respectively, and $q$ and $q^\vee$ for their corresponding vector of quanta. From here it is easy to see that
\be\label{IIBvsKK}
\frac{\calt_{\rm mem}^{\rm RR}}{M^2_{\rm P}}\simeq  \frac{|q|}{|h| \Im\tau} \frac{\rho_{\rm min}}{\rho}\, m_{\rm KK}\,, \quad \quad \frac{\calt_{\rm mem}^\vee}{M^2_{\rm P}}\simeq  \frac{|q^\vee|}{|h|}\frac{\rho_{\rm min}}{\rho}\, m_{\rm KK}\,,
\ee
and so both kinds of membranes satisfy the condition \eqref{caltcond} for $\Lambda_{\rm UV} \simeq m_{\rm KK}$ in the large volume, weak coupling region in which we are working. In fact, as we will discuss in \ref{sec:IIAscales}, one can see \eqref{caltcond} with $\Lambda_{\rm UV} \simeq m_{\rm KK}$ as a definition of the compactification flux lattice $\Gamma$. 

It is also obvious that in this region of field space  R-R membranes are much lighter than NS-NS membranes, because $\calt_{\rm mem}^{\rm RR}/\calt_{\rm mem}^\vee \simeq (\Im \tau)^{-1} \ll 1$. This relation provides a simple  energetic justification of our choice \eqref{IIBGamma} of the sublattice $\Gamma_{\rm EFT}$ of fluxes, versus an alternative choice of isotropic sublattice. One may then wonder to which cut-off scale does this choice of EFT flux lattice correspond to. For this notice that
\be\label{IIBvsmphi}
\frac{\calt_{\rm mem}^{\rm RR}}{M^2_{\rm P}}\simeq  \frac{|q|}{|h| \Im\tau} \, m_\phi\,, \quad \quad \frac{\calt_{\rm mem}^\vee}{M^2_{\rm P}}\simeq  \frac{|q^\vee|}{|h|}\,  m_\phi\,.
\ee
Therefore, the energetic condition including R-R membranes and leaving out the NS-NS membranes is
\be\label{RRcond}
\frac{\calt_{\rm mem}}{M^2_{\rm P}}\lesssim m_\phi\,.
\ee

In other words, for this class of type IIB compactifications the choice of EFT flux lattice \eqref{IIBGamma} corresponds to \eqref{caltcond} with a cut-off scale $\Lambda_{\rm UV}$ just above $m_\phi$. One may interpret this as follows. In the three-form effective field theory, the fluxes that are fixed to background values $\caln^{\rm bg}$ already set a mass scale $m_\phi$ for the otherwise moduli of the compactification, and in particular for the gauged linear multiplets. The lattice of dynamical fluxes then corresponds to those membranes whose flux-transition scales are small compared to $m_\phi$, and do not change significantly the flux-induced mass spectrum. Therefore, with the above three-form potential formulation, one should be able to describe a mini-landscape of flux vacua in which the flux-induced masses are kept at a given scale. Notice that, since $m_\phi$ is a moduli-dependent quantity, this will in practice restrict the region of field space that our EFT can access. In fact, \eqref{RRcond} will select a bounded region of the initial EFT lattice $\Gamma_{\rm EFT}$, given by
\be\label{RRcond2}
\frac{|q|}{|h| \Im \tau}  \lesssim 1\, ,
\ee
setting a region of validity of the EFT description within $\Gamma_{\rm EFT}$. Notice that such region will have a minimal radius set by $|h|\,\Im\tau_{\rm min}$. Therefore for large values of this quantity one may effectively work with a lattice of fluxes.

Interestingly, a very similar condition to \eqref{RRcond2} is obtained by considering an effective string of charge $e$ coupled magnetically to $\tau$. Indeed, by using \eqref{KIIB} and the general formulas \eqref{stringT} and \eqref{elldef}, we can evaluate  its tension
\be\label{IIBstringT}
\calt_{\rm string}= |e|M^2_{\rm P} \,\ell\simeq\, \frac{|e|M^2_{\rm P}}{\Im\tau}\,,
\ee
which  agrees with what one gets by wrapping a probe D7-brane on the complete internal space. The condition\footnote{The quantities \eqref{caltcond} and \eqref{caltscond} measure the strengths of the gravitational energy scales associated with  membranes and strings, which should be small in EFT units set by the cut-off scale $\Lambda_{\rm UV}$. $\Lambda_{\rm UV}$ does not appear in \eqref{caltscond} since strings are codimension-two and have logaritmic backreaction.}
\be\label{caltscond}
\frac{\calt_{\rm string}}{M^2_{\rm P}}\lesssim 1
\ee
reads exactly as \eqref{RRcond2} for $q = e h$ membranes. Recall however that this is exactly the number of R-R membranes that should be attached to the otherwise anomalous string. This matching of conditions can be interpreted as the fact that including the axionic strings coupled to $\tau$ in the EFT is energetically equivalent to including the  R-R membranes attached to them, as expected from the consistency of the approach.

As a final remark, notice that  the consistency of an EFT including {\em semiclassical} membranes and strings also requires that
\be\label{mmbound}
\frac{\calt_{\rm mem}}{\Lambda^3_{\rm UV}}\gtrsim 1\quad,\quad \frac{\calt_{\rm string}}{\Lambda^2_{\rm UV}}\gtrsim 1 \,.
\ee
In the IIB models under consideration, from \eqref{caltcond0} and \eqref{IIBstringT}  we obtain the estimates
\be
\frac{\calt_{\rm mem}}{\Lambda_{\rm UV}^3} \simeq \frac{|q|\|\Pi\|}{\rho^3(\Im\tau)^2}\left(\frac{M_{\rm P}}{\Lambda_{\rm UV}}\right)^3\quad,\quad  \frac{\calt_{\rm string}}{\Lambda_{\rm UV}^2} \simeq \frac{|e|}{\Im\tau} \left(\frac{M_{\rm P}}{\Lambda_{\rm UV}}\right)^2\,,
\ee
One can then check that, in the above range of $\Im\tau$ and $\rho$, \eqref{mmbound} are always satisfied. That is, in this parametric regime the  effective membranes and strings never become light enough to cause a breakdown of  the EFT  --  see also \cite{Font:2019cxq} for a detailed discussion of energy scales in various perturbative regimes of simple concrete models.

To sum up, we have shown that this class of string models exhibits a natural self-consistent selection mechanism of the sublattice $\Gamma_{\rm EFT}\subset \Gamma$, dictated by the parametric regime one is working at. Of course, different parametric regimes would select different sublattices $\Gamma_{\rm EFT}$ within $\Gamma$. An infinite family of different possible choices is obviously obtained by applying a IIB SL$(2,\mathbb{Z})$ duality transformation, which  `rotates' the choice of three-forms $A^A_3$,  background fluxes, and  the corresponding electric and magnetic membranes.\footnote{One should take into account that $V_{\rm E}$ is invariant under IIB SL$(2,\mathbb{Z})$ dualities, and the $V_{\rm s}=(\Im\tau)^{-\frac32} V_{\rm E}$ is not. } These other choices naturally arise in the broader context of F-theory compactifications, which we now turn to discuss.

\subsection{Moving to F-theory}
\label{sec:Ftheory}

In the previous examples we have seen how, in  a weak-coupling regime with only O3-planes and D3-branes, there is a natural choice of the isotropic sublattice $\Gamma_{\rm EFT}\subset \Gamma$. In more general flux compactifications, the choice of $\Gamma_{\rm EFT}$ is less obvious. This is indeed what happens if D7-branes wrapping holomorphic surfaces are present \cite{Gomis:2005wc,luca1} and we include their world-volume fluxes into $\Gamma$, or we go to a strongly coupled F-theory regime. In order to illustrate this point let us consider an F-theory compactification on a smooth elliptically fibred  Calabi-Yau four-fold $Y$ -- see \cite{Denef:2008wq,Weigand:2010wm,Weigand:2018rez} for reviews. We can then identify $\Gamma$ with the lattice of `transversal' fluxes $G_4 \in H^4(Y;\mathbb{Z})_{\rm T}$ \cite{Dasgupta:1999ss},\footnote{For smooth elliptically fibered Calabi-Yau four-folds there is no half-integral correction to the flux quantization \cite{Collinucci:2010gz}.} that is, whose Poincar\'e dual 4-cycle has vanishing intersection number with any pair of divisors of $Y$. By introducing an appropriate basis of transversal cocycles  $\alpha^\cala=H^4(Y;\mathbb{Z})_{\rm T}$, we can expand 
\be
G_4=\caln_\cala\,\alpha^\cala\, ,
\ee
and introduce the symmetric pairing
\be\label{Fpairing}
\cali^{\cala\calb}\equiv \int_Y\alpha^\cala\wedge\alpha^\calb\, .
\ee
The D3-charge tadpole condition then takes the form \eqref{4charge}, with $\tilde Q^{\rm bg}=-\frac1{24}\chi(Y)$, where $\chi(Y)$ is the Euler characteristic of $Y$. Furthermore, the flux-induced superpotential is given by
\be\label{Ftsup}
\calw(z)=\caln_\cala \calv^\cala(z) \quad~~~~\text{with}\quad~~~~ \calv^\cala(z) \equiv \frac{\pi}{\ell^6_{\rm M}}\int_Y \Omega_4\wedge \alpha^\cala\, ,
\ee
where  $\Omega_4$ is the (dimensionless) holomorphic (4,0)-form on $Y$ and $\ell^6_{\rm M}$ is the M-theory Planck length. The chiral fields $z^a$ ($a=0,\ldots,h^{3,1}$) parametrize $Y$-complex structure moduli and the overall Weyl compensator.  

Let us now pick an isotropic sublattice $\Gamma_{\rm EFT}\subset \Gamma$. By splitting $\caln_\cala$ as in \eqref{Nsplit}, and defining $\calv^A(z)\equiv v^A_\cala\calv^\cala(z)$, we can write the superpotential \eqref{Ftsup} in the form
 \be\label{Ftsupb}
 \calw(z)=N_A\,\calv^A(z)+\hat\calw(z)\quad~~~~\text{with}\quad~~~~ \hat\calw(z)\equiv \caln^{\rm bg}_\cala\,\calv^\cala(z)\,.
\ee
Notice that, by making an Hodge decomposition of $H^4(Y;\mathbb{C})_{\rm T}$, one realises that the pairing \eqref{Fpairing} has signature $(2+h^{2,2}_{\rm T}, 2h^{3,1})$. Then the dimension of any maximal isotropic sublattice $\Gamma_{\rm EFT} \subset \Gamma$  is {\em at most} $2h^{3,1}$. This means there are {\em always} enough complex scalars $z^a$ to allow for a supersymmetric dualization of the flux quanta $N_A$ spanning $\Gamma_{\rm EFT}$ to three-form potentials $A^A_3$. 

As a concrete example, leaving a more detailed discussion of other settings to the future, we can consider the model spelled out in  \cite{Collinucci:2008pf},  for which $\dim \Gamma=23320$, $h^{3,1}=3878$, $h^{2,2}_{\rm T}=h^{2,2}-2=15562$. In  this case, following our general prescription, one would obtain an EFT with $\dim \Gamma_{\rm EFT}=7756$ three-form potentials $A^A_3$ and $\dim (\Gamma/\Gamma_{\rm EFT})=15564$ non-dynamical fluxes. The $7756$ three-form potentials may be accommodated, together with the Weyl compensator and $3877=h^{3,1}-1$ of the $Y$-complex structure moduli,  into  $3878$ double three-form multiplets. Notice that there does not appear any `natural' choice of the sublattice $\Gamma_{\rm EFT}$ and of the  $Y$-complex structure modulus which is excluded from these double three-form multiplets. Analogously, differently from what we observed in the previous subsection, there is no obvious  hierarchy between the energy scales of the potential and the membrane tensions.

It is instructive to observe how such `democracy' is removed by going to weak coupling.  In this limit, the total amount of fluxes  $23320$ splits into $2\times 300$ bulk R-R plus NS-NS three-form fluxes and $22720$ D7-brane fluxes, while the moduli $z^a$ include the overall Weyl compensator, the axio-dilaton $\tau$, $149$ bulk complex-structure moduli, and $3728$ D7-brane geometric moduli. The $149$ complex-structure moduli and the Weyl compensator can combine with 300 R-R bulk fluxes into double three-form multiplets, as we did in the previous subsections. The axio-dilaton $\tau$ can again be dualized to a linear multiplet which is gauged, with charge defined by the $300$ NS-NS fluxes. The  $7756-300=7456$ D7-brane fluxes in $\Gamma_{\rm EFT}$, can  then be  accommodated, together with the   D7-brane moduli,  into other 3728 double three-form multiplets. The remaining   15264 D7-brane fluxes stand non-dynamical and contribute to $\hat\calw$ in  \eqref{Ftsupb}.

Leaving a more detailed study of the three-form formulation of the EFT for general F-theory compactifications -- as well as their weak-coupling limits --  to the future, we close this section by briefly discussing the microscopic origin of the three-form gauging 
\be\label{F3gauge}
A^A_3\rightarrow A^A_3-Q^A\Lambda_3\quad~~~~\text{with}\quad~~~~ Q^A=\cali^{\cala\calb}v_{\cala}^A\caln_\calb^{\rm bg}\, ,
\ee
see \eqref{lin3gauge} and \eqref{3Q}. This can be understood from the perspective of the dual  M-theory compactified to three dimensions on an elliptically fibred Calabi-Yau four-fold $Y$ and the derivation is similar to the weakly-coupled IIB case discussed in section \ref{tadpole1}. One can start from the eleven-dimensional Bianchi identity
\be
\d F_7-\frac12 F_4\wedge F_4=(\text{M2-brane charge})\, .
\ee
By appropriately reducing  $F_7$, $F_4$ and the corresponding potentials to three dimensions one gets a set of gauge-invariant three-forms which are dual to the F-theory field-strengths $\hat F^A_4=\d A^A_3+Q^A C_4$ -- we leave the details to the reader.  These are gauge invariant under $C_4\rightarrow C_4+\d\Lambda_3$ provided that $A^A$ transform as in \eqref{F3gauge}. In the weak-coupling limit, this generalizes the gauging of three-forms discussed in the previous subsections by including a sector supported on the D7-branes.

\section{Type IIA models}
\label{sec:IIAstringy}

IIA orientifold models are in many aspects similar to the IIB models considered above. We  will then be briefer and concentrate on the distinguishing features. Let us start by reviewing the standard EFT of these models, see for instance \cite{Grimm:2004ua,Grimm:2011dx,Kerstan:2011dy,Carta:2016ynn,Ibanez:2012zz}, in a form  which can be immediately upgraded to an EFT including three-form potentials.  
We start from an internal space of the form $X={\rm CY}_3/\calr$, where $\calr$  refers to an  O6-involution. Assuming for the moment that there are no D6-branes, the spectrum contains $b_2^-$ chiral fields $\phi^i$, which include K\"ahler structure and internal $B_2$ moduli, and $b_3^+$ chiral fields $t_\Lambda$, which combine the internal $C^{\rm RR}_3$ axions, the complex structure moduli and the dilaton. In the large volume limit, we can identify
\be
\phi^i=\frac{1}{\ell_{\rm s}^2}\int_{C^i}(B_2+\ii J)\quad,\quad t_\Sigma=\frac{1}{\ell_{\rm s}^3}\int_{S^+_\Lambda}(C^{\rm RR}_3+\ii e^{-\phi}\Re\Omega)\, ,
\ee
where $C^i$ and $S^+_\Lambda$ provide a basis of odd 2-cycles and even 3-cycles respectively. It is natural to combine the fields $\phi^i$ with the Weyl compensator $u$ into $b_2^-+1$ chiral $z^a$, $a=0,\ldots, b_2^-$, with
\be
z^0\equiv u \ , \quad z^i\equiv u \phi^i\, .
\ee
On the other hand, at weak string coupling, it will be convenient to keep using $t_\Sigma\simeq t_\Sigma+1$ as elementary fields with vanishing Weyl dimension.  

In general we can write the relevant periods $\calv^A$ in the form \eqref{maxrank}, where we can locally set $\calg_a=\del_a\calg(z)$, with $\calg(z)$ being the degree-two homogeneous prepotential associated with the underlying special geometry. In the large-volume limit we can set
\be
\calg(z)=-\frac{1}{6z^0}\,\kappa_{ijk}z^iz^kz^k+\frac12 a_{ab}\,z^az^b+\calo(e^{2\pi\ii \phi})\, ,
\ee
where $\kappa_{ijk}$ are triple-intersection numbers and the terms $a_{ab}z^az^b$ and $\calo(e^{2\pi\ii \phi})$ encode the possible perturbative and non-perturbative $\alpha'$-corrections, respectively. Away from the large volume limit, $\calg(z)$ can have a more general form. 

We are now in position to write the flux-induced superpotential. For generality, we include the effect of all kinds of (standard, geometric and non-geometric) fluxes. This will also make more manifest similarities and differences  with respect to the IIB models discussed above. At weak coupling or, more precisely, in the $\Im t_\Sigma\gg 1$ regime, the homogeneous superporpotential can be written in the form \cite{Taylor:1999ii,Kachru:2004jr,Grimm:2004ua,Koerber:2007xk,Herraez:2018vae}
\be\label{IIAsup}
\calw=N_A\calv^A(z)- c^\Lambda_A\, t_\Lambda\calv^A(z)\, ,
\ee
where $N_A$, $A=1,\ldots,2+2b_2^-$, represent the contribution of the internal R-R  fluxes, while   $c^\Lambda_A$ encodes the remaining NS-NS ordinary, geometric and non-geometric fluxes. More explicitly, by using the natural large volume splitting $z^a=(z^0,z^i)$, we can decompose $\calv^A(z)=(z^0,z^i,\calg_0,\calg_j)$ and correspondingly
\be\label{IIARR}
N_A=(e_a,m^a)= (e_0,e_i,m^0, m^j)\, ,
\ee
with $e_0,e_i,m^j, m^0$ representing the R-R 6-, 4-, 2- and 0-form fluxes respectively. We can then decompose the NS-NS fluxes into
\be\label{IIANS}
c^\Lambda_A= (c^\Lambda_0,c^\Lambda_i,c^{\Lambda 0}, c^{\Lambda j})\equiv  (h^\Lambda, \omega^\Lambda_i, R^\Lambda, Q^{\Lambda j})\, ,
\ee
where 
$h^\Lambda$ counts the  $H_3$ flux quanta, while $\omega^\Lambda_i, R^\Lambda, Q^{\Lambda j}$ represent the geometric and non-geometric fluxes -- see for instance \cite{Aldazabal:2006up,Wecht:2007wu,Ibanez:2012zz,Gao:2017gxk}.

\subsection{Three-form formulation}

The superpotential \eqref{IIAsup} looks very similar to the IIB superpotential \eqref{IIBsupE}, with the difference that $\tau$ is replaced by the $b_3^-$  fields $t_\Lambda$ and the IIB fluxes $h_A$ are replaced by the fluxes $c^\Lambda_A$. Also the number of tadpole conditions changes, since we now have $b_3^-$ conditions
\be\label{D6tad}
\calq^\Lambda\equiv I^{AB}N_A c^\Lambda_B +\calq^\Lambda_{\rm bg}=0\, ,
\ee
where $\calq^\Lambda_{\rm bg}$ represents the (negative) D6-charge carried by the O6-planes and
$I^{AB}$ represents the natural anti-symmetric Mukai pairing between even cohomology classes.
These can be written in the form \eqref{4charge}  (with $\tilde Q_I^\cala=0$) by making the index change $(\ldots)_I\rightarrow (\ldots)^\Lambda$, grouping the fluxes into  $\caln_\cala=(N_A,c^\Lambda_B)\in\Gamma$ 
and correspondingly decomposing 
\be\label{IIAcali}
(\cali^\Lambda)^{\cala\calb}\equiv\left(\begin{array}{cc}
    (\cali^\Lambda)^{AB} & (\cali^\Lambda)^{A}{}_\Theta^D\\   (\cali^\Lambda)^C_\Sigma{}^{B}
     & (\cali^\Lambda)^C_\Sigma{}_{\Theta}^{D}
\end{array}\right)=\left(\begin{array}{cc}
    0 & \delta_\Theta^\Lambda I^{AD}\\  -\delta^\Lambda_\Sigma  I^{CB}
     & 0
\end{array}\right)\, .
\ee

We can now apply our general prescription to promote the above EFT to a theory with three-form potentials. First of all, we must select the isotropic sublattice $\Gamma_{\rm EFT}\subset \Gamma$ of fluxes that will be dualized to three-forms. As anticipated by the above notation, a  possibility is to take 
\be\label{IIAGamma}
\Gamma_{\rm EFT}=\{\text{sublattice of R-R fluxes $N_A$}\}.
\ee
In the weak-coupling limit $\Im t_\Lambda\gg 1$ this choice is also natural from the energetic point of view, for reasons similar to those discussed in subsection \ref{sec:IIBlowE}. We can then pass to a description in terms of $2+2b^-_2$ three-forms $A^A_3$. These are part of the double-three form multiplets of the same kind introduced already in \cite{Farakos:2017jme}, which are completely defined by the periods $\calv^A(z)$.

At the three-form level, the tadpole conditions are implemented by introducing $b_3^-$ four-form potentials $C_{4\Lambda}\simeq C_{4\Lambda}+\d\Lambda_{3\Lambda}$,  gauging the three-forms as in \eqref{lin3gauge}. In the present case such gauging reads
\be
A_3^A\rightarrow A^A_3-Q^{\Lambda A}\Lambda_{3\Lambda}\, ,
\ee
and the tadpoles are implemented by adding a term like 
\be
\calq^{\Lambda}_{\rm bg}\int C_{4\Lambda}\,,
\ee
cf.~\eqref{C4cc}. These correspond to the choice of background fluxes $\caln^{\rm bg}_\cala=(0,c^\Lambda_B)$. Indeed, the charges $Q^{\Lambda A}$ can be computed by inserting \eqref{IIAcali} into \eqref{3Q3}, obtaining
\be\label{IIA3charge}
Q^{\Lambda A}=I^{AB}c_B^\Lambda\, .
\ee
Furthermore, with the choice \eqref{IIAGamma} it is clear that the superpotential \eqref{IIAsup}, once Weyl-fixed, takes precisely the form \eqref{gpsup} (with $\hat W=0$). 
This means that in the three-form formulation we can dualize the  $b_3^-$  chiral fields  $t_\Sigma$ into linear multiplets  gauged  under the three-form gauge symmetry $A^A_3\rightarrow A^A_3+\d\Lambda_2^A$. The gauging acts on the linear multiplet two-forms $\calb^\Lambda$ as in \eqref{2gauge}, where the charges are  as in  \eqref{IIANS}. Hence, we see that all the three-form and two-form charges originate from the NS-NS fluxes, which we have chosen as non-dynamical fields. Furthermore, since $I^{AB}=-I^{BA}$, from \eqref{IIA3charge} it is clear that these charges automatically satisfy the constrain \eqref{23cond}, which in the present case reads $c^\Lambda_AQ^{\Sigma A}\equiv 0$.\footnote{The gauging activated by the internal $H_3$ flux quanta $h^\Lambda$
can be understood from a ten-dimensional viewpoint as in the IIB models considered in section \ref{sec:IIBstringy}. 
 The  gaugings induced by the other NS-NS geometric and non-geometric fluxes follow by the same duality chain which motivates their introduction -- see for instance \cite{Derendinger:2004jn,DallAgata:2005fjb,Shelton:2006fd,Wecht:2007wu}.} 
 
We then see that, as in the weakly-coupled IIB models of section \ref{sec:weakIIB},  the  EFT is  completely specified by the periods  $\calv^A(z)$, the K\"ahler potential (or rather, as we saw in section \ref{sec:2gauging}, by its Legendre transform) and by the gauging charges assigned by the internal NS-NS fluxes. The ten-dimensional origin of the effective membranes and strings that can be coupled to the three-forms $A^A_3$ and the two-form $\calb^\Lambda_2$ is the completely analogous  `mirror' counterpart of  the IIB ones discussed in section \ref{sec:weakIIB} and can be easily worked out.  

\subsection{Inclusion of the open string moduli}

The D6-brane moduli sector can also be introduced without many difficulties.  Let us for simplicity turn off the geometric and non-geometric fluxes, restricting $\Gamma$ so that
\be\label{IIACYcond}
\omega^\Lambda_i= R^\Lambda= Q^{\Lambda j}=0\,,
\ee 
in order to deal with proper Calabi-Yau orientifold compactifications. The chiral moduli $\chi_\alpha\simeq \chi_\alpha+1$ of a D6-brane wrapping a special Lagranian 3-cycle $S$ (including D6 and image D6) combine Wilson lines and geometric moduli, both labelled by $\alpha=1,\ldots, b^-_1(S)$.  These couple to the bulk moduli  through a contribution to the superpotential \cite{Marchesano:2014iea,luca2}.   In our Weyl-invariant notation, this contribution takes the form\footnote{In the presence of metric fluxes D6-branes also develop superpotentials quadratic in $\chi_\alpha$, see e.g. \cite{Marchesano:2006ns}.}
\be\label{D6coupling}
\calw_{\rm D6}=-n^\alpha_ a\,\chi_\alpha z^a\,.
\ee
Here $n^\alpha_ 0\in\mathbb{Z}$ correspond to the D6-brane world-volume flux quanta, which take values into $H^2_-(S;\mathbb{Z})$ and are then also labelled by $\alpha=1,\ldots, b^-_1(S)$. The remaining quantized coupling constants are given by the intersection numbers $n^\alpha_i=\int_S\omega_i\wedge \eta^\alpha$, where $\omega_i$ is the basis of $H^2_-(X;\mathbb{Z})$ used to identify  the complexified K\"ahler moduli $\phi^i$, while $\eta^\alpha$ is the basis of $H^1_-(S)$ used to identify the world-volume moduli $\chi_\alpha$ \cite{Marchesano:2014iea}. Comparing \eqref{D6coupling} with \eqref{gpsup0}, it is clear that also the coupling \eqref{D6coupling} can be interpreted as produced from a two-form gauging of the linear multiplets dual to the world-volume chiral fields  $\chi_\alpha$. To be more precise, let us denote by $\hat\calb_2^\alpha$ the corresponding two-form potentials and split $A^A_3$ into $(A^a_3,\tilde A_{3 a})$, in correspondence with the decomposition \eqref{maxrank}. Then, \eqref{D6coupling} corresponds to a gauging $\hat\calb_2^\alpha\rightarrow \hat\calb_2^\alpha-n^\alpha_a\Lambda^a_2$ under the transformation $A^a_3\rightarrow A^a_3+\d\Lambda^a_2$. Notice that this implies that the fluxes $n^\alpha_a$ must be treated as non-dynamical in the EFT. This fits well with the observation made in \cite{Herraez:2018vae} that the matrix $T^{AB}$ becomes non-invertible in the presence of such fluxes.

Notice that each D6-brane (including its image) wrapping $S$ also contributes to the tadpole charges $\calq^\Lambda$ introduced in \eqref{D6tad} by a term $\mu^\Lambda$ which identifies the Poincar\'e dual cohomology class $[S]\in H^3(X;\mathbb{Z})_+$.\footnote{The restriction of the internal $H_3$ three-form to $S$ must be trivial in cohomology.  This implies that $\mu^\Lambda$ must satisfy the condition $\tilde I_{\Lambda\Sigma}\mu^\Lambda h^\Sigma=0$, where $\tilde I_{\Lambda\Sigma}$ is the appropriate intersection number between 3-cycles. As pointed out in \cite{Camara:2005dc}, this guarantees the gauge invariance of the flux superpotential.} The D6-branes may end on  D8-branes wrapping the entire internal space, providing an example of effective 3-brane/membrane system described in section \ref{sec:sm3}. However, the presence of a D-brane moduli sector as for instance the one described above, complicates the formulation of the EFT in terms of effective 3-branes, which now contain adjoint matter. We will not try to address this interesting issue in the present paper, leaving the inclusion of non-trivial sectors supported on the effective branes  to the future.   

\subsection{Scale estimates and strong coupling}
\label{sec:IIAscales}

As already stressed several times, the  choice of $\Gamma_{\rm EFT}\subset \Gamma$ is not unique but depends on the perturbative regime in which one is computing the EFT.  In fact, different perturbative regimes may be actually characterized by different choices of the larger $\Gamma$ itself. This effect is exemplified by moving from the weak to the strong coupling regime of the above IIA models because, from the effective supergravity viewpoint, we perceive them as different classes of flux compactifications.

\subsubsection*{Weak coupling}

In order to have a reasonable control over the microscopic structure, let us again consider a purely Calabi-Yau setting with flux lattice $\Gamma$ restricted by \eqref{IIACYcond}. Furthermore, for simplicity, we will ignore the possible presence of mobile D6-branes.  
 One may then repeat an analysis similar to that of section \ref{sec:IIBlowE}, starting from our EFT formulas obtained by using the explicit form of the K\"ahler potential and of the relevant periods. However, the upshot can be more easily understood from a simple estimate of the appropriate scales. Let us follow the schematization introduced in \cite{Hertzberg:2007wc}, isolating the moduli   
 \be
 \rho\equiv V^{\frac13}_{\rm s}\, ,\quad\quad
 \sigma\equiv \frac1{g_{\rm s}}V^{\frac12}_{\rm s}\,.
 \ee
 which have diagonal kinetic terms
\be 
-\frac34 M^2_{\rm P}\frac{\del_\mu\rho\del^\mu\rho}{\rho^2}- M^2_{\rm P}\frac{\del_\mu\sigma\del^\mu\sigma}{\sigma^2}\, ,
 \ee
and denoting the remaining moduli with $\chi^\alpha$.
The effective potential can then be written in the form \cite{Hertzberg:2007wc}
\be\label{IIApot}
V=M^4_{\rm P}\left[\frac{ A_3(\chi)}{\rho^3\sigma^2}+ \sum_{n=0,2,4,6}\frac{ A_{n}(\chi)}{\rho^{n-3}\sigma^4}+\frac{ A_{\rm D6}(\chi)-A_{\rm O6}(\chi)}{\rho^3\sigma^3}\right]\, ,
\ee
where $A_3(\chi)$  comes from the NS-NS internal flux and then scales as $|h|^2$.
In the weak-coupling limit $g_{\rm s}\ll 1$ the first term of \eqref{IIApot} dominates and one gets the following estimate of the moduli mass
\be\label{mphiIIA}
m_\phi\simeq \sqrt{\frac{ A_3(\chi)}{\rho^3\sigma^2}} M_{\rm P}=\frac{g_{\rm s}}{\rho^3}\sqrt{A_3(\chi)} M_{\rm P}\,,
\ee
which agrees with its IIB counterpart given in \eqref{Lflux}. The Kaluza-Klein mass scale $m_{\rm KK}$ can again be estimated by \eqref{LKK}. Hence, the arguments leading to \eqref{taurange} (with $\Im\tau=\frac1{g_{\rm s}}$) can be applied to the IIA case as well. 

The membranes charged under the R-R fluxes $N_A$ come from D$p$-branes wrapped along internal $(p-2)$-cycles (with $p\geq 2$) and generate jumps of the internal R-R $n$-form fluxes, with $n=8-p$. On the other hand, a jump in the NS-NS fluxes $h^{\Lambda}$ is generated by a membrane obtained by wrapping an NS5-brane along an internal 3-cycle. For the corresponding tensions $\calt^{(n)}_{\rm mem}$ and $\calt^{\vee}_{\rm mem}$, respectively, one can easily estimate
\be\label{IIAmemT}
\calt^{(n)}_{\rm mem}\simeq \frac{g^2_{\rm s}\, \cala^{(n)}(\chi)}{\rho^{\frac12(n+3)}}
\,M^3_{\rm P}\, ,\quad   \quad \calt^{\vee}_{\rm mem}\simeq \frac{g_{\rm s}\, \cala^\vee(\chi)}{\rho^3} \,M^3_{\rm P}\, ,
\ee
where $\cala^{(n)}(\chi)$ and $\cala^\vee(\chi)$ scale linearly with the corresponding membrane charges $|q_{(n)}|$ and $|q^\vee|$.  
We then see that, in the limit $g_{\rm s}\rightarrow 0$ with $V^{\frac13}_{\rm s}$ fixed, there is a clear hierarchy $\calt^{(n)}_{\rm mem}\ll \calt^{\vee}_{\rm mem}$.  Furthermore, we have that
\be\label{RRcondIIA}
\frac{\calt_{\rm mem}^{(n)}}{M^2_{\rm P}}\simeq \frac{g_s}{\rho^{\frac12(n-3)}} \frac{|q_{(n)}|}{|h|}\,  m_\phi\,, \quad \quad \frac{\calt_{\rm mem}^\vee}{M^2_{\rm P}}\simeq  \frac{|q^\vee|}{|h|}\,  m_\phi\,,
\ee
and so the condition \eqref{RRcond} is satisfied for a parametrically large fraction of R-R membranes at sufficiently weak coupling, while it essentialy excludes the NS-NS membranes. Therefore, once again the criterion  \eqref{RRcond} matches the choice of EFT flux lattice, c.f. \eqref{IIAGamma}.

\subsubsection*{Strong coupling}

It is also clear that as we move to strong coupling, such hierarchy changes. Take first the case of moderately strong coupling $g_{\rm s}\simeq 1$ and apply it to \eqref{IIAmemT}. Then, one sees that
\be 
\frac{\calt_{\rm mem}^{(0)}}{M^2_{\rm P}} \simeq \frac{\cala^{(0)}(\chi)}{\rho^{\frac32}}\, M_{\rm P} \simeq  \cala^{(0)}(\chi) \rho^\frac12 \, m_{\rm KK} > m_{\rm KK}  \,. 
\ee
This shows that the effective membranes charged under  $m^0$ (alias, the Romans mass) violate the EFT condition \eqref{caltcond}, with $\Lambda_{\rm UV} \simeq m_{\rm KK}$. This fact is just a manifestation of the usual strong coupling obstruction for massive IIA, see for instance \cite{Aharony:2010af} for a recent discussion. In our language, we can interpret this result as stating that those membranes for which $\frac{\calt_{\rm mem}}{M^2_{\rm P}} > m_{\rm KK}$ not only do not correspond to elements of $\Gamma_{\rm EFT}$, but in fact must be excluded from the larger flux lattice $\Gamma$. This is to be expected, in the sense that if one works in the 10d supergravity approximation such flux lattice is defined at the compactification scale, being different for each compact manifold.

Notice that, with the choice \eqref{IIACYcond} and excluding the Romans mass from $\Gamma$, the flux contribution to the tadpole conditions \eqref{D6tad} disappears, and so does the obstruction to dualize all the remaining fluxes to three-form potentials.  Even though we do not know how to compute the K\"ahler potential and therefore the EFT at $g_{\rm s}\simeq 1$, the structure of the three-form multiplets is dictated just by the holomorphic periods, and so it is expected to enjoy some protection mechanism against perturbative corrections. One can then apply our general procedure and dualize all remaining fluxes to three-form potentials, and work out the details of the resulting description from the above formulas. In short, the fluxes $(e_i,m^j)$ are dualized  into three-form potentials $(A^i_3,\tilde A_{3j})$ which can be accommodated into $b_2^-$ double three-form multiplets, while the remaining fluxes $e_0$ and $h^\Lambda$ can be dualized to three-form $A^0_3$ and $\hat A_{3\Lambda}$ which are part of single three-form multiplets. 

\subsubsection*{M-theory regime}

We finally consider the very strong coupling regime $g_{\rm s}\gg 1$, in which the IIA description is no longer suitable and one must rather formulate the setup in terms of M-theory compactifications. The eleven-dimensional M-theory metric $\d s^2_{11}$ is related to the IIA string frame metric $\d s^2_{10}$  by 
\be\label{Mstrel}
\d s^2_{11}=e^{-\frac23\phi}\d s^2_{\rm 10}+\ell^2_{\rm M} e^{\frac43\phi}(\d y+C_1)^2\, ,
\ee
where $y\simeq y+1$ and we have chosen a parametrization such that the M-theory Planck length $\ell_{\rm M}$ coincides with the string length, $\ell_{\rm M}\equiv\ell_{\rm s}$. We can  consider a limit in which the internal seven-dimensional space $\hat X$ is large in natural $\ell_{\rm M}$-units, $ V_{\rm M}\gg 1$. For simplicity, we also assume that $\hat X$ is approximately isotropic and homogeneous. Then, from  \eqref{Mstrel} we get the relations $\rho= V_{\rm M}^{\frac37}$ and $g_{\rm s}=\langle e^\phi\rangle = V_{\rm M}^{\frac3{14}}=\rho^\frac12$. The estimates \eqref{IIAmemT}  can then be  expressed in terms of $\rho$ only:
\be\label{IIAmemTb}
\calt^{(n)}_{\rm mem}\simeq \frac{\cala^{(n)}}{\rho^{\frac12(n+1)}}\, M^3_{\rm P}\, , \quad\quad \calt^{\vee}_{\rm mem}\simeq \frac{\cala^\vee}{\rho^\frac52}M^3_{\rm P}\,.
\ee
In the M-theory regime, with metric \eqref{Mstrel}, the KK-scale becomes
\be
m^{\rm M}_{\rm KK}=\frac{M_{\rm P}}{\rho^\frac32}\,.
\ee
and therefore we obtain that
\be
\frac{\calt^{(n)}_{\rm mem}}{M^2_{\rm P}}\simeq \cala^{(n)}\rho^{\frac12(2-n)} \, m_{\rm KK}^{\rm M}\, , \quad\quad  \frac{\calt^{\vee}_{\rm mem}}{M^2_{\rm P}}\simeq \cala^{(n)}\rho^{-1} \, m_{\rm KK}^{\rm M}\, .
\ee
Hence, in the geometric regime $\rho\gg 1$, both $\calt^{(0)}_{\rm mem}$ and $\calt^{(2)}_{\rm mem}$ violate the KK scale condition $\frac{\calt_{\rm mem}}{M^2_{\rm P}}  \lesssim m_{\rm KK}$. According to our criterion above the corresponding fluxes, namely the Romans mass and the IIA R-R two-form fluxes, must not be included. Notice that the latter correspond, from the M-theory perspective,  to geometric fluxes that vanish on $G_2$-holonomy spaces.

The flux lattice $\Gamma$ in this regime is parametrized by the former type IIA fluxes $e_0,e_a$ and $h^\Lambda$. In M-theory language, $e_0$ is identified with the internal $G_7$-flux over the entire $\hat X$, while $e_a,h^\Lambda$ recombine into the flux quanta $n_I$ of  $G_4\in H^4(\hat X;\mathbb{Z})$. The associated membranes correspond to M2-branes and M5-branes on 3-cycles, respectively. On the other hand, the chiral fields $z^0,z^i$ and $ t_\Lambda$ recombine into $b_3(\hat X)+1$ chiral fields $z^0,\hat z^I$ (including the Weyl compensator), which combine the M-theory $C_3$-axions and the moduli of the $G_2$-holonomy associative three-form.
The (restricted form of the) superpotential \eqref{IIAsup} can then be uplifted to the  superpotential 
\be
\calw=e_0 z^0+n_I \hat z^I\, .
\ee
See, for example, \cite{Beasley:2002db} for a  discussion on the EFT of M-theory flux compactifications on $G_2$ spaces, and \cite{Andriolo:2018yrz} for a recent discussion to more general IIA/M-theory compactifications.

Interestingly, with this restricted choice of $\Gamma$ there are no tadpole conditions, and so in principle one may take $\Gamma_{\rm EFT} = \Gamma$. More precisely, one can see that $(e_0,n_I)$ can be dualized to three-form potentials $A^0_3,A^I_3$ which can be incorporated, together with $z^0$ and $\hat z^I$ respectively, in the single three-form multiplets defined in section \ref{sec:bosduality}. One may then see if this choice is compatible with the hierarchy of corresponding membrane tensions.  By using  the K\"ahler potential of \cite{Beasley:2002db} in the one-modulus case, one can obtain a simple estimate $m_\phi\sim M_{\rm P}\rho^{-\frac52}$ of the scaling behaviour of the moduli masses. One then finds that
\be
\frac{\calt^{(6)}_{\rm mem}}{M^2_{\rm P}} \simeq \rho^{-1}\, m_\phi\, , \quad \quad  \frac{\calt^{(4)}_{\rm mem}}{M^2_{\rm P}}\simeq \frac{\calt^{\vee}_{\rm mem}}{M^2_{\rm P}}\simeq \, m_\phi\, .
\ee
Therefore, to set $\Gamma_{\rm EFT} = \Gamma$ one must take a cut-off scale  such that $m_\phi \ll \Lambda_{\rm UV} \lesssim m^{\rm M}_{\rm KK}$.  As a result, the corresponding EFT has the attractive feature of including flux transitions that change significantly the masses of the would-be moduli, unlike in previous examples.

While this is a perfectly consistent low-energy effective action, it does not incorporate anomalous axionic strings  in its spectrum of  fundamental extended objects, as well as the associated gauging of the two-forms dual to the corresponding axions. 
Indeed, such strings are given by M5-branes wrapping four-cycles of the compactification manifold. If the integral of the internal four-form flux does not vanish over the M5-brane, a Freed-Witten anomaly will be generated on its worldvolume \cite{Witten:1999vg}, which will then be cured by M2-branes ending on the corresponding 4d axionic string. In terms of the EFT Lagrangian, we will have a series of non-trivial gaugings of the 4d two-forms dual to the $C_3$-axions by the three-form potential $A_3^0$ coupling to the M2-branes. On the other hand, the gauging coefficients are nothing but the four-form fluxes $n_I$ and, by the reasoning of section \ref{sec:2gauging}, we could adopt the dual description in terms of two-forms, in which the four-form fluxes are considered as part of the background fluxes $\caln^{\rm  bg}_\cala$ and specify the two-form gaugings.  In this case, 
we identify $\Gamma_{\rm EFT}$ with the one-dimensional lattice parametrized by $e_0$.

It seems that this class of compactifications allow for two different, complementary descriptions in terms of the three-form Lagrangians of section \ref{sec:genEFT}. Either we describe a 4d EFT containing $b_3 + 1$ classes of membranes, or we have an EFT with one class of membranes and one anomalous strings. It would be interesting to consider further examples of compactifications of this sort, and to understand whether the obstruction we find in incorporating all of these ingredients simultaneously is fundamental or can be overcome by adopting some so far unknown alternative scheme.

\section{Supersymmetric three-form actions}
\label{sec:susy}

In the following we will show that the bosonic EFTs described in section \ref{sec:genEFT}, and then the string theory models of sections \ref{sec:IIBstringy} and \ref{sec:IIAstringy} admit a manifestly supersymmetric extension, featuring, what we dub, {\em master} three-form multiplets. As anticipated in section \ref{sec:weyl}, we will use a super-Weyl invariant superspace formulation, whose main features are summarized in Appendix \ref{app:SWL}. There should exist also an equivalent superconformal formulation, but the superspace approach will allow us to naturally couple to these EFTs strings, membranes and 3-branes in a manifestly supersymmetric way.

We  start from the ordinary formulation for the $n+1$ homogeneous chiral multiplets $Z^a$ introduced in section \ref{sec:weyl}.
The corresponding supersymmetric effective Lagrangian takes the form\footnote{\label{foot:supconv}We mostly adopt the conventions of \cite{Wess:1992cp}. The main difference is that for the bosonic formulas involving differential forms  we keep using the more standard conventions of ordinary differential geometry,  reviewed for instance in \cite{Eguchi:1980jx}. See footnote \ref{foot:sForms} for more details.}
\be\label{SUGRA}
\call= \int\d^4\theta\, E\, \calk(Z,\bar Z)+\left(\int\d^2\Theta\,2\cale\,\calw(Z)+\text{c.c.}\right)\,.
\ee
For convenience, let us remind the form of the superpotential
\be\label{wvw2}
\calw(Z)=N_A\calv^A(Z)+\hat\calw(Z).
\ee
By expanding \eqref{SUGRA} in components and integrating out all auxiliary fields one gets the potential \eqref{pot1}, which was the starting point to dualize the constants $N_A$ to three-form potentials in section \ref{sec:bosduality}. Let us now see how this dualization can be performed at a manifestly supersymmetric level, generalizing the strategy followed in \cite{Farakos:2017jme,Bandos:2018gjp}. 

In order to streamline the presentation, the following formulas will be written in rigid superspace. Hence, in \eqref{SUGRA} we will set $E=1=2{\cal E}$ and, strictly speaking, the formulas will hold for a  rigid theory for $n+1$ chiral superfields $Z^a$, with  (non-necessarily homogeneous) general super- and K\"ahler potentials $\calw(Z)$ and $\calk(Z,\bar Z)$, with  $\calw(Z)$  of the form \eqref{wvw2}.  However, since we are using a super-Weyl invariant approach, all the steps can be immediately generalized to the locally supersymmetric case. One must just appropriately covariantize all the quantities,  recall that the homogeneous chiral multiplets $Z^a$ can be split  as in \eqref{Zsplit} into $n$ physical chiral multiplets $\Phi^i$ and one compensator $U$,  and restrict to homogeneous $\calw(Z)$ and  $\calk(Z,\bar Z)$. These are related  to the standard Einstein-frame  super- and K\"ahler   potentials $W(\Phi)$ and $K(\Phi,\bar\Phi)$, and one goes to the Einstein frame formulation as outlined in section \ref{sec:weyl}. In appendix \ref{app:SWL} we collect some comments  and formulas about superspace and super-Weyl invariance, which may be useful to translate from global to local Einstein-frame supersymmetry.  

\subsection{Master three-form multiplets from duality}
\label{sec:susy1}

As a starting point, we observe that the three-form potential $A^A_3$ can be identified as  a component of a real superfield $P^A$. The two-form gauge transformations \eqref{2gauge} complete to the following superfield gauge transformations  
\be\label{Sgauge}P^A\rightarrow P^A+R^A\, ,
\ee
where $R^A$ are  arbitrary real linear multiplets (such that $ D^2R^A=\bar D^2R^A=0$), which contain the gauge  two-form parameters $\Lambda^A_2$. More precisely, the component expansion of $P^A$ reads\footnote{Strictly speaking, the component expansions \eqref{VectorB} and \eqref{RealLM} hold for multiplets of rigid supersymmetry, with the  components of the corresponding superfields defined by taking the $D_\alpha$ and $\bar D_{\dot\alpha}$ derivatives of the latter. In locally supersymmetric theories the components of a multiplet are defined by acting on the corresponding superfield with curved covariant derivatives $\mathcal D_\alpha$ and $\bar {\mathcal D}_{\dot\alpha}$. See \cite{Wess:1992cp,Buchbinder:1995uq} for the definition of the components of real multiplets in the locally supersymmetric case and \cite{Bandos:2018gjp} for the definition of superspace real potentials containing gauge three-forms.} 
\be
\begin{split}
P^A =& \, p^A + \ii \theta \rho^A  - \ii \bar\theta \bar\rho^A + {\ii} \theta^2 \bar s^A -  {\ii} \bar\theta^2 s^A 
+ \theta \sigma^m \bar\theta (*A^A_3)_m  
\\
    & +\ii\theta^2\bar{\theta} \left( \bar{\chi}^A 
    +\frac{\ii}{2}\bar{\sigma}^m\partial_m \rho^A\right)-\ii\bar{\theta}^2 \theta \left( \chi^A+\frac{\ii}{2}{\sigma}^m\partial_m \bar{\rho}^A\right)+ \frac12 \theta^2\bar\theta^2 \left(d^A -\frac12\Box p^A \right) \, ,
\label{VectorB}
\end{split}
\ee
while the component expansion of the real linear superfields is
\be
\begin{split}
R^A = &r^A + \ii \theta \eta^A -\ii \bar\theta \bar\eta^A + \frac12 \theta \sigma_m \bar\theta \varepsilon^{mnpq} \partial_{[n}\Lambda^A_{pq]} 
\\
&+ \frac12 \theta^2\bar\theta \bar\sigma^m \partial_m \eta^A - \frac12 \theta \bar\theta^2 \sigma^m \partial_m \bar\eta^A -\frac14 \theta^2\bar\theta^2 \Box r^A \, .
\end{split}
\label{RealLM}
\ee
This shows that, at the component level, the gauge transformation \eqref{Sgauge} incorporates \eqref{2gauge}, and furthermore  $p^A\rightarrow p^A+l^A$ and $\rho^A\rightarrow \rho^A+\eta^A$. Hence,  $A^A_3$ appearing in \eqref{VectorB} can indeed be interpreted as a  three-form potential while the components $p^A$ and $\rho^A$ are not physical and may be set to zero by imposing a Wess-Zumino gauge. The only  gauge-invariant degrees of freedom contained in $P^A$ are  the complex scalars $s^A$, the Weyl fermions $\chi^A$ and the real scalars $d^A$. Together with the field-strengths
\be
F_4^A\equiv \d A^A_3
\ee
they can be combined in the chiral  superfield
\cite{Gates:1980ay}:
\be\label{defSA}
 S^A\equiv -\frac{\ii}{4}\bar D^2P^A=s^A+\sqrt{2}\theta\chi^A+\frac12 \theta^2( *F^A_4+\ii d^A   )+\ldots
 \ee
Hence, we can  interpret each $P^A$ as an {\em elementary} three-form multiplet and $S^A$ as its field-strength multiplet.  
 
 We can now repeat, at a superspace level, the dualization procedure described in section \ref{sec:bosduality}. As a first step, one promotes the constants $N_A$ appearing in the superpotential \eqref{wvw} to chiral superfields $X_A$, whose lowest components contain as real parts the scalar fields $y_A$. The parent Lagrangian can then be obtained by replacing the second term in  \eqref{SUGRA} (with $2\cale=1$) with  
 \be\label{newterm}
2\int\d^4\theta\, P^A\Im X_A +\Big[\int\d^2\theta\, X_A\calv^A(Z)+\int\d^2\theta\,\hat\calw(Z)+\text{c.c.}\Big]\, .
\ee
Ignoring for a moment a subtle role of boundary terms discussed in Appendix \ref{app:bdcontr}, we observe that the $P^A$ appear linearly in the action and can then be integrated out exactly. Indeed, their equations of motion are $\Im X_A=0$ which, for $X_A$ being chiral, has the unique solution 
\be\label{XN}
X_A=N_A\, ,
\ee
with $N_A$ real constants.\footnote{Of course, the discussion of section \ref{sec:bosduality}  on the quantization of the constants $N_A$ extend to the complete supersymmetric case.} It is then clear that, by plugging \eqref{XN} back into \eqref{newterm}, we get  the original theory with the superpotential \eqref{wvw2}.
As in the bosonic case discussed in section \ref{sec:bosduality}, the well-defined variation principle for $P^A$ and the gauge invariance of the action under \eqref{Sgauge} are ensured by taking as boundary conditions $X_A|_{\rm bd}=N_A$. 


We can now make the crucial step to get the dual formulation. 
Since the chiral multiplets $X_A$ appear linearly, they can be  integrated out  too. This produces the constraints 
$$
\calv^A(Z)=-\frac{\ii}{4}\bar D^2P^A\, ,
$$
which are the superspace analog of the bosonic \eqref{Fdual} and the same ones appearing in the definition \eqref{defSA} of the field-strength multiplets $S^A$. We therefore dub as {\em master} three-form multiplets those  constrained chiral multiplets which satisfy these conditions. Denoting them by $\calz^a$, we set 
\be\label{chircalvconstr}
 \calv^A(\calz)\equiv S^A
\ee 
 and then, by plugging  $Z^a=\calz^a$ into the parent Lagrangian, we get the dual effective Lagrangian   
\be\label{3formlagr1}
\int\d^4\theta\, \calk(\calz,\bar\calz)+\left(\int\d^2\theta\,\hat\calw(\calz)+\text{c.c.}\right)+\call_{\rm bd}\, ,
\ee
where $\call_{\rm bd}$ collects the boundary terms, whose structure is described in Appendix \ref{app:bdcontr}. 

We see that the bulk terms of the effective Lagrangian \eqref{3formlagr1}  depend on the field-strength multiplets $S^A$ only through the master multiplet $\calz^a$. In order to better understand the constraints \eqref{chircalvconstr} and the structure of the $\calz^a$'s,  
let us expand \eqref{chircalvconstr} in components. By using \eqref{defSA} and the analogous expansion
\be
\calz^a=z^a+\sqrt{2}\,\theta \psi^a+\theta^2 f^a+\ldots
\ee
we see that \eqref{chircalvconstr} translates into 
\begin{subequations}\label{conVS}
\begin{align}
\calv^A(z)&=s^A\, ,\label{conVS1}\\
\calv^A_a(z)\psi^a&=\chi^A\, ,\label{conVS2}\\
\calv^A_a(z)f^a&= \frac12 \left(*F_4^A+\ii d^A \right)\, . \label{conVS3}
\end{align}
\end{subequations}
Here \eqref{conVS1}, \eqref{conVS2} and  the real part of \eqref{conVS3}   simply fix  $s^A$, $\chi^A$ and $d^A$ in terms of  $z^a$, $\psi^a$ and $f^a$. The key nontrivial condition is provided by the real part of \eqref{conVS3}, which 
constrains the $f^a$'s to depend in a specific  way on the field-strengths $F_4^A=\d A^A_3$.  In other words, part of  $f^a$'s should be considered as composites of $F_4^A$ and not as auxiliary fields anymore. 

We can be more precise about this latter point, in parallel with the discussion on the invertibility of the three-form kinetic matrix $T^{AB}$ of appendix \ref{app:TAB}. For fixed $z^a$, the number  of ($z^a$-dependent) linear combinations of field-strengths $F^A_4$ that appear in $f^a$ through \eqref{conVS3} is given by the rank of the matrix $(\calv^A_a,\bar\calv^A_{\bar b})$. Hence, the incorporation of $k$ three-forms $A^A_3$ into $\calz^a$ is non-degenerate only if  $k={\rm rank}(\calv^A_a,\bar\calv^A_{\bar b})$. Clearly, this can happen only if  $k\leq 2(n+1)$, where $n+1$ is number of complex scalars $z^a$. We then find the same conditions as obtained by imposing the invertibility of $T^{AB}$.     

 By appropriately covariantizing the above formulas, integrating out the auxiliary fields and  Weyl-fixing the theory to the Einstein frame,  one can get the component action with the physical fields, now including also the three-form potentials $A^A_3$. Appendix \ref{app:SWL} contains some details about this procedure, showing in particular how the bosonic action discussed in section \ref{sec:bosduality} is recovered.

A few  comments are in order:
\begin{itemize}
    \item The dualization procedure is independent of $\calk$ and $\hat\calw$, which are completely arbitrary. Furthermore, one can easily include spectator multiplets without  affecting it. In particular, spectator chiral multiplets may  be considered as the part of the $\calz^a$ that, because of some  degeneracy of the matrix $\calv^A_a$, are not affected by the constraint \eqref{chircalvconstr} -- see  page 21 for an example.  
    
    \item The important  distinguishing features characterising the way in which the three-form multiplets enter the EFT are encoded in the periods  $\calv^A(z)$.  Notice that the periods $\calv^A(z)$ are part of the `holomorphic' structure characterizing the EFT and as such are expected to be protected quantities. After appropriate covariantization to local supersymmetry, this holomorphic structure is manifest in the Weyl-invariant formulation and becomes hidden after Weyl-fixing to the Einstein frame. 
   
      \item   It may be practically convenient to use the physical fields $z^a,\psi^a$ and $A^A_3$ in the effective action.
      Since the constrain \eqref{chircalvconstr} is manifestly supersymmetric, the component effective action is guaranteed to be  supersymmetric as well. However, one should keep in mind that $z^a,\psi^a$ and $A^A_3$ are supersymmetrically related  only through the non-linear constraint \eqref{chircalvconstr}.

   \item The boundary term $\call_{\rm bd}$ in \eqref{3formlagr1} plays a key role in the duality, since it non-trivially contributes to the action.  This  should be perhaps clearer from the bosonic discussion of section \ref{sec:bosduality}.   
   
   \item The invariance under the EFT dualities which acts linearly on the periods, observed at the bosonic level in section \ref{sec:dualities}, immediately extends to the complete superspace formulation. In particular, combining \eqref{perduality}, \eqref{chircalvconstr} and \eqref{defSA} it is clear that the elementary three-form multiplets $P^A$ must transform as the periods under the duality: $P^A\rightarrow R^A{}_B P^B$. The supersymmetric lagrangian \eqref{3formlagr1} is then automatically invariant under such duality tranformations. 

\end{itemize}

 In general, an explicit  solution of the constraint \eqref{chircalvconstr} may be non-obvious. We now discuss two extreme cases, corresponding to the two cases of bosonic EFTs considered in section \ref{sec:bosduality}, in which an explicit solution actually exists at the superspace level. In these cases \eqref{chircalvconstr} reduces to the  single and double three-form multiplets discussed in \cite{Farakos:2017jme, Bandos:2018gjp}. 

 \subsubsection*{The single three-form case}
 
 Suppose for instance that $k=n+1$, so that we can identify the indices, e.g.\ $A=a$. Assuming  the non-degeneracy of $\calv_a^A$,  locally in field space one can make a field redefinition $\calv^a(Z)\rightarrow Z^a$,  so that the original superpotential contains the linear term $N_a Z^a$. In this case the constraint \eqref{chircalvconstr} reduces to 
 \be
 \calz^a\equiv S^a =-\frac\ii4\bar D^2 P^a\, .
 \ee
That is, the master multiplets $\calz^a$ can be directly identified with the field-strength multiplets of the elementary three-form multiplets $P^a$. In such a case, \eqref{3formlagr1} reduces to the Lagrangian for elementary three-form multiplets (plus possible spectators).

 \subsubsection*{The double three-form case}
 \label{sec:double3form}

Suppose now that $k=2n+2$. By a field-redefinition, we may (locally in field space)  set
\be
\calv^A(\calz)\equiv\left(\begin{array}{c} 
\calz^a \\ \calg_b(\calz)\end{array}\right)\, , \quad \quad P^A\equiv\left(\begin{array}{r} 
\calp^a \\ \tilde\calp_b\end{array}\right)\, .
\ee
 The constraint \eqref{chircalvconstr} then splits as follows: 
\begin{subequations}\label{PhiGcond}
\begin{align}
\calz^a&=\cals^a\equiv-\frac{\ii}{4}\bar D^2\calp^a\, ,\label{PhiGcond1}\\
\calg_{a}(\calz)&=\tilde\cals_a=-\frac{\ii}{4}\bar D^2\tilde\calp_a\, .
\end{align}
\end{subequations}
Assume now that we can write $\calg_a(\calz)=\calg_{ab}(\calz)\calz^b$, which is indeed the case in the locally supersymmetric case by the homogeneity of $\calv^A$, with invertible $\calm_{ab}\equiv \Im\calg_{ab}$.
By substituting the first condition into the second, we get
\be
\bar D^2\big[\calg_{ab}(\calz)\calp^b-\tilde\calp_a\big]=0\, .
\ee
Hence, the above constraints can be explicitly solved by setting
\be\label{phiSigma}
\calg_{ab}(\calz)\calp^b-\tilde\calp_a\equiv -2\Sigma_a\, ,
\ee
where $\Sigma_a$ is an arbitrary complex linear multiplet, i.e.\ such that $\bar D^2\Sigma_a=0$. 
We can then invert \eqref{phiSigma} into 
\be\label{PPsol}
\calp^a=-2\calm^{ab}\Im\Sigma_b\,, \quad \quad \tilde\calp_a=-2\Im(\bar\calg_{ab}\calm^{bc}\Sigma_c)\, .
\ee
That is, we can consider $\Sigma_a$ as elementary superfields. By using the first of \eqref{PPsol} into \eqref{PhiGcond1} we get the following explicit form of the master three-form multiplets
\be
\label{doubleS}
\calz^a=\frac\ii2\bar D^2\left(\calm^{ab}\Im\Sigma_b\right)\,,
\ee
and by plugging this into \eqref{3formlagr1} we recover the  double three-form multiplet Lagrangian constructed in \cite{Farakos:2017jme,Bandos:2018gjp}.

Notice that the definition of the three-form constraints \eqref{PhiGcond} is invariant under possible symplectic/duality-like transformations, and makes manifest the duality properties of the double three-form multiplets of \cite{Farakos:2017jme,Bandos:2018gjp}. In particular, in the case of special K\"ahler structures, the constraints \eqref{PhiGcond} do not require the existence of a special geometry prepotential.

\subsection{Tadpole and three-form gauging}
\label{sec:susy2}

 In section \ref{sec:tadpole} it was explained how to implement flux tadpole conditions of the form \eqref{4charge}   at the EFT level. One must split the complete set of fluxes $\caln_\cala$ characterizing the EFT as in \eqref{Nsplit}, where $\caln^{\rm bg}_\cala$ denote some   non-dynamical background fluxes, while $N_A$ are the fluxes that will eventually  be dualized  to  dynamical three-forms potentials. Then the quadratic tadopole conditions \eqref{4charge} become  linear in $N_A$, as in \eqref{lineartad}, and can be implemented by coupling the theory to  four-form potentials $C^I_4$ in the EFT and gauging the potentials $A^A_3$ as in \eqref{lin3gauge} under the  three-form gauge transformations \eqref{C4gauge}. In this section we discuss the supersymmetric completion of this mechanism.

The four-form potentials $ C_4^I$ can be regarded as components of  chiral superfields $\Gamma^I$ \cite{Gates:1980ay}: 
 \be
 C_4^I=\frac12\Im(D^2\Gamma^I)|\,\d x^0\wedge \d x^1\wedge \d x^2\wedge \d x^3\, ,
 \ee
and the gauge transformations \eqref{C4gauge} and \eqref{lin3gauge} admit the following supersymmetric completion  
 \begin{subequations}\label{s3gauge}
 \begin{align}
 \Gamma^I&\rightarrow \Gamma^I+\frac14\bar D^2\Xi^I\, ,\\
 P^A&\rightarrow P^A-Q^A_I\Xi^I\, ,
 \end{align}
 \end{subequations}
where $\Xi^I_3$ are real superfield which admit an expansion similar to \eqref{VectorB},  with $A_3^A$ replaced by $\Lambda_3^I$. The first term of \eqref{newterm}, which appear in the parent Lagrangian, is not gauge-invariant anymore and must then be supplemented by
\be
\ii\int\d^2\theta\,\Gamma^I(Q^A_IX_A-\tilde\calq^{\rm bg}_I)+\text{c.c.}\, ,
\ee
where $\tilde\calq^{\rm bg}_I$ and $Q^A_I$ are defined in \eqref{3Q}. Notice that indeed the $\Gamma^I$ equations of motion, combined with the solution \eqref{XN} of the $P^A$ equations of motion, imply the tadpole condition \eqref{lineartad}. One can then integrate $X_A$ out, getting a dual theory for the three-form multiplets which is formally identical to \eqref{3formlagr1}. 
The only difference is that the constraints \eqref{chircalvconstr} defining the master three-form multiplet $\calz^a$ must be modified into 
\be\label{gicon}
\calv^A(\calz)=\hat S^A\equiv S^A-\ii Q^A_I\Gamma^I\, ,
\ee
which are  invariant under the gauge transformations \eqref{s3gauge}. The chiral superfields $\hat S^A$ defined in \eqref{gicon} provide the multiplets containing the modified field-strengths $\hat F^A_4$ introduced in  \eqref{tildeF4}.  Hence, by expanding the supespace action in components and integrating the auxiliary fields, one gets formally the same action for $z^a,\psi^a$, with  $F^A_4$ replaced by $\hat F^A_4$. This is discussed in appendix \ref{app:SWL} at the bosonic level. The resulting bosonic Lagrangian  indeed coincides with the bosonic theory obtained in section \ref{sec:tadpole}.  

\subsection{Gauged linear multiplets and axion monodromy}
\label{sec:gaugedlin}

Suppose now that the spectator sector contains a number of real linear multiplets $L^\Lambda$ (such that $D^2 L^\Lambda=\bar D^2 L^\Lambda=0$), which have the component expansion  like \eqref{RealLM}:
\be
L^\Lambda=l^\Lambda+\ldots +\theta\sigma^m\bar\theta (*\calh^\Lambda_3)_m+\ldots
\ee
where, as in section \ref{sec:2gauging}, $\calh^\Lambda_3=\d \calb^\Lambda_\Sigma$.
It is well known \cite{Lindstrom:1983rt} that these linear multiplets can be dualized to a set of chiral multiplets $T_\Lambda=t_\Lambda+\ldots$ and that the resulting EFT will be symmetric under constant shift of the axions $a_\Lambda\equiv \Re\, t_\Lambda$, which are dual to the two-form potentials $\calb^\Lambda_2$. 

As in section \ref{sec:2gauging}, we would instead like to reconsider the linear/chiral multiplet duality in presence of a two-form gauging of the form \eqref{2gauge}. This admits the following  supersymmetrization
\be\label{Lgauge}
L^\Lambda\quad\rightarrow\quad  L^\Lambda-c^\Lambda_AR^A\, ,
\ee
where $R^A$ are the linear multiplets introduced in \eqref{RealLM} that define the  gauge transformations \eqref{Sgauge}. One can then construct the gauge-invariant real superfields 
\be\label{hatL}
\hat L^\Lambda\equiv L^\Lambda+c^\Lambda_AP^A\, ,
\ee
which satisfy the modified `Bianchi identity'
\be\label{conhatL}
-\frac{\ii}{4}\bar D^2\hat L^\Lambda=c^\Lambda_A S^A\, ,
\ee
and provide the supersymmetric completion of the gauge-invariant field-strengths \eqref{tildeH3}. Ignoring for the moment tadpole conditions, in order to  include the above gauged linear multiplets  we can generalize \eqref{3formlagr1} to
\be\label{32formlagr}
\call=\int\d^4\theta \calf(\calz,\bar\calz,\hat L)+\left(\int\d^2\theta\,\hat\calw(\calz)+\text{c.c.}\right)+\call_{\rm bd}\, ,
\ee
where $\calf(\calz,\bar\calz,\hat L)$ is a (real) kinetic function which substitutes the K\"ahler potential $\calk$ in presence of linear multiplets, and the $\call_{\rm bd}$ can be computed using the procedure explained in Appendix \ref{app:bdcontr}. Starting from the superspace Lagrangian \eqref{32formlagr} one can in principle obtain its component form straightforwardly -- see  appendix \ref{app:SWL}, which focuses on the bosonic components. 

We can now combine the dualization described in section \ref{sec:susy1} and the  standard linear/chiral duality \cite{Lindstrom:1983rt} to relate \eqref{32formlagr} to a description in terms of ordinary chiral multiplets only. 
As a first step, one must relax the constraint \eqref{conhatL} and complete the parent Lagrangian which was used in section \ref{sec:susy1} 
as follows:
\be\label{plagr}
\begin{aligned}
\call=&\int\d^4\theta \calf(Z,\bar Z,\hat L)+2\int\d^4\theta\, P^A\left(\Im X_A+c^\Lambda_A\Im T_\Lambda\right){-} 2\int\d^4\theta\,\hat L^\Lambda\Im T_\Lambda\\
&\quad~~~+\Big[\int\d^2\theta X_A\calv^A(Z)+\int\d^2\theta\,\hat\calw(Z)+\text{c.c.}\Big]\, ,
\end{aligned}
\ee 
with $T_\Sigma$ a chiral superfield and $\hat L^\Lambda$ must be considered as an unconstrained real superfield. 
Notice that this Lagrangian is invariant under \eqref{Sgauge}, as it should. 
On the one hand, if we integrate out  $T_\Lambda$ we get the constraint \eqref{conhatL}, so that we can write  $L^\Lambda$ as in \eqref{hatL}, with $L^\Lambda$ being linear multiplets. We may then integrate out $X_A$ to get \eqref{chircalvconstr} and obtain \eqref{32formlagr}.
On the other hand, starting back from \eqref{plagr} and integrating out $P^A$, one  gets the identification
\be
X_A= N_A-c_A^\Lambda T_\Lambda
\ee
for constant $N_A$. 
One may then integrate out $\hat L^\Lambda$ to get a Lagrangian for the chiral multiplets $Z^a$ and $T_\Sigma$, with superpotential 
\be\label{extSup}
\calw(Z,T)=N_A\calv^A(Z)-c^\Lambda_A\, T_\Lambda\calv^A(Z)+\hat \calw(Z)\, ,
\ee
and K\"ahler potential
\be\label{neqchiral}
\calk=\calf -2 \hat L^\Lambda\Im T_\Lambda \, .
\ee 
Here $\hat L^\Lambda$'s must be considered as the functions of the chiral multiplets obtained by inverting 
\be\label{Tdef}
\Im T_\Lambda=  \frac12\frac{\del\calf}{\del \hat L^\Lambda}\, .
\ee
We may now dualize back $N_A$ to four-form field strengths, as in the previous sections.
The resulting Lagrangian takes the form
\be\label{susyZT}
\int\d^4\theta\,\calk(\calz,\bar\calz,T,\bar T)+\left(\int\d^2\theta\,\hat\calw'(T,\calz)+\text{c.c.}\right)+\call_{\rm bd}\, ,
\ee
where $\calz^a$ are master multiplets, we have introduced
\be\label{WZT}
\hat \calw'(\calz,T)\equiv -c^\Lambda_A T_\Lambda \calv^A(\calz)+\hat\calw(\calz)\,,
\ee
and $\call_{\rm bd}$ is given by \eqref{BT_susybd}  and \eqref{BT_Thetadef}, with $\hat\calw(\calz)$  replaced by $\hat\calw'(\calz,T)$ in \eqref{BT_Thetadef}.
In some circumstances, we may also regard $c_A^\Lambda$ as flux quanta, which may be dualized to corresponding four-form field-strengths. In our string theory examples of section \ref{sec:IIBstringy} and \ref{sec:IIAstringy},  $c_A^\Lambda$ can be associated with various kinds of ordinary, geometric and non-geometric fluxes.

It is clear that the effect of the gauging \eqref{Lgauge}  is dual to the appearance of the mixed term $c^\Lambda_A\, T_\Lambda\calv^A(Z)$
in the superpotential \eqref{extSup}. This term breaks the shift-symmetry under constant real shifts \be\label{Tshift}
T_\Lambda\quad\rightarrow\quad T_\Lambda +n_\Lambda 
\ee
On the other hand, if $n_\Lambda\in\mathbb{Z}$, that is, \eqref{Tshift} are multiple of the axionic  periods of 
$T_\Lambda$, 
then \eqref{Tshift} can be reabsorbed by the shift $N_A\rightarrow N_A+c_A^\Lambda n_\Lambda$, hence exhibiting an axion monodromy structure. In the corresponding formulations in terms of three-form multiplets, these spurionic symmetries are upgraded to proper symmetries, as anticipated at the bosonic level in section \ref{sec:dualities}.  Indeed, the shift \eqref{Tshift} induces a shift $\Delta\hat\calw'=-n_\Lambda c^\Lambda_A\calv^A$, whose contribution to the Lagrangian \eqref{susyZT} is exactly cancelled by the variation of the boundary term $\call_{\rm bd}$.

\section{Supersymmetric 3-branes, strings and membranes}
\label{sec:susy3}
In supersymmetric theories branes can be described by an embedding of the brane bosonic world-volumes to superspace. In this way one can construct quite generic brane actions which however preserve  supersymmetry only at the non-linear level, hence spontaneously breaking the complete bulk supersymmetry. Having a brane that locally preserves part of the bulk supersymmetry strongly constrains the form of the action. As we show in this section, this is what happens also in our models, in which we can couple strings, membranes and 3-branes to the bulk two-, three- and four-form potentials. As anticipated in section \ref{sec:sm3}, the corresponding supersymmetric actions are completely fixed by their charges. In particular, this fixes the field-dependence of their tensions. 

In order to avoid ambiguities, in the following we will abandon the sloppy rigid-superspace notation adopted so far, going back to the more precise fully curved superspace notation.  

\subsection{Effective super 3-branes}

Let us start from the 3-branes. Such objects do  not have a standard brane dynamics in four dimensions and are then qualitatively different from  membranes and strings. However, as discussed in section \ref{sec:sm3} they can end on membranes and lead to interesting tadpole-changing effects.\footnote{Typically, 3-branes also support dynamical world-volume fields. For simplicity, in this paper we will not consider this possibility. }

Let us consider the bosonic WZ term \eqref{3braneWZ}. This admits the following supersymmetrization
\be\label{sWZ3}
\mu_I\int_\cals {\bf C}^I_4\, ,
\ee
where $\cals$ denotes the 3-brane world-volume which is embedded in superspace and defines the pull-back of the integrated super four-form\footnote{\label{foot:sForms} We use the following definition of superspace exterior derivative: given a super $p$-form ${\bf A}_p$, its exterior derivative is given by $\d {\bf A}_p=(-)^p\d_{\rm WB}{\bf A}_p$, where $\d_{\rm WB}$ is the exterior derivative used in \cite{Wess:1992cp}. In this way, the exterior derivative of the lowest component $A_p\equiv {\bf A}_p|_{\theta-\bar\theta=0}$ is the usual bosonic one -- see footnote \ref{foot:supconv}. Furthermore, given the embedding of a $(p+1)$-dimensional bosonic manifold $\Sigma_{p+1}$ into superspace, we use orientation conventions such that Stokes' theorem reads $\int_{\Sigma_{p+1}}\d {\bf A}_p=\int_{\del\Sigma_{p+1}}{\bf A}_p$.  
}   
\begin{eqnarray} \label{C4}   {\bf C}^I_{4}   &=&    E^b\wedge E^a \wedge \bar E^{\dot\alpha} \wedge \bar  E^{\dot\beta }\bar{\sigma}_{ab\; \dot{\alpha}\dot{\beta}} \Gamma^I {+} E^b\wedge E^a \wedge E^\alpha \wedge E^\beta \sigma_{ab\; \alpha\beta}\bar{ \Gamma}^I \qquad \nonumber \\  && {-} \frac{1}6  E^c\wedge E^b\wedge E^a \wedge \bar E^{\dot\alpha} \epsilon_{abcd} \sigma^d_{\alpha\dot\alpha} {\cal D}^{\alpha}{ \Gamma^I}  +\frac{1}6 E^c\wedge E^b\wedge E^a \wedge E^\alpha \epsilon_{abcd} \sigma^d_{\alpha\dot\alpha} \bar{{\cal D}}^{\dot\alpha}\bar{ \Gamma}^I \nonumber \\  &&  +\frac{\ii}{96} E^{d} \wedge E^c \wedge E^b \wedge E^a \epsilon_{abcd} \left(({\cal D}^2-24\bar{\calr})
{ \Gamma^I} -(\bar{{\cal D}}^2-24\calr) 
\bar{ \Gamma}^I \right) 
\,, \qquad
\end{eqnarray}
where $\Gamma^I$ are the chiral superfields introduced in section \ref{sec:susy2}.
This  is the unique closed (but not exact) super four-form which has the bosonic four-form potential $C_4^I$ as bosonic component. 
This property implies that, as long as $\cals$ has no boundary, the WZ-term \eqref{sWZ3} is left invariant under any superdiffeomorphism. Hence, it certainly does not spontaneously break any bulk supersymmetry.

One may try to add  a supersymmetric Nambu-Goto like term, associated with a tension for the 3-brane. In presence of such a term, the 3-brane would become a `goldstino brane' \cite{Bandos:2015xnf,Bandos:2016xyu}, realizing supersymmetry only non-linearly and hence always spontaneously breaking it completely. Even though such contributions could be important in string compatifications, for instance in presence of microscopic anti-branes, the resulting EFT becomes quite unconstrained \cite{Bandos:2016xyu} and  one probably needs to work it out case-by-case or understand some other organizing principle.
We emphasize that, at the effective four-dimensional level, nothing seems to correlate the sign of charges $\mu_I$ with the presence of a goldstino brane term in the action. This is in contrast with what happens in typical string compactifications, in which such correlation is quite  universal and the EFT supersymmetry is linearly realized only for specific signs of the $\mu_I$'s. It would then be interesting to see whether such correlation could be understood at the EFT level only.

Postponing these interesting problems to the future, in this paper we just consider  contributions of the form \eqref{sWZ3} to the EFT and  no additional goldstino brane terms. The signs of the $\mu_I$ charges can be then considered as free discrete parameters to be fixed by matching with the microscopic supersymmetric models.

The action \eqref{sWZ3} is basically topological and then, apart from contributing to the total four-form charge $\calq_I$ by $\mu_I$, does not seem to have much physical content. However, as discussed in section \ref{sec:sm3}, things become more interesting if we allow the world-volume $\cals$ to have non-trivial boundary $\del\cals$.  In section \ref{sec:sm3} we have discussed, at the bosonic level, how to use these  open 3-branes to  cancel the potential anomalies of  membrane. In the following subsection we will see that this mechanism can indeed be made manifestly supersymmetric.


\subsection{Effective supermembranes}

We now show how to supersymmetrically  add  membranes by extending the results of \cite{Bandos:2018gjp}.  
A membrane couples to the bosonic three-form potentials $A^A_3$ through a bosonic WZ-term $q_A\int_\Sigma A^A_3$, where $\Sigma$ denotes the membrane world-volume and $q_A$ its charges. This WZ-term  can be supersymmetrized by promoting the bosonic embedding of $\Sigma$ to an embedding in the complete superspace and by completing $A^A_3$ to the super three-forms
\be\label{super3form}
\begin{aligned}
{\bf A}^A_{3}=&\,  { -}2 {\ii} E^a \wedge E^\alpha \wedge \bar E^{\dot\alpha}  \sigma_{a\alpha\dot\alpha}P^A + {\frac 12}  E^b\wedge E^a \wedge  E^\alpha
\sigma_{ab\; \alpha}{}^{\beta}{\mathcal D}_{\beta}P^A \\ &+{\frac 12}  E^b\wedge E^a \wedge  \bar E^{\dot\alpha}
\bar\sigma_{ab}{}^{\dot\beta}{}_{\dot\alpha}\bar{\mathcal D}_{\dot\beta}P^A  
\\&+\frac {1} {24} 
  E^c \wedge E^b \wedge E^a \epsilon_{abcd} \,\left(\bar{\sigma}{}^{d\dot{\alpha}\alpha}
  [{\mathcal D}_\alpha, \bar {\mathcal D}_{\dot\alpha}]P^A+8G^d P^A\right)
 \, , \qquad
\end{aligned}
\ee
where $P^A$ are the (appropriately covariantized) real superfields introduced in section \ref{sec:susy1}. Notice that the lowest component of ${\bf A}^A_{3}$ is indeed  $A_3^A$. The super three-forms ${\bf A}^A_{3}$  are  defined up to a gauge transformation, but their super field-strengths ${\bf F}^A_{4}=\d {\bf A}^A_{3}$ are uniquely determined in terms of the composite chiral multiplets $S^A$ defined in \eqref{defSA} -- see appendix D of \cite{Bandos:2018gjp}. 

By itself, the resulting  WZ-term spontaneously breaks all four generators of  the bulk $\caln=1$ supersymmetry, which always shifts the  fermionic  world-volume fields defined by the embedding. The usual strategy to overcome this problem is to add a term to the  WZ-term such that the resulting action enjoys a fermionic gauge symmetry, the so-called $\kappa$-symmetry.  This redundancy allows part of the bulk supersymmetry transformations  to induce unphysical  world-volume fermionic shifts  and then to be locally preserved. In the present case,  this requirement fixes the following form of the membrane action
\be\label{memaction}
S_{\rm mem}=-2\int_\Sigma \d^3\zeta\,|q_A \hat S^A|\sqrt{-\det {\bf h}}+q_A\int_\Sigma{\bf A}^A_3\,,
\ee 
where $\zeta^i$ denote some local coordinates on $\Sigma$ and ${\bf h}_{ij}\equiv \eta_{ab}E^{a}_iE^{b}_j$,  with $E^{a}_i$  the pull-back of the bulk super-vielbein to $\Sigma$. By using  \eqref{super3form} and the local supersymmetry counterpart of \eqref{VectorB}, keeping only the lowest bosonic components, one reduces \eqref{memaction}  to \eqref{bosmem}.

Postponing for the moment the discussion of $\kappa$-symmetry, we first recall that  superfields $\hat S^A$ appearing in \eqref{memaction} have been defined in \eqref{gicon} and are gauge-invariant under the three-form gauge transformations \eqref{s3gauge}. Hence, the first Nambu-Goto-like  term appearing on the r.h.s\ of \eqref{memaction} is manifestly gauge invariant as well. On the other hand,  the WZ-term appearing in  \eqref{memaction} is  generically invariant under the super-form completion of three-form gauge transformations \eqref{lin3gauge} {\em only} if $q_AQ^A_I=0$. (Notice also that in such a case the combination $q_A\hat S^A$ appearing in \eqref{memaction} reduces to $q_A S^A$.)  This supersymmetric completion is given by 
\begin{subequations}\label{sgforms}
\begin{align}
{\bf C_4}^I&\quad\rightarrow\quad {\bf C_4}^I+\d {\bf \Lambda}^I_3\,,\\
{\bf A}^A_3&\quad\rightarrow\quad {\bf A}^A_3-q_AQ^A_I{\bf \Lambda}^I_3\,,
\end{align}
\end{subequations}
 with
\be\label{s3gaugeb}
\begin{aligned}
{\bf \Lambda}^I_3=&\,  { -}2 {\ii} E^a \wedge E^\alpha \wedge \bar E^{\dot\alpha}  \sigma_{a\alpha\dot\alpha}\Xi^I + {\frac 12}  E^b\wedge E^a \wedge  E^\alpha
\sigma_{ab\; \alpha}{}^{\beta}{\cald}_{\beta}\Xi^I \\ &+{\frac 12}  E^b\wedge E^a \wedge  \bar E^{\dot\alpha}
\bar\sigma_{ab}{}^{\dot\beta}{}_{\dot\alpha}\bar{\cald}_{\dot\beta}\Xi^I
\\&+\frac {1} {24} 
  E^c \wedge E^b \wedge E^a \epsilon_{abcd} \,\left(\bar{\sigma}{}^{d\dot{\alpha}\alpha}
  [\cald_\alpha, \bar{\cald}_{\dot\alpha}]\Xi^I+8 G^d\Xi^I \right)
 \, . \qquad
\end{aligned}
\ee
At this point the open 3-brane mentioned at the end of the previous subsection come to the rescue. Indeed, the bosonic discussion of section \ref{sec:sm3} extends immediately to the superspace level by using the super-form gauge tranformations \eqref{sgforms}. Hence,  the membrane anomaly can be cancelled by attaching to it  an open  3-brane of charges $\mu_I=Q^A_Iq_A$. 

Let us now pass to the $\kappa$-symmetry. In this section we adopt the standard notation \cite{Wess:1992cp} in which the superspace coordinates are denoted  by $z^M$, which should not be confused with the bulk complex fields $z^a$! The $\kappa$-tranformation is defined by
\be\label{kappasymm}
\delta z^M (\zeta)=\kappa^\alpha(\zeta) E^M_\alpha(z(\zeta))+\bar\kappa_{\dot\alpha}(\zeta) E^{M\dot\alpha}(z(\zeta))\, ,
\ee
where the local fermionic parameter $\kappa^\alpha(\zeta)$ (with $\bar\kappa^{\dot\alpha}(\zeta)\equiv \overline{\kappa^\alpha}(\zeta)$) satisfies the projection condition 
\be\label{kappaproj}
\kappa_\alpha=-\frac {q_A \hat S^A}{|q_A\hat S^A|}\Gamma_{\alpha\dot\alpha}\bar\kappa^{\dot\alpha} , 
\ee
and
\be\label{kappagamma}
\Gamma_{\alpha\dot\alpha}\equiv \frac{\ii\epsilon^{ijk}}{3!\sqrt{-\det h}}\epsilon_{abcd} E^b_iE^c_j E^d_k\,\sigma^a_{\alpha\dot\alpha}. 
\ee

In the non-anomalous case $q_AQ^A_I=0$, the  invariance of the action \eqref{memaction} follows immediately from the proof given in \cite{Bandos:2018gjp}. In the case  $q_AQ^A_I\neq 0$, the same proof shows that \eqref{memaction} is not invariant anymore, but rather produces a non-vanishing contribution\, .
\be\label{kappanomaly}
q_AQ^A_I\int_\Sigma \delta z^M\iota_M {\bf C}_4^I 
\ee
However, the open 3-brane that need to be attached to membrane in order to cancel its three-form anomaly comes to the rescue once again. Indeed, it is sufficient to extend $\delta z^M(\zeta)$ to the 3-brane world-volume in an arbitrary way. By using the condition $\mu_I=q_A Q^A_I$ and the fact that ${\bf C}_4^I $ is closed  \eqref{kappasymm}, it is then easy to see that the corresponding variation of the 3-brane topological action \eqref{sWZ3} localizes on its boundary $\Sigma$ and perfectly cancels \eqref{kappanomaly}.


\subsection{Effective superstrings}

We finally consider strings. The logic will be completely analogous to the membrane case, hence we will proceed more quickly. 
The following results can be easily justified by appropriately adapting and extending the results of \cite{Bandos:2003zk,Bandos:2019qok,Bandos:2019lps}.

By imposing $\kappa$-symmetry and the appropriate WZ-coupling to the bulk two-form potentials $\calb_2^\Lambda$, one arrives at the following unique  action for a string of charges $q_\Lambda$
\be\label{sstring}
-\int_\calc\d^2\zeta |e_\Lambda \hat L^\Lambda|\sqrt{-\det {\bf h}}  + e_\Lambda\int_\calc {\bf B}^\Lambda_2\, ,
\ee
where $\hat L^\Lambda$ are defined in \eqref{hatL}, and ${\bf B}^\Lambda_2$ is a super-two-form whose field-strength super-three-form ${\bf H}^\Lambda_3$ is given by 
\be\label{tildecalh}
\begin{aligned}
{\bf H}^\Lambda_3=&\, \d {\bf B}^\Lambda_2= \,  - 2{\ii} E^a \wedge E^\alpha \wedge \bar E^{\dot\alpha}  \sigma_{a\alpha\dot\alpha}{ L}^\Lambda \\ & + {\frac 12}  E^b\wedge E^a \wedge  E^\alpha
\sigma_{ab\; \alpha}{}^{\beta}{\mathcal D}_{\beta}{ L}^\Lambda +{\frac 12}  E^b\wedge E^a \wedge   \bar E^{\dot\alpha}
\bar\sigma_{ab}{}^{\dot\beta}{}_{\dot\alpha}\bar{\mathcal D}_{\dot\beta}{ L}^\Lambda
\\&+\frac {1} {24}
  E^c \wedge E^b \wedge E^a \epsilon_{abcd} \,(\bar{\sigma}{}^{d\dot{\alpha}\alpha}
  [{\mathcal D}_\alpha, \bar{\mathcal D}_{\dot\alpha}]+8G^d){ L}^\Lambda
 \, .
\end{aligned}
\ee
This is the unique closed super-three-form that can be constructed from the linear multiplets and  that has $\calh^\Lambda_3=\d\calb^\Lambda_2$ as the lowest component  \cite{Bandos:2003zk}. 
It is also clear that \eqref{sstring} reduces to \eqref{bostring} once we restrict to the bosonic components. 

The $\kappa$-transformation is given by \eqref{kappasymm} with $\kappa^\alpha$ satisfying the projection condition
\be\label{kappastring}
\kappa_\alpha=-\frac{e_\Lambda\hat L^\Lambda}{|e_\Lambda\hat L^\Lambda|}\Gamma_{\alpha}{}^\beta\kappa_\beta\, ,
\ee
with  
\be
\Gamma_{\alpha}{}^\beta\equiv \frac{1}{2\sqrt{-\det{\bf h}}}\epsilon^{ij}E^a_iE^b_j(\sigma_{ab})_\alpha{}^\beta\, .
\ee
The invariance of the string action under the bulk gauge transformations \eqref{Lgauge} and world-sheet $\kappa$ transformations  holds in a way similarly to the membrane case. Namely, if $e_\Lambda c^\Lambda_A=0$, then the string action \eqref{sstring} is invariant under both kinds of transformations. If $e_\Lambda c^\Lambda_A\not =0$, it is not, but both anomalies can be cancelled by attaching to the string an open membrane with charges 
$q_A=e_\Lambda c^\Lambda_A$. The combined action for the membrane and the string will be invariant under $\kappa$-transformations whose parameters are subject to two projections \eqref{kappaproj} and \eqref{kappastring} implying that this system may in general preserve  
the maximum of 1/4 bulk supersymmetry.

Clearly, the above actions can be combined to obtain supersymmetric effective actions for  more complicated networks of 3-branes, membranes and strings, like those considered in e.g. \cite{Evslin:2007ti,BerasaluceGonzalez:2012zn}. A detailed treatment of these more involved configurations is left for future work.


\section{Conclusions}
\label{sec:conclu}

In this work we have extended the construction of 4d supersymmetric three-form Lagrangians initiated in \cite{Farakos:2017jme,Bandos:2018gjp} to accommodate, from an EFT viewpoint, the most general superpotentials found in string theory compactifications with fluxes. The key technical development is the inclusion of a more general class of three-form multiplets, dubbed {\em master} multiplets, defined in terms of the periods entering the flux-induced superpotential in its Weyl-invariant formulation --  see sections \ref{sec:genEFT} and \ref{sec:susy}. As a direct application one can construct the $\kappa$-symmetric actions for 4d extended objects like 3-branes, membranes and strings, which have a tension-to-charge ratio in agreement with dimensional reduction expectations --  see sections \ref{sec:sm3} and \ref{sec:susy3}.

This improvement can be easily combined with other important ingredients that appear in three-form Lagrangians, like the different $p$-form gaugings encoding discrete, topological data of the EFT. A relatively familiar kind of gauging is that of a two-form potential by a three-form \cite{Dvali:2005ws,Dvali:2005an,Kaloper:2008fb,Kaloper:2011jz,Dvali:2013cpa}. This gauging appears in certain compactification regimes in which a continuous shift symmetry is developed at the level of the chiral field kinetic terms, but is broken at the level of the superpotential. In other words, this gauging signals the presence of an axion in the EFT superpotential \cite{Marchesano:2014mla}, which translates into a multi-branched effective scalar potential. In terms of 4d defects, this feature manifests as the presence of an anomalous axionic string, in which certain membranes must end to cure the anomaly.

This is however not the only kind of gauging involving three-forms. As we have shown, implementing the tadpole conditions of a string compactification at the EFT level results in gaugings of three-form potentials by four-forms. Indeed, the presence of 4d four-forms has so far been essentially ignored from the EFT viewpoint, but it is easy to see that each of these forms is related to a different tadpole condition of the compactification. As for the previous gaugings, there is a counterpart of their presence in terms of 4d defects, this time in terms of certain anomalous membranes, in which 3-branes must end to cure their anomaly. There is a consistency condition ensuring the compatibility of both kinds of gaugings, which prevents extended objects that are boundaries to have boundaries themselves.

Bringing all these ingredients together, one finds several obstructions to dualize to three-forms the whole set of fluxes appearing in the superpotential. More precisely, if $\Gamma$ describes the lattice of compactification fluxes, one needs to select a sublattice $\Gamma_{\rm EFT}$ of `dynamical' fluxes, that can then be dualized to a set of three-form multiplets. As a result, the corresponding EFT is only able to describe membrane-mediated flux transitions within $\Gamma_{\rm EFT}$, unlike one may have initially thought. The elements of the quotient $\Gamma/\Gamma_{\rm EFT}$ are to be thought  as fixed discrete parameters, and in order to vary them one should change the EFT itself.  

In this framework, the obstructions to identify $\Gamma_{\rm EFT}$ with $\Gamma$ are a priori all different, and arise independently from each of the three EFT ingredients mentioned above: 

\begin{itemize}

\item[-] \underline{Supersymmetry}: The number of dynamical fluxes must be such that $n+1 \leq {\rm dim}\, \Gamma_{\rm EFT} \leq 2n+2$, where $n$ is the number of chiral fields entering the superpotential terms generated by the dynamical fluxes. This comes from the upper and lower bound on the number of three-forms per scalar in master multiplets.

\item[-] \underline{Tadpoles}: Fluxes typically contribute to tadpole conditions quadratically, by means of symmetric bilinear forms $\cali_I$. To implement tadpoles as three-form gaugings, $ \Gamma_{\rm EFT}$ must be an isotropic sublattice of $\Gamma$ with respect to each of these parings. As a result, at the level of the EFT tadople cancellation appears as a set of linear conditions on the dynamical fluxes. 

\item[-] \underline{Axion-monodromy}: In certain regimes of the compactification anomalous axionic strings appear at low energies, which means that membranes can nucleate holes in their worldvolume \cite{Evslin:2007ti,BerasaluceGonzalez:2012zn}. At the EFT level it makes sense to include the two-forms coupled electrically to the strings, together with a gauging encoding the anomaly. The gauging parameters will be fluxes that cannot belong to $ \Gamma_{\rm EFT}$.

\end{itemize}

We have applied these criteria to different instances of string compactifications. In each of them we have compared the most natural choice of $\Gamma_{\rm EFT}$ solving the above obstructions with the spectrum of membrane tensions. We have found that, in all cases, one can choose an EFT cut-off scale $\Lambda_{\rm UV}$ such that $\Gamma_{\rm EFT}$ is selected as the sublattice of membranes satisfying
\be\label{GEFTcond}
\frac{\calt_{\rm mem}}{M^2_{\rm P}}\lesssim \Lambda_{\rm UV} \,,
\ee
so that the notion of dynamical flux acquires a more precise energetic meaning. Particularly interesting are weakly-coupled type IIB orientifold compactifications, where $\Lambda_{\rm UV}$ lies just above the mass scale induced by NS-NS three-form fluxes, and $\Gamma_{\rm EFT}$ is given by the lattice of R-R membranes, parametrically lighter than their NS-NS counterparts. Another very illustrative example is type IIA compactifications, where one can see that, due to the spectrum of membrane tensions, even the definition of $\Gamma$ changes as one goes from the weakly-coupled to the M-theory regime.

In fact, the condition \eqref{GEFTcond} does not select the full sublattice $\Gamma_{\rm EFT}$, but a large region whose vectors have their norm bounded from above. This is reminiscent of certain EFT criteria like the Swampland Distance Conjecture and generalizations thereof \cite{Ooguri:2006in,Baume:2016psm,Klaewer:2016kiy}, which claim that the same EFT should be valid only up to displacements of a maximal distance in field space. Indeed, intuitively one expects that flux transitions with larger norms in $\Gamma_{\rm EFT}$ correspond to jumps between vacua related by larger displacements in field space. Therefore, the SDC implies that the whole  $\Gamma_{\rm EFT}$ should not be accessible to a single EFT, in agreement with our scheme. We leave a more detailed discussion of how our results combine with this and other conjectures of the Swampland Program for the future. 

Our findings can be applied and generalized in different directions. For instance, one interesting technical development would be to include world-volume matter supported on the  effective 3-branes, membranes and strings that we have considered. Such a matter content is expected from the string theory constructions, like 4d gauge bosons and adjoint chiral fields on 3-branes, and it would be interesting to see how the EFT treats this sector upon membrane-mediated transitions that change the number of 3-branes. Moreover, the presence of world-volume matter would allow for the study of possible novel swampland criteria, along the lines of \cite{Kim:2019vuc}.  

Moreover from our results it follows that, in appropriate parametric regimes, three-forms, membranes and strings are in fact needed to provide the {\em complete} low-energy EFT of flux compactifications. More traditional formulations with fixed flux quanta can only access part of the low-energy dynamical phenomena. The price to pay for this more complete description is that the EFT cut-off scale $\Lambda_{\rm UV}$ must be fixed at a certain energy range compatible with $\Gamma_{\rm EFT}$. We have found that, in several instances, such cut-off scale lies just above the flux-induced mass scale. It is thus natural to wonder about the phenomenological features of this EFT, and in particular about the physics that can be extracted from the sub-ensemble of vacua dynamically connected through membrane nucleations within $\Gamma_{\rm EFT}$. One may for instance ask whether, having the cut-off scale as low as $m_{\rm flux}$, physical observables like the cosmological constant, Yukawa couplings, soft terms, etc. vary significantly along this EFT landscape or not. Then, for those couplings that are effectively scanned over by the EFT, one may attempt to see if any physically relevant information can be extracted from a statistical analysis \cite{Douglas:2003um,Denef:2007pq}. In fact, it could well be that these three-form EFTs give us a simple framework to understand how the statistical method of analysis of vacua and the Swampland Program intertwine with each other. If that was the case, they could become crucial for developing a new scheme to extract predictions out of string theory vacua. 




\bigskip

\bigskip

\centerline{\bf Acknowledgments} 

\bigskip

We would like to thank Igor Bandos, Andreas Braun, Sergei Kuzenko, Ruben Monten, Miguel Montero, Toine Van Proeyen, Irene Valenzuela, Thomas Van Riet and Roberto Volpato for useful discussions.
S.L. is deeply grateful to Instituto de F\'isica Te\'orica UAM-CSIC, Madrid for the warm hospitality while this work was initiated. 
L.M.'s work was supported in part by the MIUR-PRIN contract 2017CC72MK\_003. 
Work of D.S. was supported in part by the Australian Research Council project No. DP160103633. D.S. is grateful to the School of Physics and Astrophysics, University of Western Australia for hospitality during an intermediate stage of this work. 
The work of F.M. is supported through the grants SEV-2016-0597, FPA2015-65480-P and PGC2018-095976-B-C21 from MCIU/AEI/FEDER, UE.
F.M. gratefully acknowledges support from the Simons Center for Geometry and Physics, Stony Brook University and from a grant from the Simons Foundation at the Aspen Center for Physics. Some of the research for this paper was performed at both institutions. 

\bigskip

\newpage

\centerline{\LARGE \bf Appendices}
\vspace{0.5cm}

\begin{appendix}


\section{Geometrical meaning of three-form kinetic matrix}
\label{app:TAB}

In section \ref{sec:bosduality} we have shown that the kinetic matrix of the three-forms $A^A_3$, with index $A=1,\ldots, k$, is given by 
\be\label{TABapp}
T^{AB}\equiv 2\Re\left(\calk^{a\bar b}\calv^A_a\bar\calv^B_{\bar b}\right)\, ,
\ee
where $\calv^A(z)$ are homogeneous  periods, which depend holomorphically on the homogeneous fields $z^a$ (including the Weyl compensator), with index $a=0,\ldots, n$, and $\calk^{a\bar b}$ is the inverse of the  hermitian metric $\calk_{a\bar b}$.

Let us now decompose $e^{\ii\theta}\calv^A(z)$, where $e^{\ii\theta}$ is an arbitrary phase, into real and imaginary parts 
\be\label{calvdec}
e^{\ii\theta}\calv^A(z)={\cal X}^A+\ii{\cal Y}^A\quad~~~~ {\cal X}^A,{\cal Y}^A\in\mathbb{R}\, .
\ee 
By holomorphy $e^{\ii\theta}\calv_a^A=2{\cal X}^A_a$ and then we can rewrite \eqref{TABapp} in the form
\be\label{TABapp2}
T^{AB}\equiv 4\left(\calk^{a\bar b}{\cal X}^A_a{\cal X}^B_{\bar b}+\calk^{\bar a b}{\cal X}^A_{\bar a}{\cal X}^B_{ b}\right)\, .
\ee
In section \ref{sec:susy1} we have also seen  that the periods $\calv^A(z)$  define the homogeneous embedding
\be
s^A=\calv^A(z)\, ,
\ee
 into the field-space parametrized by the complex fields $s^A$ which appear in the elementary three-form multiplets as in  \eqref{VectorB} and \eqref{defSA}. The fields $s^A$ parametrize a $k$-dimensional complex plane $\calm_S\simeq \mathbb{C}^k$ and
 $\Re(e^{\ii\theta}s^A)$ parametrize a $k$-dimensional real subplane $\calm_{\cal X}\subset\calm_S$, with $\calm_{\cal X}\simeq \mathbb{R}^{k}$. 
 Then the real functions ${\cal X}^A(z)$ define a map  from the field space $\calm_{\calz}$ parametrized by $z^a$ to $\calm_{\cal X}$:
 \be\label{zxmap}
 \iota:\calm_{\calz}\rightarrow \calm_{\cal X}\,.
 \ee

We arrive at the following interpretation of $T^{AB}$.  $4\calk^{a\bar b}$ defines a metric on the contangent bundle $T^*\calm_{\calz}$. Then $T^{AB}$  represent the  push-forward of this metric to the cotangend bundle  $T^*\calm_{\cal X}$ restricted to the image of \eqref{zxmap}. Since $\dim_{\mathbb{R}}\calm_{\calz}=2n+2$, it is clear that $T^{AB}$ has at most rank $2n+2$ and can  then be non-degenerate only if $k\leq 2n+2$.  More precisely, if $k\leq 2n+2$ then the rank of $T^{AB}$  is given by the rank of the  matrix $\left(\frac{\del{\cal X}^A}{\del x^a},\frac{\del{\cal X}^A}{\del y^a}\right)$, where $x^a+\ii y^a\equiv z^a$ and $T^{AB}$ is invertible if and only if this rank equals $k$.
By using the holomorphicity of $\calv^A(z)$, we can write it in the following alternative ways
\be\label{rankT}
\text{rank}(T^{AB})=\text{rank}\left(\Re\calv^A_a,\Im\calv^A_b\right)=\text{rank}_{\mathbb{C}}\left(\calv^A_a,\bar\calv^A_{\bar b}\right)\, .
\ee
The formula \eqref{rankT} allows us to write the invertibility  condition $\text{rank}(T^{AB})=k$ in terms of the periods.

\section{Super-Weyl invariant Lagrangians}
\label{app:SWL}

Throughout this work, we have extensively used the super-Weyl invariant formalism \cite{Howe:1978km,Kaplunovsky:1994fg} (for a review see e.g. \cite{Buchbinder:1995uq}). Here we provide a very brief summary on how to construct super-Weyl invariant Lagrangians in supergravity for the cases of interest of string/M-theory models here considered. After showing, in section \ref{app:SWLa}, how to recover the usual Lagrangian with only chiral multiplets from the super-Weyl invariant approach, in section \ref{app:SWLb} we illustrate how to extract the bosonic components of Lagrangians which embed both chiral and linear multiplets. Finally, in section \ref{app:SWLc} we present the derivation of the full bosonic Lagrangian which embeds three-form multiplets as well as gauged linear multiplets.

\subsection{With only chiral multiplets}
\label{app:SWLa}

Consider a set of $N$ dimensionless chiral multiplets $\Phi^i$, whose bosonic components are
\be
\label{Phimult}
\Phi^i = \{\phi^i, F^{i}_{\Phi}\}\,,\quad \text{with}\quad i=1,\ldots,n\,,
\ee
where $\phi^i$  are the lowest component complex scalar fields and $F^{i}_{\Phi}$ are the highest component auxiliary complex scalar fields. At the core of the super-Weyl invariant formalism is the introduction of an unphysical, chiral compensator $U$, which we choose to transform as
	\be
	\label{Ytransfb}
	U  \rightarrow  e^{-6 \Upsilon} U
	\ee
under super-Weyl transformations. We recall that these
act on the super-vielbeins as \cite{Howe:1978km}
	\be\label{EW}
E^a_M\rightarrow e^{\Upsilon+\bar\Upsilon}E^a_M\,,\quad E^\alpha_M\rightarrow e^{2\bar\Upsilon-\Upsilon}
	\left(E^\alpha_M-\frac{\ii}{4}E^a_M\sigma^{\alpha\dot\alpha}_a\bar\cald_{\dot\alpha}\bar\Upsilon\right).
	\ee
where $(a,\alpha)$ are flat superspace indices, $M=(m,\mu)$ are curved indices and $\Upsilon$ is an arbitrary chiral superfield parameterizing the super-Weyl transformation. Instead the dimensionless superfields $\Phi^i$ are  invariant under super-Weyl tranformations.  
Combining $\Phi^i$ and the compensator $U$, we introduce new chiral superfields $Z^a$
	\be
	\label{Zmult}
	Z^a = \{z^a,\, F^a_Z\}\,\quad \text{with}\quad a=1,\ldots,n+1\,,
	\ee
where $z^a$ and $F_Z^a$ are understood to be functions of the components of $\Phi^i$ and  $U$ and transforming as $U$ under super-Weyl transformations.  In order to isolate the physical fields, we assume that we can single out the compensator $U$ as
	\be
	\label{homZ}
	Z^a = U g^a (\Phi)\,,
	\ee
where $g^a$ are functions of the physical fields only and are inert under super-Weyl transformations.
	
The most general supergravity Lagrangian that we can build out of the $Z^a$ multiplets is
	\be
	\label{SugraSWa}
	\begin{split}
		\call = \int \d^4\theta\, E\, \calk (Z,\bar Z) + \left(\int \d^2 \Theta\, 2\cale\, \calw (Z) +{\rm c.c.}\right)
	\end{split}
	\ee
where $\calk(Z,\bar Z)$ is the kinetic potential and $\calw(Z)$ the superpotential. Additionally, however, we require that they satisfy the following homogeneity conditions
	\be\label{homKW}
	\calk(\lambda  Z,\bar\lambda\bar {Z})=|\lambda|^{\frac 23}\calk({ Z},\bar { Z})\,,\qquad
	\calw(\lambda  Z)=\lambda\calw(Z)\,,
	\ee
with $\lambda$ an arbitrary chiral superfield. 
	
In order to recover the ordinary K\"ahler potential $K(\Phi, \bar \Phi)$ and superpotential $W(\Phi)$,  we isolate the compensator $U$ as
	\be
	\label{homKWb}
	\calk({ Z},\bar { Z})=|U|^{\frac23}e^{-\frac13 K(\Phi,\bar\Phi)}\,, \qquad 	\calw(Z)= U\,W(\Phi)\,,
	\ee
where $K(\Phi,\bar\Phi)\equiv -3\log\left[-\frac13 \calk(g(\Phi),\bar g (\bar\Phi))\right]$ and $W(\Phi)\equiv \calw (g(\Phi))$. Such homogeneity properties of $\calk$ and $\calw$ make the Lagrangian  \eqref{SugraSWa} manifestly invariant under super-Weyl transformations.  In particular, \eqref{EW} implies that
	\be
	E \rightarrow e^{2 \Upsilon + 2 \bar \Upsilon} E\,, \quad  \d^2 \Theta\, 2 \cale \rightarrow  e^{6\Upsilon} \d^2 \Theta\, 2 \cale\,.
	\ee

Indeed, the Lagrangian \eqref{homKW} is also independently invariant under K\"ahler transformations. In fact, the split \eqref{homZ} is clearly not unique, since  we may redefine
	\be\label{transfarb}
	U\rightarrow e^{h(\Phi)}U\,,\quad g^a(\Phi)\rightarrow e^{-h(\Phi)} g^a(\Phi)\,.
	\ee
with $h(\Phi)$ an arbitrary holomorphic function of $\Phi^i$. In turn, this redefinition corresponds to an ordinary K\"ahler transformation \be
\label{Ktransf}
K (\Phi, \bar \Phi) \rightarrow K (\Phi, \bar \Phi) + h(\Phi) + \bar h (\bar \Phi)\,,\quad 	W(\Phi) \rightarrow e^{-h(\Phi)} W(\Phi)\,.
\ee

The bosonic components of the Lagrangian \eqref{SugraSWa} acquire a very simple form
	\be
	\label{SugraSWacomp}
	\begin{split}
		e^{-1} \call_{\rm bos} &= -\frac16\calk\, R - \calk_{a\bar b} D_\mu z^a \bar D^\mu \bar z^b + \calk_{a\bar b}  f^a \bar f^b + \left(\calw_a f^a + {\rm c.c.}\right)\,.
	\end{split}
	\ee
Here we have redefined
    \be
    \label{SW_fdef}
    f^a \equiv  \bar M z^a - F^a_Z
    \ee
and introduced the $U(1)$-covariant derivatives
	\be
	D_\mu z^a = \del_\mu z^a + \ii A_\mu z^a\,\qquad {\rm with}\quad	A_\mu = \frac{3\ii}{2\calk} (\calk_a \del_\mu z^b - \calk_{\bar b} \del_\mu \bar z^b)\,.
	\ee
The auxiliary fields $f^a$ may be easily integrated out from \eqref{SugraSWacomp}, leading to the Lagrangian
	\be
	\label{SugraSWacompb}
	\begin{split}
		e^{-1} \call_{\rm bos} &= -\frac16\calk\, R - \calk_{a\bar b} D_\mu z^a \bar D^\mu \bar z^b - \calk^{a\bar b}  \calw_a \bar \calw_{\bar b}\,.
	\end{split}
	\ee
In order to pass to the Einstein frame, we isolate the compensator $ u \equiv	U |_{\theta=\bar\theta=0}$ and gauge-fix the super-Weyl invariance by setting 
	\be
	\label{SWgf}
  u = M_{\rm P}^2\, e^{\frac12{K(\phi, \bar \phi)}}\quad\Rightarrow\quad \calk=-3M^2_{\rm P}\,.
	\ee
For simplicity, in the following formulas we  will set  $M_{\rm P}=1$ (an eventual dependence on the Planck mass may be easily reinstated by dimensional analysis). Exploiting the homogeneity properties \eqref{homKWb}, along with the condition \eqref{SWgf}, we arrive at the gauge-fixed Lagrangian
	\be
	\label{SugraCLag}
	\begin{split}
		e^{-1} \call_{\rm bos} &= \frac{1}2 R - K_{i\bar \jmath} \del_\mu \phi^i \del^\mu \bar \phi^{\bar \jmath}- e^{K} \left( K^{\bar\jmath i} D_i W \bar D_{\bar \jmath} \bar W - 3 |W|^2 \right) \;,
	\end{split}
	\ee
with a canonically normalized Einstein-Hilbert term and where the last term is nothing but the well-known Cremmer et al. potential \cite{Cremmer:1982en}.

\subsection{With chiral and linear multiplets}
\label{app:SWLb}
	
Now consider, along with the chiral superfields \eqref{Zmult}, the linear multiplets
	\be
	\label{Lmult}
	L^\Lambda = \{ l^\Lambda,  \calh_3^\Lambda = \d \calb_2^\Lambda\}\,,\quad \text{with}\quad \Lambda=1,\ldots,M\,,
	\ee
with $l^\Lambda$ real scalar fields and $ \calh_3^\Lambda$ real field strengths of gauge two-forms $\calb_2^\Lambda$.  These transform under the super-Weyl transformations as $L^\Lambda \rightarrow e^{-2 \Upsilon -2 \bar \Upsilon} L^\Lambda$. The most general Lagrangian that we can build out of the chiral multiplets \eqref{Zmult} and linear multiplets \eqref{Lmult} is
	\be
	\label{SWLina}
	\begin{split}
		\call &= \int \d^4\theta\, E\, \calf (Z,\bar Z, L) + \left(\int \d^2 \theta\, 2\cale\, \calw (Z) +{\rm c.c.}\right)\,.
	\end{split}
	\ee
As in \eqref{neqchiral}, the kinetic function $ \calf (Z,\bar Z, L)$ is related to the kinetic function $ \calk (Z,\bar Z, \Im T)$ of the dual chiral  formulation by the Legendre tranform
\be\label{KFleg}
\calf (Z,\bar Z, L)=\calk+2L^\Lambda\Im  T_\Lambda
\ee
with $L^\Lambda=-\frac12\frac{\del\calk}{\del \Im T_\Lambda}$.  In analogy with \eqref{homKWb}, $\calf$ and $\calw$ satisfy the homogeneity conditions  
    \be
    \label{homFW}
    \calf(\lambda Z, \bar \lambda \bar Z, |\lambda|^\frac23 L) = |\lambda|^\frac23 \calf (Z,\bar Z, L)\,,\qquad
	\calw(\lambda  Z)=\lambda\calw(Z)\,,
    \ee
with $\lambda$ an arbitrary chiral superfield.

The bosonic components of \eqref{SWLina} can be extracted by using the method explained in \cite{Wess:1992cp}\footnote{We refer to \cite{Adamietz:1992dk,Binetruy:2000zx} for the definition of the components of the linear multiplet in supergravity.} 
	\be
	\label{SW_GF_Lag_LC}
	\begin{split}
		e^{-1} \call_{\rm bos} &= -\frac16\tilde \calf\, R - \calf_{a\bar b} D_\mu z^a \bar D^\mu \bar z^b  + \frac1{4} \calf_{\Lambda \Sigma} \del_\mu l^\Lambda \del^\mu l^\Sigma + \frac1{4\cdot 3!} \calf_{\Lambda \Sigma} \calh_{\mu\nu\rho}^\Lambda \calh^{\Sigma\mu\nu\rho}
		\\
		&\quad\,+ \left(\frac\ii{2\cdot 3 !} \calf_{\bar a \Sigma} \varepsilon^{\mu\nu\rho\sigma}\calh^{\Sigma}_{\nu\rho\sigma} D_\mu \bar z^a+ {\rm c.c.}\right) + \calf_{a\bar b} f^a \bar f^b +\left(\calw_a f^a+{\rm c.c.}\right) 
	\end{split}
	\ee
where we have redefined the auxiliary fields as in \eqref{SW_fdef} and defined $\tilde \calf = \calf - l^\Lambda \calf_\Lambda$. Now, the $U(1)$-covariant derivative is given by
	\be
	\label{SW_LC_cov}
	\begin{split}
		D_\mu z^a &= \del_\mu z^a + \ii A_\mu z^a\,,
		\\
		&\quad {\rm with}\quad A_\mu = \frac{3}{2(\tilde \calf -\tilde \calf_\Lambda l^\Lambda)} \left[\ii(\tilde\calf_{a}  \del_\mu z^a -  \bar{\tilde \calf}_{\bar a} \del_\mu \bar z^a ) + \frac{1}{3!} \tilde \calf_\Lambda \varepsilon_{\mu\nu\rho\sigma}\calh^{\Sigma\nu\rho\sigma} \right]\,.
	\end{split}
	\ee
The integration of the auxiliary fields $f^a$ is immediate and gives
	\be
	\label{SW_GF_Lag_LCb}
	\begin{split}
		e^{-1} \call_{\rm bos} &= -\frac16\tilde \calf\, R - \calf_{a\bar b} D_\mu z^a \bar D^\mu \bar z^b  + \frac1{4} \calf_{\Lambda \Sigma} \del_\mu l^\Lambda \del^\mu l^\Sigma + \frac1{4\cdot 3!} \calf_{\Lambda \Sigma} \calh_{\mu\nu\rho}^\Lambda \calh^{\Sigma\mu\nu\rho}
		\\
		&\quad\,+ \left(\frac\ii{2\cdot 3 !} \calf_{\bar a \Sigma} \varepsilon^{\mu\nu\rho\sigma}\calh^{\Sigma}_{\nu\rho\sigma} D_\mu \bar z^{\bar a}+ {\rm c.c.}\right) - \calf^{a\bar b} \calw_a \bar\calw_{\bar b} 
	\end{split}
	\ee
In order to fix the super-Weyl invariance, it is convenient to introduce a function $F(\phi,\bar\phi,\ell)$ which is the Legendre transform of the K\"ahler potential $K(\phi,\bar\phi,\Im t)$ related to the dual kinetic function $\calk$ as in \eqref{homKWb}, that is:
\be
F(\phi,\bar\phi,\ell)=K+2\ell^\Lambda \Im t_\Lambda
\ee
with $\ell^\Lambda=-\frac12\frac{\del K}{\del \Im T_\Lambda}$. More directly, the variables $(z^a,l^\Lambda)$ are related to the new variables $(u,\phi^i,\ell^\Lambda)$ by
    \be
    z^a = u g^a(\phi)\,,\qquad  l^\Lambda = |u|^\frac23 e^{-\frac13{\tilde F(\phi, \bar \phi, \ell)}} \ell^\Lambda\,,
    \ee
where 
\be
\tilde F \equiv F - \ell^\Lambda F_\Lambda\,.
\ee (Notice that it equals $K$.) The direct relation between $\calf$ and $F$ is somewhat convoluted and is given by:
\be
\calf(z,\bar z,L)=-3|u|^{\frac23}e^{-\frac13\tilde F}\left(1-\frac13\ell^\Lambda F_\Lambda\right)\,.
\ee
Notice that, after the above field redefinitions, the dual kinetic function $\calk$ can be identified with  $\tilde\calf=-3|u|^{\frac23}e^{-\frac13 \tilde F}$ and indeed the Einstein frame condition \eqref{SWgf} now becomes
	\be
	\label{SWgfb}
 u = e^{\frac12 \tilde F(\phi, \bar \phi,\ell)}\quad\Rightarrow \quad \tilde\calf=-3\,.
	\ee
We finally arrive, after integrating out the auxiliary fields $\tilde f^a$, at
	\be
	\label{SW_GF_Lag_LChb}
	\begin{split}
		e^{-1} \call_{\rm bos} &= \frac{1}2 R -F_{i\bar \jmath}  \del \phi^i \bar \del \bar \phi^{\bar\jmath} + \frac14 F_{\Lambda\Sigma}  \left( \del_\mu \ell^\Lambda \del^\mu \ell^\Sigma + \frac1{ 3!}  \calh_{\mu\nu\rho}^\Lambda \calh^{\Sigma\mu\nu\rho}\right)
		\\
		&\quad\,+ \left\{\frac\ii{2\cdot 3 !} F_{\bar \imath \Sigma} \varepsilon^{\mu\nu\rho\sigma}\calh^{\Sigma}_{\nu\rho\sigma} \del_\mu \bar \phi^{\bar\imath} + {\rm c.c.}\right\} - e^{\tilde F} \left[ F^{\bar\jmath i} D_i W \bar D_{\bar \jmath} \bar W- (3 - \ell^\Lambda \tilde F_\Lambda) | W|^2 \right]
	\end{split}
	\ee
where the K\"ahler covariant derivatives are now given by
\be
D_i = \del_i + \tilde F_i\,.
\ee	
The Lagrangian \eqref{SW_GF_Lag_LChb} matches with that of \cite{Grimm:2004uq} and is a generalization of the Lagrangian computed in \cite{Adamietz:1992dk} in presence of a single linear multiplet.\footnote{This kind of supergravity Lagrangians with both chiral and linear multiplets can also be obtained by using the tensor calculus in the superconformal approach. Our Lagrangian \eqref{SW_GF_Lag_LChb} matches with the one obtained in this alternative way. We are grateful to Ruben Monten and Toine Van Proeyen for discussions on this point.}

\subsection{With chiral and gauged linear multiplets}
\label{app:SWLc}

Finally, let us consider the more general case where some chiral superfields are endowed with gauge three-forms, namely they are constrained as in \eqref{chircalvconstr}, and also linear multiplets are present. The linear multiplets can also be gauged by the three-form potentials as in \eqref{Lgauge}. The most general Lagrangian including these ingredients (plus other hidden multiplets) is given by \eqref{32formlagr}.
However, in order to obtain its expression in bosonic components, it is convenient to start from the parent Lagrangian \eqref{plagr}. After integrating out $T_\Sigma$ from \eqref{plagr}, we arrive at a parent Lagrangian of the form\footnote{With respect to \eqref{plagr}, we have slightly modified the way in which the (anti-)chiral projectors and Grassmannian integrations appear. See \cite{Farakos:2017jme,Farakos:2017ocw,Bandos:2018gjp} and Appendix \ref{app:bdcontr} for more details about singling out this particular combination of projections and integration.} 
\be\label{SW_32formlagr}
\begin{split}
\call&=\int\d^4\theta\, E \calf(Z,\bar Z,\hat L)+\left(\int\d^2\Theta\,2\cale\,\hat\calw(Z)+\text{c.c.}\right)
\\
&\quad\,+ \left[\int \d^2\Theta\, 2\cale\, X_A\calv^A(Z)+\frac\ii8\int \d^2\Theta\, 2\cale\, (\bar \cald^2 - 8 \calr)  (X_A-\bar X_A)P^A  +\text{c.c.}\right]\,.
\end{split}
\ee
We recall that $X_A$ is a chiral superfield, with bosonic components $\{x_A, F^{(X)}_{A}\}$ and $\calv^A$ are homogeneous of degree one as in \eqref{homVA}. The homogeneity properties for $\calf$ and $\hat W$ are the same as in \eqref{homFW}. 

The bosonic components of the Lagrangian \eqref{SW_32formlagr} are 
	\be
	\label{SW_CLgauged}
	\begin{split}
		e^{-1} \call_{\rm bos} &= -\frac16\tilde \calf\,R - \calf_{a\bar b} D z^a \bar D \bar z^b  + \frac1{4} \calf_{\Lambda \Sigma} \del_\mu l^\Lambda \del^\mu l^\Sigma + \frac1{4\cdot 3!} \calf_{\Lambda \Sigma} \hat\calh_{\mu\nu\rho}^\Lambda \hat\calh^{\Sigma\mu\nu\rho}
		\\
		&\quad\,+ \left(\frac\ii{2\cdot 3 !} \calf_{\bar a \Sigma} \varepsilon^{\mu\nu\rho\sigma}\hat\calh^{\Sigma}_{\nu\rho\sigma} D_\mu \bar z^{\bar a}+ {\rm c.c.}\right) +
		\\
		&\quad\,+\calf_{a\bar b} f^a \bar f^b+\calf_{\Lambda \Sigma} c_A^\Lambda c_B^\Sigma \calv^A \bar \calv^B
		\\
		&\quad\,+\left\{ -\frac{\ii}{2} \calf_\Lambda c_A^\Lambda \calv^A_a f^a - \ii \calf_{\Lambda b}f^b c_A^\Lambda \calv^A+\hat\calw_a f^a+{\rm c.c.}\right\}
		\\
		&\quad\, + \Big[\left(F^{(X)}_{A}- \bar M x_A\right) \left(- s^A + \calv^A_a z^a\right) \\
		&\quad\quad\,+ x_A \calv^A_a F^a_Z - \frac\ii2 x _A d^A 
		- \frac1{2 \cdot 3! e} \varepsilon^{\mu\nu\rho\sigma} \del_\mu x_A  A^A_{\nu\rho\sigma}  - x_A \Re (\bar M s^A) + {\rm c.c.} \Big]\,,
	\end{split}
	\ee
where now $\hat \calh^\Lambda_{\mu\nu\rho} = \calh^\Lambda_{\mu\nu\rho} + c^\Lambda_A A_{\mu\nu\rho}^A$ and with the same covariant derivative as \eqref{SW_LC_cov}, modulo the replacement $\calh^\Lambda \to \hat \calh^\Lambda$. 

We now proceed as follows. First, we integrate out the real auxiliary fields $d^A$, which constrain $x_A$ to be real, and $F^{(X)}_{A}$, identifying $\calv^A(z) = s^A$ as in \eqref{conVS1}. Then, \eqref{SW_CLgauged} becomes
	\be
	\label{SW_CLgaugeda}
	\begin{split}
		e^{-1} \call_{\rm bos} &=- \frac16\tilde \calf\,R - \calf_{a\bar b} D z^a \bar D \bar z^b  + \frac1{4} \calf_{\Lambda \Sigma} \del_\mu l^\Lambda \del^\mu l^\Sigma + \frac1{4\cdot 3!} \calf_{\Lambda \Sigma} \hat\calh_{\mu\nu\rho}^\Lambda \hat\calh^{\Sigma\mu\nu\rho}
		\\
		&\quad\,+ \left(\frac\ii{2\cdot 3 !} \calf_{\bar a \Sigma} \varepsilon^{\mu\nu\rho\sigma}\hat\calh^{\Sigma}_{\nu\rho\sigma} D_\mu \bar z^{\bar a}+ {\rm c.c.}\right)
		\\
		&\quad\,+\calf_{a\bar b} f^a \bar f^b+\calf_{\Lambda \Sigma} c_A^\Lambda c_B^\Sigma \calv^A \bar \calv^B
		\\
		&\quad\,+\left\{ -\frac{\ii}{2} \calf_\Lambda c_A^\Lambda \calv^A_a f^a - \ii \calf_{\Lambda b}f^b c_A^\Lambda \calv^A+\hat\calw_a f^a+{\rm c.c.}\right\}
		\\
		&\quad\, + \Big[ x_A \calv^A_a f^a 
		- \frac1{2 \cdot 3! e} \varepsilon^{\mu\nu\rho\sigma} \del_\mu x_A\,  A^A_{\nu\rho\sigma} + {\rm c.c.} \Big]\,.
	\end{split}
	\ee
 Then, we further integrate out the auxiliary fields $f^a$ of the chiral multiplets $Z^a$ via
	\be
		\bar f^b = - \calf^{a\bar b} \left(x_A \calv^A_a - \frac{\ii}{2} \calf_\Lambda c_A^\Lambda \calv^A_a - \ii \calf_{a\Lambda} c_E^\Lambda \calv^E+\hat\calw_a  \right)
	\ee
and, subsequently, the real Lagrange multipliers $x_A$. We then arrive at the super-Weyl invariant Lagrangian
	\be
	\label{SW_CLgaugedb}
	\begin{split}
		e^{-1} \call_{\rm bos} &= -\frac16\tilde \calf\,R - \calf_{a\bar b} D z^a \bar D \bar z^b  + \frac1{4} \calf_{\Lambda \Sigma} \del_\mu l^\Lambda \del^\mu l^\Sigma + \frac1{4\cdot 3!} \calf_{\Lambda \Sigma} \hat\calh_{\mu\nu\rho}^\Lambda \hat\calh^{\Sigma\mu\nu\rho}
		\\
		&\quad\,+ \left(\frac\ii{2\cdot 3 !} \calf_{\bar a \Sigma} \varepsilon^{\mu\nu\rho\sigma}\hat\calh^{\Sigma}_{\nu\rho\sigma} D_\mu \bar z^{\bar a}+ {\rm c.c.}\right)+ e^{-1} \call_{\text{three-forms}}\,,
	\end{split}
	\ee
where the three-form Lagrangian $\call_{\text{three-forms}}$ can be recast as in \eqref{dualF4lagr}, with
		\begin{subequations}\label{ThVdefgl}
			\begin{align}
				T^{AB}(z,\bar z)&\equiv 2\,\Re\left(\calf^{\bar b a}\,\calv_a^A\bar\calv_{\bar b}^B\right)\,,\label{ThVdef1gl}
				\\
				h^A(z,\bar z)&\equiv 2\Re \left[\calf^{\bar b a} \left(\bar{\hat\calw}_{\bar b} + \frac{\ii}{2} \calf_\Lambda c_{B}^\Lambda \bar\calv^B_{\bar b} + \ii \calf_{\Lambda \bar b} c_F^\Lambda \bar \calv^{\bar F} \right) \calv_{a}^A \right]\,,\label{ThVdef2gl}
			\\
			\begin{split}
				\hat V(z,\bar z)&\equiv \calf^{\bar b a} \left(\hat\calw_a - \frac{\ii}{2} \calf_\Lambda c_A^\Lambda  \calv^A_a - \ii \calf_{\Lambda a} c_E^\Lambda \calv^E \right) \left(\bar{\hat\calw}_{\bar b} + \frac{\ii}{2} \calf_\Lambda c_{B}^\Lambda \bar\calv^B_{\bar b} + \ii \calf_{\Lambda \bar b} c_F^\Lambda \bar \calv^{\bar F} \right)
				\\
				&\quad\, - \calf_{\Lambda \Sigma} c_A^\Lambda c_B^\Sigma \calv^A \bar \calv^B\,.\label{ThVdef3gl}
			\end{split}
			\end{align}
		\end{subequations}
We can now proceed to gauge-fixing the super-Weyl invariance by following the same procedure described in the previous subsection. After having imposed the Einstein frame condition  \eqref{SWgfb}, we arrive at
	\be
	\label{SW_GF_Lag_LChg}
	\begin{split}
		e^{-1} \call_{\rm bos} &= \frac{1}2 R  -F_{i\bar \jmath}  \del \phi^i \bar \del \bar \phi^{\bar\jmath} + \frac14 F_{\Lambda\Sigma}  \left( \del_\mu \ell^\Lambda \del^\mu \ell^\Sigma + \frac1{ 3!}  \hat\calh_{\mu\nu\rho}^\Lambda \hat\calh^{\Sigma\mu\nu\rho}\right)
		\\
		&\quad\,+ \left\{ \frac\ii{2\cdot 3 !} F_{\bar \imath \Sigma} \varepsilon^{\mu\nu\rho\sigma}\hat\calh^{\Sigma}_{\nu\rho\sigma} \del_\mu \bar \phi^{\bar\imath} + {\rm c.c.}\right\} + e^{-1} \call_{\text{three-forms}}\,,
	\end{split}
	\ee
where the three-form Lagrangian has the same form as \eqref{dualF4lagr} with
\begin{subequations}\label{ThVdefgl_wf}
		\begin{align}
			T^{AB}&\equiv 2e^{\tilde F}\,\Re\left({F}^{i \bar \jmath}\,D_i  \Pi^A \bar D_{\bar \jmath} \bar\Pi^B-(3 - \ell^\Lambda \tilde F_\Lambda) \Pi^A\bar\Pi^B\right)\,,\label{ThVdef1gl_wf}
			\\
			\begin{split}
				h^A&\equiv 2e^{\tilde F}\Re \Bigg[{F}^{i \bar \jmath}\,\left( D_{\bar \jmath}  \bar{\hat W} + \frac{\ii}{2} F_\Lambda c_{B}^\Lambda D_{\bar \jmath} \bar \Pi^B + \ii F_{\Lambda \bar \jmath} c_F^\Lambda \bar \Pi^{\bar F}\right) D_i\Pi^A 
				\\
				&\qquad\qquad\, - (3 - \ell^\Lambda \tilde F_\Lambda) \left(\overline{\hat W} + \frac\ii2 F_\Lambda c_D^\Lambda \bar\Pi^D\right) \Pi^A - \ii \tilde F_{\Lambda} c_F^\Lambda \bar \Pi^{\bar F} \Pi^A\Bigg]\,,\label{ThVdef2gl_wf}
			\end{split}
			\\
			\begin{split}
				\hat V&\equiv e^{\tilde F} {F}^{\bar \jmath i} \left( D_{i}  {\hat W} - \frac{\ii}{2} F_\Lambda c_{A}^\Lambda D_{i} \Pi^A - \ii F_{\Lambda i} c_F^\Lambda \Pi^{F} \right) \left(D_{\bar \jmath}  \bar{\hat W} + \frac{\ii}{2} F_\Lambda c_{B}^\Lambda D_{\bar \jmath} \bar \Pi^B + \ii F_{\Lambda \bar \jmath} c_F^\Lambda \bar \Pi^{\bar F} \right)
				\\
				&\quad\,- (3 - \ell^\Lambda \tilde F_\Lambda) e^{\tilde F} \left|\hat W - \frac\ii2 F_\Lambda c_A^\Lambda \Pi^A \right|^2 - e^{\tilde F} F_{\Lambda \Sigma} c_A^\Lambda c_B^\Sigma \Pi^A \bar \Pi^B
				\\
				&\quad\,+e^{\tilde F}\left[ \ii c_F^\Lambda \tilde F_\Lambda \Pi^F \left( \bar{\hat W} + \frac\ii2 F_\Lambda c_A^\Lambda \bar\Pi^A \right)+{\rm c.c.}\right] .\label{ThVdef3gl_wf}
			\end{split}
		\end{align}
\end{subequations}

\section{Boundary terms and component actions}
\label{app:bdcontr}

As stressed in \cite{Brown:1987dd,Groh:2012tf,Farakos:2017jme}, the actions which contain gauge three-forms $A_3^A$ should be properly equipped with boundary terms. These are necessary to set a consistent variational principle for the gauge three-forms, with the gauge invariant boundary conditions $\delta F_4^A|_{\rm bd} = 0$. In \cite{Farakos:2017jme,Bandos:2018gjp} it was shown how to compute these boundary contributions from the Lagrangian term \eqref{newterm}. 
To this end let us rewrite it as
\be\label{LPhiXr}
\begin{aligned}
\int \d^2\theta\, X_A\calv^A(Z)+\frac\ii8\int \d^2\theta \bar D^2  (X_A-\bar X_A)P^A  +\text{c.c.}
\end{aligned}
\ee
After imposing \eqref{chircalvconstr} and using \eqref{defSA}, the Lagragian \eqref{LPhiXr} becomes a pure boundary term 
    \be\label{BT_susybd}
    \call_{\rm bd}=-\frac\ii8\left(\int\d^2\theta\bar D^2-\int\d^2\bar\theta D^2\right)X_AP^A+\text{c.c.}
    \ee
In this formula $X_A$ should be regarded as composite chiral superfields determined by the equations of motion of $Z^a$ obtained from \eqref{SUGRA} and \eqref{newterm} 
    \be\label{BT_Thetadef}
    X_A\calv^A_a(\calz)=\left(\frac14\bar D^2\calk_a(\calz,\bar \calz)-\hat\calw_a(\calz)\right)\,.
    \ee
In order to explicitly show the structure of the boundary terms in \eqref{BT_susybd}, let us only consider the bosonic components. Then, \eqref{BT_susybd} reduces to
    \be
    \call_{\rm bd} = \frac18 \sigma_{\alpha \dot \alpha}^n \del_n \left\{ [D^\alpha, \bar D^{\dot\alpha}]  \left( X_A P^A  \right) \right\} + {\rm c.c.}
    \ee
where it is understood that the components are evaluated at $\theta=\bar\theta=0$ and all fermionic fields are set to zero. These are manifestly boundary terms, but they include many more terms in addition to those which are necessary for the correct definition of the variational problem with respect to the three-forms $A_3^A$. The only relevant terms are those which \emph{explicitly} include $A_3^A$ in the boundary Lagrangian contained in the $\theta$-expansion \eqref{VectorB} of $P^A$
    \be
    \call_{\rm bd} = \frac18 \sigma_{\alpha \dot \alpha}^n \del_n \left( X_A [D^\alpha, \bar D^{\dot\alpha}]  P^A \right) + {\rm c.c.}  = -  \del_n \left[ \Re X_A (* A_3^A)^n \right] \,, 
    \ee
because these are the only ones which do not trivially vanish at the boundary. By reinserting the model-dependent explicit expression for $X_A$ in \eqref{BT_Thetadef}
    \be\label{BT_Thetadefb}
    X_A\calv^A_a(\calz)=-\calk_{a \bar b}(z,\bar z) \bar f^{\bar b}-\hat\calw_a(z)\,,
    \ee
we may fully re-express the boundary contributions in terms of the known EFT quantities.


\end{appendix}






\providecommand{\href}[2]{#2}\begingroup\raggedright\endgroup


\end{document}